\documentclass[fleqn,usenatbib]{mnras}

\usepackage{newtxtext,newtxmath}

\usepackage[T1]{fontenc}

\DeclareRobustCommand{\VAN}[3]{#2}
\let\VANthebibliography\thebibliography
\def\thebibliography{\DeclareRobustCommand{\VAN}[3]{##3}\VANthebibliography}

\usepackage{graphicx}	
\usepackage{amsmath}
\usepackage{subcaption}
\usepackage{hyperref}
\usepackage{color}
\usepackage{array,xcolor,arydshln,tabularx}

\newcommand\VRule[1][\arrayrulewidth]{\color{lightgray}\vrule width #1\color{black}}

\title[Searching for dust spirals]{Searching for planet-driven dust spirals in ALMA visibilities} 

\author[E. T. Stevenson et al.]{
Edward T. Stevenson,$^{1}$\thanks{E-mail: es833@cam.ac.uk}
Álvaro Ribas,$^{1}$
Jessica Speedie,$^{2}$
Richard A. Booth$^{3}$
and Cathie J. Clarke$^{1}$
\\
$^{1}$Institute of Astronomy, University of Cambridge, Madingley Road, Cambridge CB3 0HA, UK\\
$^{2}$Department of Physics and Astronomy, University of Victoria, Victoria, BC V8P 1A1, Canada\\
$^{3}$School of Physics and Astronomy, University of Leeds, Leeds LS2 9JT, UK
}

\date{Accepted XXX. Received YYY; in original form ZZZ}

\pubyear{2023}

\begin{document}
\label{firstpage}
\pagerange{\pageref{firstpage}--\pageref{lastpage}}
\maketitle


\begin{abstract}
ALMA (Atacama Large Millimetre/submillimetre Array) observations of the thermal emission from protoplanetary disc dust have revealed a wealth of substructures that could evidence embedded planets, but planet-driven spirals, one of the more compelling lines of evidence, remain relatively rare. Existing works have focused on detecting these spirals using methods that operate in image space. 
Here, we explore the planet detection capabilities of fitting planet-driven spirals to disc observations directly in visibility space.  
We test our method on synthetic ALMA observations of planet-containing model discs for a range of disc/observational parameters, finding it significantly outperforms image residuals in identifying spirals in these observations and is able to identify spirals in regions of the parameter space in which no gaps are detected. 
These tests suggest that a visibility-space fitting approach warrants further investigation and may be able to find planet-driven spirals in observations that have not yet been found with existing approaches.
We also test our method on six discs in the Taurus molecular cloud observed with ALMA at 1.33\,mm, but find no evidence for planet-driven spirals. We find that the minimum planet masses necessary to drive detectable spirals range from \(\approx 0.03\) to \(0.5 \, M_{\text{Jup}}\) over orbital radii of 10 to 100\,au, with planet masses below these thresholds potentially hiding in such disc observations.
Conversely, we suggest that planets $\gtrsim$ 0.5 to 1\,$M_{\text{Jup}}$ can likely be ruled out over orbital radii of \(\approx 20\) to \(60 \, \text{au}\) on the grounds that we would have detected them if they were present.  
\end{abstract}



\defcitealias{speedieSpirals2022}{S22}


\section{Introduction}

During its formation, a planet gravitationally interacts with its natal protoplanetary disc and generates substructure that is potentially observable. For instance, every planet drives a wake in its disc, which is sheared into spiral arms by the disc's differential rotation \citep{kleyPlanetDiskInteraction2012}. For larger planets, these spiral arms steepen into shocks as they propagate and can eventually lead to the opening of annular gaps in the disc \citep{linTidalInteraction1986,goodmanPlanetaryTorques2001}. To what extent these substructures can be identified in observations has been explored across various regimes, including: near-IR scattered light images (e.g., \citealp{dongObservationalSignatures2015,juhaszSpiralArms2015, dongHOWBRIGHT2017}), (sub-)mm dust thermal emission \citep{weberPredictingObservational2019,speedieSpirals2022,binkertThreedimensionalDust2023}, and CO emission lines \citep{teagueKinematicalDetection2018, teagueEvidenceVertical2018, baeObservationalSignature2021}. 
Through detection of these substructures, there is the possibility of inferring the presence of young planets and constraining their demographics, which, in turn, will serve to test planet formation theories. 

Exoplanet surveys have made it clear that planets are abundant \citep{zhuExoplanetStatistics2021}, and spirals are a robust prediction of planet-disc interactions \citep{goldreichDisksatelliteInteractions1980}. Therefore, one would expect a large number of planet-driven spirals in protoplanetary discs. This theoretical expectation has not, however, translated into a large number of clear detections. The majority of spiral detections to date have come from scattered light observations of sub-$\mu\text{m}$ grains (e.g., \citealp{fukagawaSpiralStructure2004, mutoDiscoverySmallscale2012}). Spiral features have also more recently been observed in CO emission lines that trace gas kinematics (e.g., \citealp{christiaensSpiralArms2014, tangPlanetFormation2017, teagueSpiralStructure2019}). Continuum dust emission, which has revealed plenty of annular structure, i.e., gaps and rings (e.g., \citealp{almapartnership2015, andrewsRINGEDSUBSTRUCTURE2016, andrewsDiskSubstructures2018, clarkeHighresolutionMillimeter2018, huangDiskSubstructures2018, huangMultifrequencyALMA2020, kudoSpatiallyResolved2018, kepplerHighlyStructured2019, longGapsRings2018, maciasCharacterizingDust2021}), has additionally delivered a small number of spiral detections. Among these, the discs around single stars that display spiral patterns in dust emission include Elias 2-27, IM Lupi, WaOph 6, and HD 143006, which have been found to contain large-scale dust spirals \citep{perezSpiralDensity2016, huangDiskSubstructures2018a, andrewsLimitsMillimeter2021}, and a smaller scale dust spiral found in MWC 758 \citep{dongEccentricCavity2018}. These spirals have not been definitively attributed to planets, with \citet{mawetDirectImaging2012} finding no evidence for an embedded companion in their direct imaging analysis of IM Lupi. In the case of Elias 2-27, \citet{meruOriginSpiral2017} suggested that the spirals could be consistent with gravitational instability or a $10\text{--}13\,M_{\text{Jup}}$ companion located at a large orbital radius external to the spirals (between $ \approx 300\text{--}700\,\text{au}$), but considered the spirals unlikely to be due to a planet internal to the spirals. Similarly, \citet{brown-sevillaMultiwavelengthAnalysis2021} found the spirals in WaOph 6 to require a planet of minimum mass $10\,M_{\text{Jup}}$ exterior to the spirals (at orbital radii $>100\,\text{au}$), and \citet{dongSpiralArms2018} found the spirals in MWC 758 could be consistent with a planet of a few $M_{\text{Jup}}$ exterior to the spirals at $\approx 100\,\text{au}$. The faint spiral in HD 143006 reported by \citet{andrewsLimitsMillimeter2021} lends support for the $10\text{--}20\,M_{\text{Jup}}$ embedded companion already suggested by other forms of evidence such as CO kinematics \citep{perezDiskSubstructures2018} and asymmetries in scattered light \citep{benistyShadowsAsymmetries2018}. Therefore, the few spirals found could only have a planetary origin for extremes of mass or orbital radius, and no spirals attributable to more moderate planet masses have yet been found in the mm continuum. 

The extent to which this scarcity of planet-driven dust spirals is genuine, or a result of observational limitations, is currently poorly understood.
In addition, it is unclear how reliably dust spirals could be detected across the disc parameter space and interferometric set-ups. \citet{speedieSpirals2022} (henceforth \citetalias{speedieSpirals2022}) shed light on these uncertainties by using image residuals and contouring to analyse disc asymmetries, applied to synthetic ALMA observations of planet-containing model discs. 

In this work, we build on \citetalias{speedieSpirals2022} by exploring a different approach to analysing disc asymmetries that specifically targets planet-driven spirals, and aims to test the limits of planet recovery with spirals in ALMA observations. Existing works with efforts to highlight or detect spiral structures in ALMA continuum images have used a variety of approaches: unsharp masking (e.g., \citealp{perezSpiralDensity2016, meruOriginSpiral2017}), high-pass filtering (e.g., \citealp{rosottiSpiralArms2020, norfolkOriginDoppler2022}), image contouring (e.g., \citealp{jenningsSuperresolutionTrends2022}), and axisymmetric brightness subtraction (e.g., \citealp{andrewsLimitsMillimeter2021, jenningsSuperresolutionTrends2022}; \citetalias{speedieSpirals2022}; \citealp{speedieTestingVelocity2022}). While some of these approaches perform the suppression of axisymmetric emission in image space and others in visibility space, all of them lead to the process of identifying spiral structure being done in image space.
These image-space methods therefore require transformation from the observed visibilities into an image, generally performed with the \texttt{CLEAN} deconvolution algorithm \citep{hogbomApertureSynthesis1974, clarkEfficientImplementation1980, cornwellMultiscaleCLEAN2008}. Although \texttt{CLEAN} is the standard and clearly effective technique used across radio astronomy, it suffers the well-known drawbacks of: information loss, primarily from the convolution of the \texttt{CLEAN} model with the \texttt{CLEAN} beam, and correlated noise with ill-defined uncertainties.

Fitting models via comparison between synthesised and observed visibilities is therefore preferable to model fitting in the image plane since observational data is acquired in the visibility plane and its uncertainties are well understood.
Here we develop a method that implements this approach and serves as a proof-of-concept, providing insights into its potential uses and challenges.

We test our new method on (an extended version of) the set of synthetic observations of planet-containing discs from \citetalias{speedieSpirals2022}, and compare its planet recovery ability to looking for spirals in \texttt{CLEAN} image residuals (the principal spiral detection method tested in \citetalias{speedieSpirals2022}), and looking for gaps in \texttt{CLEAN} image profiles. We also test the method on a sample of six discs (Table~\ref{tab:disc properties}) from the ALMA survey of the 1.33\,mm dust thermal emission of 32 discs in the Taurus star-forming region ($\sim 0.12''$ resolution, $50\,\mu \text{Jy}/\text{beam}$ mean sensitivity) presented in \citet{longCompactDisks2019}. These discs were chosen for their `smoothness' (relative lack of annular substructure) and for easy comparison with the analysis of this sample undertaken in \citet{jenningsSuperresolutionTrends2022}, who used both image contouring and axisymmetric brightness subtraction (performed in visibility space) to analyse the disc asymmetries. 

\S\ref{Sec:Methods} describes our method.
\S\ref{sec:results synthetic observations} and \S\ref{sec:results Taurus} present and analyse the main results of our tests of the method on the synthetic observations and Taurus disc sample respectively. 
\S\ref{sec:Discussion} discusses the significance of these results and suggests avenues for further work. 
We summarise and conclude in \S\ref{sec:Conclusions}.

\section{Method}
\label{Sec:Methods}

\begin{figure}
    \includegraphics[width=\columnwidth]{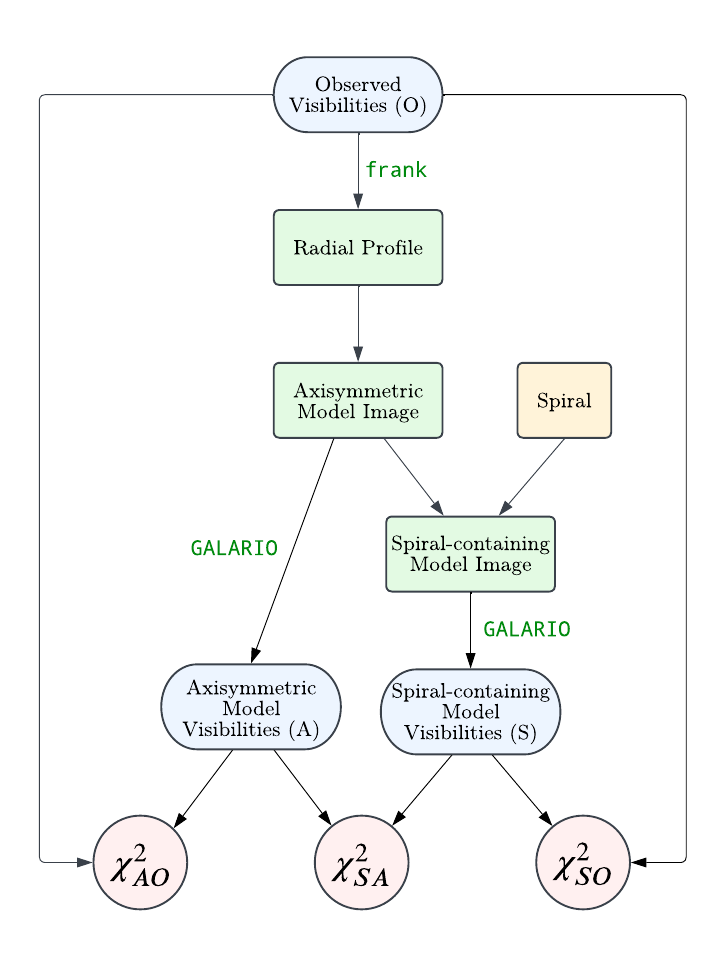}
    \caption{Flow chart showing the key steps of the method, including how the $\chi^2$ metrics are generated. The steps going from the observed visibilites to the axisymmetric model image are described in \S\ref{sec:observed Vis to axisym model}. How the spiral is combined with the axisymmetric model image to create the spiral-containing model image is described in \S\ref{sec:adding the spiral}. The transformation to visibility space and $\chi^2$ metrics are explained in \S\ref{sec:spiral fits}.}
    \label{fig:42_2}
\end{figure}

We require two components to apply our method to a given disc observation (whether real or synthetic): an axisymmetric model image of the disc (\S\ref{sec:observed Vis to axisym model}) and a spiral perturbation driven by a planet (\S\ref{sec:adding the spiral}). Combining these elements, we construct a model disc image that incorporates the planet's influence. We then transform to visibility space and use $\chi^2$ statistics to compare the axisymmetric and spiral-containing model to the observed data. Repeating this for different planet positions and spiral simulation parameters, we use the $\chi^2$ metrics to determine which parameters give the best fits to the observed visibilities and interpret the results using the Bayesian Information Criterion (BIC) and more qualitative methods (\S\ref{sec:spiral fits}). The steps of the method, including how the $\chi^2$ metrics are generated, are summarised visually in Figure~\ref{fig:42_2}. 

\subsection{Creating the axisymmetric model image} 
\label{sec:observed Vis to axisym model}

First, we use \texttt{frank} \citep{jenningsFrankensteinProtoplanetary2020} to reconstruct the disc's 1D radial profile. \texttt{frank} fits the observed visibilities directly to construct a non-parametric, axisymmetric model of the disc's brightness. 

We then generate an axisymmetric model image $I_A$ from this radial profile by interpolating it and sweeping over $2\pi$ radians, making sure to use a large enough image plane to avoid aliasing, and small enough pixels to Nyquist sample the \textit{uv}-points.


Next, we partially reintroduce the disc's on-sky projection, inclining the model but forgoing its rotation or offset. The justification for this is that the inclination of the disc affects its total observable flux; here we take the optically thick limit, in which the total flux is reduced by a factor $\cos i$ for inclination $i$. By contrast, neither rotation nor offset affect total flux, and both can be implemented in visibility space, rather than image space, with better performance and accuracy \citep{Briggs1999}. We thus defer their inclusion to the final step of generating the model visibilities.

\subsection{Adding the spiral} \label{sec:adding the spiral}

To introduce the spiral into the model, we employ an extended version of the set of 2D gas + dust hydrodynamic simulations generated in \citetalias{speedieSpirals2022}. 
In brief, the simulations were performed with \texttt{FARGO3D} \citep{benitez-llambayFARGO3DNEW2016} 
modified to compute dust dynamics \citep{rosottiMinimumMass2016, boothSmoothedParticle2015}, with a planet placed at $r_{\rm p}=50\,$ au. 
The 2D cylindrical $(r, \phi)$ domain spans $2\pi$ in azimuth and $0.1 \, r_{\rm p}$ to $3.0 \, r_{\rm p}$ in radius, with a resolution of $N_{r} \times N_{\phi} = 1100 \times 2048$ cells (spaced logarithmically and linearly in the $r$ and $\phi$ directions, respectively). 
Wave damping zones \citep{deval-borro2006} were employed to minimise wave reflections. The dust layer was given open/inflow radial boundary conditions. 
The equation of state of the gas was varied to be locally isothermal or approximately adiabatic; in the latter case, the energy equation was solved using a $\beta$-cooling prescription (see below). 
The simulations were evolved to 1500 orbits, after which the disc was radially truncated to $0.2 \, r_{\rm p}$ to $2.2 \, r_{\rm p}$ in order to remove the damping zones, and radiative transfer was performed on the outputs analytically. 
In the approximately adiabatic case, this was done directly with the dust surface density and temperature distributions, assuming $T_{\rm dust}=T_{\rm gas}$; in the isothermal case, an axisymmetric temperature distribution was generated. 
We refer the reader to \S2.1 and \S2.2 of \citetalias{speedieSpirals2022} for details.

\begin{figure}
    \includegraphics[width=\columnwidth]{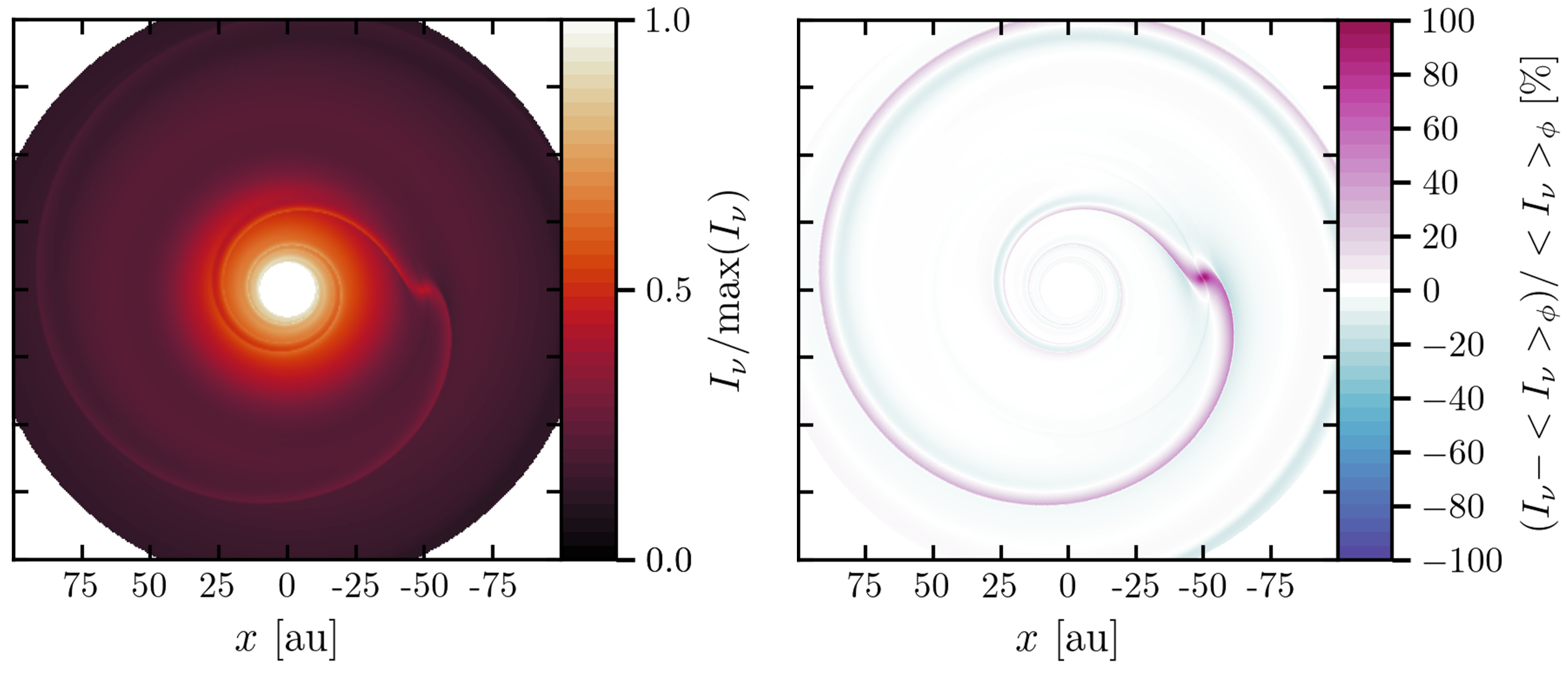}
    \caption{Simulation of a $3\,M_{\text{th}}$ planet in an approximately adiabatic ($\beta=10$), marginally optically thick ($\tau_0=1$) disc. Left: brightness map. Right: corresponding residual brightness map in percentage units of the azimuthally averaged brightness.}
    \label{fig:swirls}
\end{figure}

The result of these steps 
is a continuum brightness map 
of a disc containing a planet (left of Fig.~\ref{fig:swirls}). 
To isolate the the planet's asymmetric influence, we subtract the map's azimuthally averaged brightness and normalise by it, giving us a fractional residual brightness map (right of Fig.~\ref{fig:swirls}).\footnote{Throughout this work we refer to the planet's asymmetric influence simply as a `spiral', though note it can include additional, non-linear structure, particularly for simulations of planets above a thermal mass.}
We then interpolate this map onto the same cartesian grid as the axisymmetric model image. This allows the spiral perturbation and axisymmetric model to be superimposed, giving us our spiral-containing model disc image. Mathematically, 

\begin{equation}
I_S = I_A \left(1+ \frac{I_\text{sim}-\left<I_\text{sim}\right>_{\!\phi}}{\left<I_\text{sim}\right>_{\!\phi}} \right) = I_\text{sim} \times \frac{I_A}{\left<I_\text{sim}\right>_{\!\phi}}
\label{eq:I_S}
\end{equation}
where $I_A$ is the axisymmetric disc image from \texttt{frank}, $(I_\text{sim}-\left<I_\text{sim}\right>_{\!\phi})/\left<I_\text{sim}\right>_{\!\phi}$ is the fractional residual map of a spiral simulation, and $I_S$ is our resulting spiral-containing model disc image. This procedure is equivalent to rescaling the brightness map by our disc's axisymmetric profile, as indicated by the second equality.

\begin{figure}
    \includegraphics[width=\columnwidth]{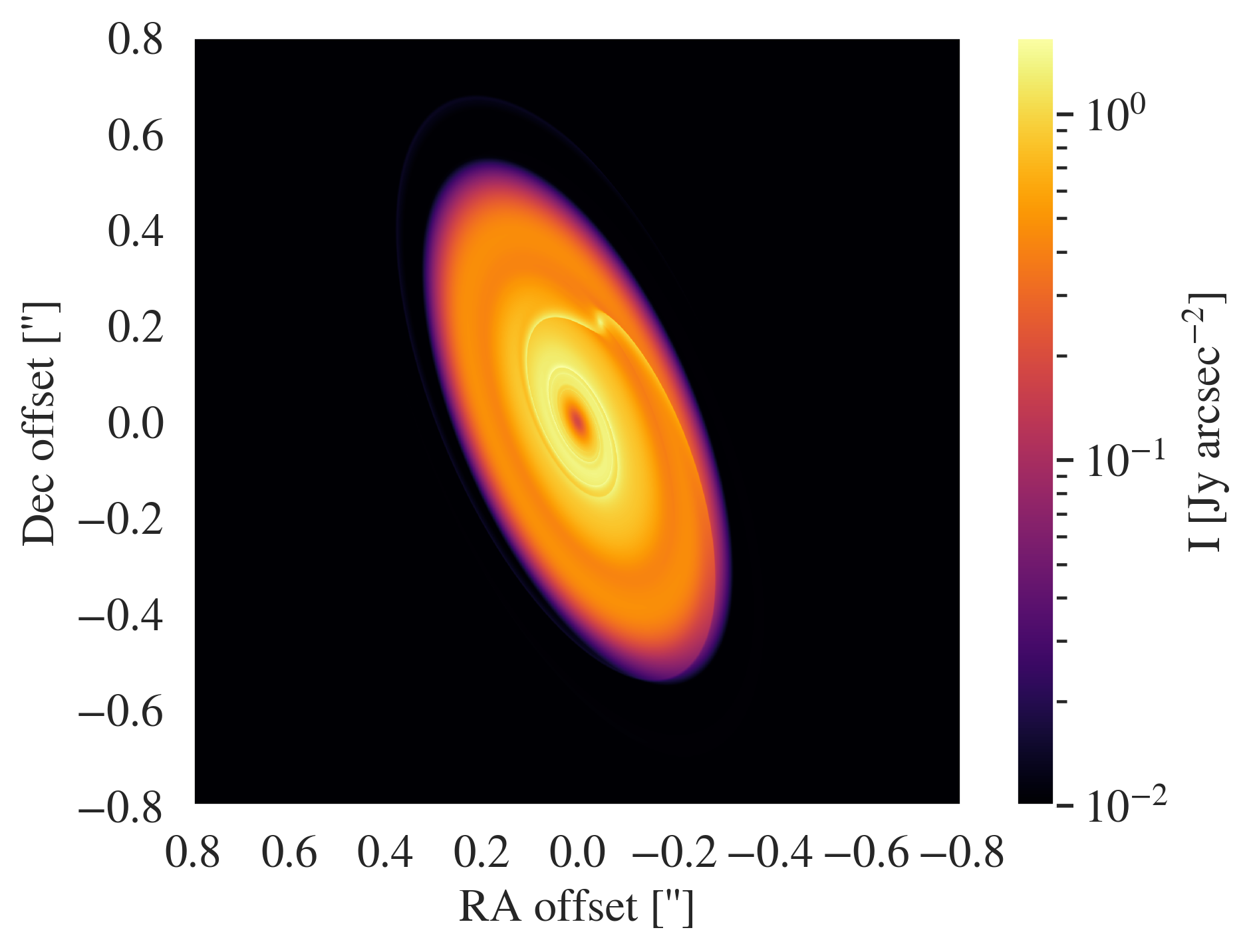}
    \caption{Projected disc model of RY Tau containing a $1\,M_{\text{th}}$ planet at $r_\text{p}=0.34''$, $\phi_\text{p}=30^\circ$. $\text{Inclination}=65.0^\circ$, $\text{position angle}=23.1^\circ$.}
    \label{fig:pretty}
\end{figure}

\paragraph*{Simulation parameters:}
\begin{enumerate}
    \item \textbf{Planet mass}, \(M_\text{p}\). This can take values \(0.1\), \(0.3\), \(1\) and \(3\,M_{\text{th}}\), where the thermal mass\footnote{Physically, the thermal mass corresponds to the mass at which the planet's Hill radius is equal to $3^{-1/3}$ times} the local scale height of the gas disc. \(M_{\text{th}} =  (H/r)_\text{p}^3 M_*\) with \(r_\text{p}\) the planet orbital radius, \(H\) the local scale height, and \(M_*\) the stellar mass. For these simulations, we fix \(r_\text{p} = 50\,\)au, with \((H/r)_\text{p} = 0.07\) and \(M_* = 0.8\,M_{\odot}\), giving a range of \(0.1\,M_{\text{th}} \approx 0.5\,M_{\text{Nep}}\) to \(3\,M_{\text{th}} \approx 0.9\,M_{\text{Jup}}\) at this radius. 
    The disc aspect ratio $H/r$ affects the spiral morphology \citep{baePlanetdrivenSpiral2018}, and therefore we investigate its effects on our method in \S\ref{sec:effect of spiral model parameters}.
    The planet mass parameter has the strongest effect on spiral morphology in these simulations.
    
    \item \textbf{Optical depth.} This is parameterised by the initial (i.e., after zero orbits) optical depth at the planet's orbital radius (50\,au) \(\tau_0\). For these simulations, $\tau_0 = 0.1$, 0.3, 1 or 3, thus ranging from optically thin through to optically thick. This optical depth is coupled to the observational wavelength, for which we consider two values in this work: $0.87$ and $1.33$\,mm. We scale the brightness of the simulations to ensure the optical depths are correct for both wavelengths.
    
    \item \textbf{Cooling time.} The gas spiral morphology and amplitude depends on the cooling timescale of the disc gas, which then affects the dust spiral via dust-gas coupling \citep{mirandaGapsRings2020, zhangEffectsDisc2020}. We parameterise the cooling timescale by $\beta$, where $t_\text{cool}=\beta/\Omega$ and $\Omega$ is the local angular speed. We use two $\beta$ values to explore the extremes: $\beta=0$ for isothermal, and $\beta=10$ for approximately adiabatic. As a shorthand, we often refer to approximately adiabatic simply as adiabatic (this terminology is consistent with \citetalias{speedieSpirals2022}).

    \item \textbf{Spiral handedness.} Because it is not generally possible tell in which direction the disc is rotating -- which determines the winding sense of the spiral -- from our observations, it is necessary to introduce a handedness parameter for the spiral sense. We define the spiral handedness such that the spirals shown in Figures~\ref{fig:swirls} and \ref{fig:pretty} are right-handed.
\end{enumerate}

\noindent In total, this gives \(4 \times 4 \times 2 \times 2 = 64\) possible spiral simulations. We then need two parameters to specify the planet's location in the disc: its orbital radius \(r_\text{p}\) and its azimuthal angle \(\phi_\text{p}\), which is defined anticlockwise from the right arm of the disc's minor axis following deprojection (assuming conventional sky coordinates). Hence, for any given spiral, we have six parameters: four simulation parameters ($M_\text{p}$, $\tau_0$, $\beta$, and handedness) and two positional parameters (\(r_\text{p}\) and \(\phi_\text{p}\)).
In the rest of this work, we refer to a given set of these six parameters as a spiral model instance. We use the term `spiral templates' to refer to spiral model instances in the context of their use fitting disc observations in our method (via their residual brightness maps).

Note that although we consider different \(r_\text{p}\) for our spiral templates, we only use spiral \textit{simulations} taken at a fixed $r_\text{p}\ (=50\,\text{au})$. We assume the simulations are scale-free, meaning that the brightness perturbation due to the spiral depends only on the ratio \(r/r_\text{p}\). This allows us to scale \(r_\text{p}\) for a template without running a new spiral simulation. However, because the magnitude of the spiral brightness perturbation scales approximately with the ratio \( M_\text{p} / M_{\text{th}} \) \citep{zhuSTRUCTURESPIRAL2015}, and we keep the brightness perturbation amplitude constant with \( r_\text{p} \) in our templates, we must also keep \( M_\text{p} / M_{\text{th}} \) constant with \( r_\text{p} \). Therefore, since the simulations assume a disc flaring of \( H/r \propto r^{1/4} \), implying \( M_{\text{th}} \propto r_\text{p}^{3/4} \), this means \( M_\text{p} \propto r_\text{p}^{3/4} \). That is, the absolute planet mass of a given template changes with \( r_\text{p} \).

As a checkpoint, by the end of this step we have generated our model image, consisting of a spiral template superimposed with an axisymmetric model of the disc (the disc is inclined but not yet rotated or offset to match the real disc observation, which is performed in visibility space). 
Figure~\ref{fig:pretty} shows a representation of one of these model discs (displayed at a higher resolution and with an on-sky geometry matching the real disc observation).

\subsection{Detecting a spiral} \label{sec:spiral fits}

The visibilities of a given model disc image are generated via the 2D Fourier transform, incorporating the necessary rotation and offset to match the real disc observation. We achieve this using \texttt{GALARIO} \citep{tazzariGALARIOGPU2018}.
For clarity, we introduce a shorthand for the three visibilities we are working with: $O$ for the Observed data, $A$ for those of the Axisymmetric model, and $S$ for those of the Spiral model instance. 

We use $\chi^2$ metrics to assess the goodness-of-fit, where the $\chi^2$ taken between a pair of visibilities $P$ and $Q$ is defined as 
\begin{equation}
\chi_{PQ}^2 = \sum_k w_k |V_{P,k} - V_{Q,k}|^2
\end{equation}
where $w_k = 1/\sigma_k^2$ is the weight of the $k$-th visibility point and $\sigma_k^2$ is the variance of its real and imaginary components. Since we can take a $\chi^2$ between any pair of visibilities, we have three relevant $\chi^2$ metrics: $\chi_{AO}^2$, $\chi_{SA}^2$, and $\chi_{SO}^2$.\footnote{$\chi_{SA}^2$ is not used directly in the method; however, it is useful for exploring the magnitude of effect on the visibilities of adding a spiral to an axisymmetric model disc (Appendix~\ref{Sec:Effect of adding a spiral}).} The weights used for all of these metrics are those of the observed visibilities.\footnote{For the synthetic observations, we use the weights computed from the median variance of the deprojected and binned visibilities via the \texttt{frank} package \citep{jenningsFrankensteinProtoplanetary2020}, as the weights returned by \texttt{CASA} \citep{McMullinCASA2007} are not statistical.}. 

For a spiral model instance that is a better fit to the observed visibilities than the axisymmetric model, we expect \(\chi_{SO}^2 < \chi_{AO}^2\). For brevity, we introduce the quantity $\Delta \chi^2 = \chi_{AO}^2 - \chi_{SO}^2$, meaning that a positive (negative) \(\Delta \chi^2\) corresponds to a model instance that is a better (worse) fit than the axisymmetric model. 

To confidently claim a spiral detection in an observation using our method, there are two tests the data must pass:

\begin{enumerate}
    \item The spiral model must be a better model than the axisymmetric model for the data. We evaluate this using the Bayesian Information Criterion (BIC) \citep{schwarzEstimatingDimension1978} defined as \begin{equation*} \text{BIC}(M) = 2\ln L_M - |M|\ln N \end{equation*} where $L_M$ is the maximised value of the likelihood of model $M$, $|M|$ is the number of parameters estimated by the model, and $N$ is the number of data points.\footnote{The BIC is well-suited to large $N$ (as is the case for visibility data), whereas $p$-value-based model selection is less reliable for such datasets \citep{rafteryBayesianModel1995}.} The change in BIC between the spiral and axisymmetric model is given by \begin{equation} \Delta \text{BIC}=\text{BIC}(S)-\text{BIC}(A) \approx \Delta \chi^2_\text{max} - 6\ln N 
    \label{eq:BIC} 
    \end{equation} which follows because the noise in visibility data is approximately Gaussian and we estimate an additional six parameters from the data for the spiral model. $\Delta \text{BIC}>0$ provides evidence for the spiral model, with $\Delta \text{BIC}>10$ typically being taken as very strong evidence \citep{kassBayesFactors1995}. For our data (\S\ref{sec:results synthetic observations},\ref{sec:results Taurus}), $\Delta \chi^2 \gtrsim 70$ would generally constitute strong evidence by this criterion.\footnote{$\Delta \text{BIC}$ approximates the Bayes factor between the models $B_{SA}$ via $\Delta\text{BIC} \approx 2\ln B_{SA}$.}

    \item $\Delta\chi^2$ must have a spatial ($r_\text{p}, \phi_\text{p}$) structure that indicates unambiguously that the improvement is caused by a spiral, not some other asymmetry. We cannot discern this with statistics alone; we must also rely on more qualitative methods. For this purpose, we introduce $\Delta\chi^2$ heatmaps (e.g., Figs.~\ref{fig:handedness test}-\ref{fig:H/r}) which show the $\Delta\chi^2$ for different planet positions in the deprojected disc. This test is needed because the spiral model is capable of fitting certain non-spiral asymmetries to the extent that it is favoured over the axisymmetric model by the BIC. This `false fitting' can therefore pass our first test and, if unchecked, would lead to incorrectly inferring a spiral. In \S\ref{sec:results synthetic observations} we will see examples of this false fitting and how to discern it from true fitting.
\end{enumerate}


\section{Results: synthetic observations} \label{sec:results synthetic observations}

Here we generate synthetic observations of model discs and then treat these as we would a real disc observation in which we are seeking evidence of spiral structure (i.e., we evaluate the difference in fits to the observed visibilities between an axisymmetric model and models including various spiral structures). We investigate five different observational set-ups, which are described in Table~\ref{tab:ALMA set-ups}.

The synthetically observed discs are based on the same set of spiral simulations as the spiral templates which we use in our method. Therefore, these observations clearly present an idealised situation for analysis with our method.
In light of this, we first explore whether an improvement in fit is only manifest for the spiral template with exactly the same parameters as those used in the model disc we observe (which we call the `matching' spiral template) or whether the spiral can be recovered, and the location of the planet constrained, for a range of spiral templates (\S\ref{sec:effect of spiral model parameters}).  
We then compare the method’s ability to identify planets to image residuals (\S\ref{sec:maps vs residuals}), and a simple gap-based method (\S\ref{sec:gaps vs spirals}).
The $\Delta\chi^2$ values and heatmaps from these synthetic observations will also serve as comparison points for the analysis of the Taurus disc sample in \S\ref{sec:results Taurus}. The effects of errors in the fitted disc geometry are considered in Appendix~\ref{sec:errors}.


The synthetically observed discs we examine in this section are all face-on, with zero rotation or offset. Their inner and outer boundaries are at 10 and 110\,au, with the planet positioned at $r_\text{p} = 50\,\text{au}$, $\phi_\text{p} = 0^\circ$, and producing a right-handed spiral (orbiting anticlockwise). They are computed with a distance to the disc of 140\,pc. The host stars are given a mass of $0.8\,M_\odot$ and luminosity of $1.5\,L_\odot$, which are the approximate medians of the sample in the DSHARP survey (Table 1 of \citealp{andrewsDiskSubstructures2018}). The chosen planet location of 50\,au aligns roughly with the orbital radius where many of the DSHARP survey's gaps and rings are observed (Fig. 7 of \citealp{huangDiskSubstructures2018}). 

To generate the radial brightness profiles for the synthetic discs we use the following \texttt{frank} hyperparameters: $N = 200$, $R_\text{out} = 1.2''$, $\alpha = 1.05$, $p_0 = 10^{-15}\,\text{Jy}^2$, and $w_\text{smooth} = 0.1$. These are within the ranges suggested by \citet{jenningsFrankensteinProtoplanetary2020} and our resulting profiles are quite insensitive to variations within those ranges.

\subsection{Effect of spiral model parameters} \label{sec:effect of spiral model parameters}

To test the method, we first examine the goodness-of-fit when a synthetic observation of a model disc is fitted with spiral templates of different parameters to that disc. For these tests, we consider discs in our fiducial set described in Table~\ref{tab:ALMA set-ups}. For spiral detection in this set, our second test tends to be more exacting than our first (\S\ref{sec:spiral fits}), meaning our main obstacle is discerning true spiral signals from false ones. In this section, the maximum $\Delta\chi^2$ in all of the heatmaps shown well-exceed the threshold for $\Delta \text{BIC}>10$.

\begin{table*} 
    \centering
    \caption{ALMA set-ups and observational parameters for the synthetic observations in this work. 
    The sensitivities given refer to \textit{requested} sensitivities with the ALMA Cycle 8 Observing Tool. The combined (i.e., summed compact and extended configuration) on-source times given are those needed to achieve these requested sensitivities. After \texttt{CLEAN}ing, the measured rms noise in all our images is $\sim 80\text{--}95\,\%$ of the requested sensitivity.
    The penultimate column gives the $\Delta\chi^2$ value for which $\Delta \text{BIC}=0$, meaning the spiral and axisymmetric models are equally likely. From Equation~\ref{eq:BIC}, this relates to the number of data points $N$ by $\Delta\chi^2(\Delta \text{BIC}=0)=6 \ln N$. Hence, a $\Delta\chi^2>\Delta\chi^2(\Delta \text{BIC}=0) + 10$ indicates a very strong preference for the spiral model. 
    The visibilities and images for all of our sets are available at \href{https://doi.org/10.6084/m9.figshare.25464109}{doi.org/10.6084/m9.figshare.25464109}. The final column indicates if the set was also analysed in \citet{speedieSpirals2022}.
    For comparison, the bottom two rows give the corresponding information for the Taurus disc sample analysed in \S\ref{sec:results Taurus} and the DSHARP survey.}
    \label{tab:ALMA set-ups}
    \renewcommand{\arraystretch}{1.5}
    \begin{tabular}{ 
        >{\raggedright\arraybackslash}p{2.55cm}   
        >{\centering\arraybackslash}p{1.7cm}  
        >{\centering\arraybackslash}p{1.7cm}  
        >{\centering\arraybackslash}p{1.7cm} 
        >{\centering\arraybackslash}p{1.7cm}
        >{\centering\arraybackslash}p{1.7cm}
        >{\centering\arraybackslash}p{1.875cm}
        >{\centering\arraybackslash}p{1.55cm}}   
        \hline
        Descriptor & Angular resolution [$''$] &  Sensitivity [$\mu\text{Jy/beam}$] & Combined on-source time [mins] & ALMA Configuration & Observational wavelength [mm]  & $\Delta\chi^2(\Delta \text{BIC}=0)$ & In \citetalias{speedieSpirals2022} \\
        \hline
        \multicolumn{8}{c}{Synthetic Observations} \\
        \hline
        Fiducial set & 0.061 & 35 & 40 & C-4 + C-7 & 0.87 & 63 & yes \\
        Higher sensitivity set & 0.061 & 10 & 480 & C-4 + C-7 & 0.87 & 78 & yes \\
        Lower sensitivity set & 0.061 & 50 & 19 & C-4 + C-7 & 0.87 & 59 & no \\
        Higher resolution set & 0.028 & 35 & 40 & C-5 + C-8 & 0.87 & 63 & yes \\
        Taurus comparison set & 0.13 & 50 & 6 & C-6 & 1.33 & 52 & no \\
        \hline
        \multicolumn{8}{c}{Real Observations} \\
        \hline
        Taurus survey \citep{longCompactDisks2019} & $\sim 0.12$ & $\sim 50$ & $\sim 6$ & C-6 & 1.33 & $\sim 60$ &--\\
        DSHARP survey \citep{andrewsDiskSubstructures2018} & $\sim 0.035$ & $\sim 20$ & $\sim 90$ &  C-5 + C-8/9 & 1.25 & -- & --\\
        \hline
    \end{tabular}
\end{table*}

\begin{figure*}
    \includegraphics[width=\textwidth]{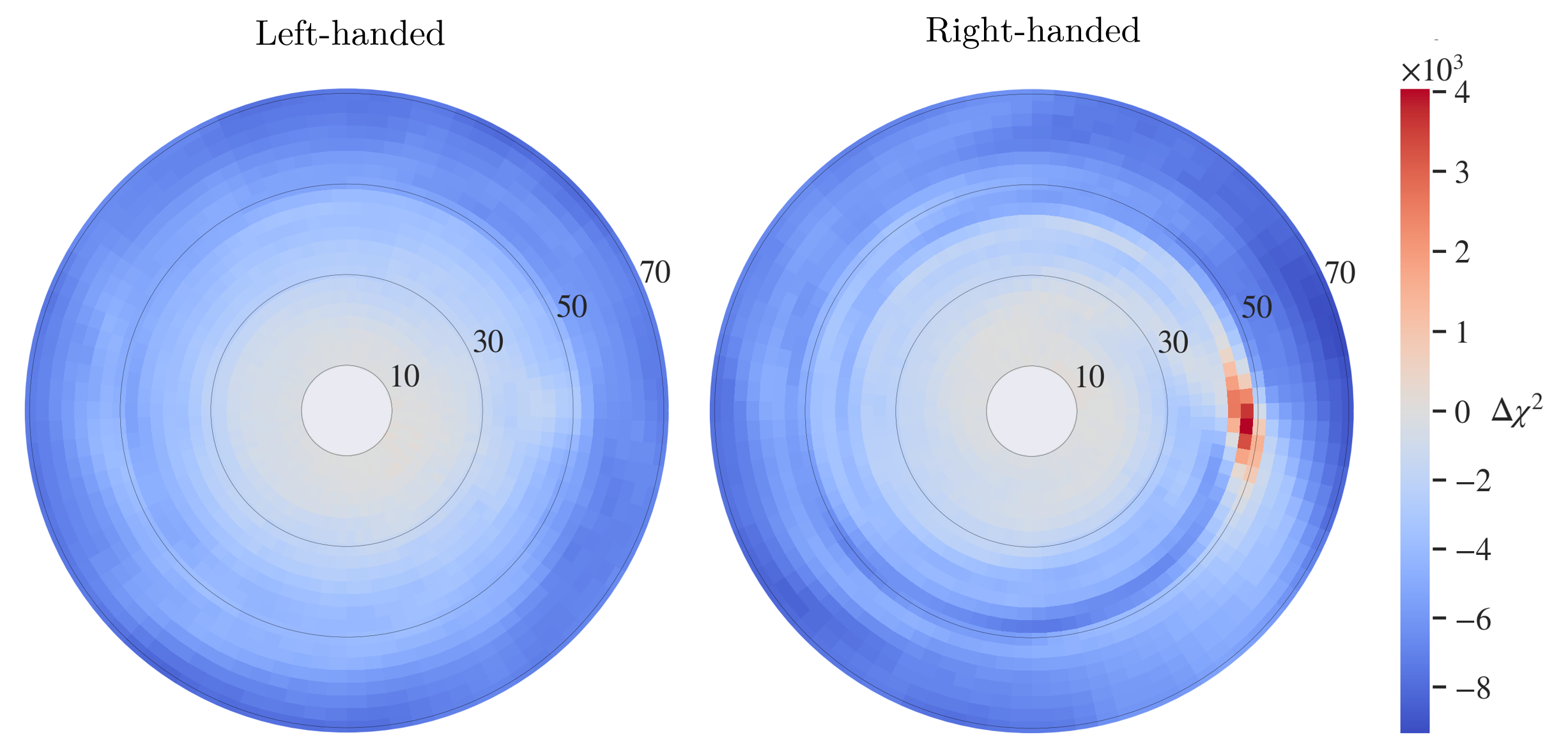}
    \caption{$\Delta \chi^2$ heatmaps of a synthetic observation from our fiducial set of a model disc ($\tau_0=3$, $\beta=10$, $M_\text{p}=1\,M_\text{th}$, right-handed spiral) for its matching right-handed spiral template (right) and a left-handed, but otherwise matching, spiral template (left).  The disc's planet is located at $r_\text{p} = 50\,\text{au}$, $\phi_\text{p} = 0$ (along the right horizontal axis). The heatmap spatially represents the disc, with each point on the heatmap corresponding to a spiral template with its planet placed at that location. Recall $\Delta\chi^2 = \chi^2_{AO} - \chi^2_{SO}$, i.e., the difference in goodness-of-fit between the axisymmetric model and spiral model instance. Positive $\Delta \chi^2$ values (red) mean the spiral model instance is a better fit, and negative values (blue) mean the axisymmetric model is a better fit. The maximum $\Delta\chi^2$ value for the right-handed template is $4.1\times 10^3$, occurring in the region around the planet position. The maximum $\Delta\chi^2$ value for the left-handed template in the same region is $-430$, indicating a strong preference for the right-handed model. The heatmaps are truncated at the disc's inner boundary of 10\,au, and an outer boundary of 70\,au (the 70-110\,au region of these discs offers little interest). The radial ticks are in au and the colourbar applies to both heatmaps to aid comparison.  Note that the colourbar diverges asymmetrically about zero.}
    \label{fig:handedness test}
\end{figure*}

\begin{figure*}
    \begin{tabular}[b]{@{}c@{}} 
        \rotatebox{90}{\parbox{4cm}{\centering\large $\beta=10$}} \\ [0.8cm]
        \rotatebox{90}{\parbox{4cm}{\centering\large $\beta=0$}} \\ 
        \hspace{0.6cm}
        
    \end{tabular}%
    \begin{minipage}[b]{0.95\linewidth}
        \begin{minipage}{0.27\linewidth}
            \centering
            \hspace{-.1cm} \large$M_\text{p}=0.1\,M_\text{th}$
        \end{minipage}\hfill
        \begin{minipage}{0.3\linewidth}
            \centering
            \hspace{1.1cm} \large$M_\text{p}=0.3\,M_\text{th}$
        \end{minipage}\hfill
        \begin{minipage}{0.43\linewidth}
            \centering
            \large$M_\text{p}=1\,M_\text{th}$
            \hspace{0.2cm}
        \end{minipage}
        
        \includegraphics[width=\linewidth]{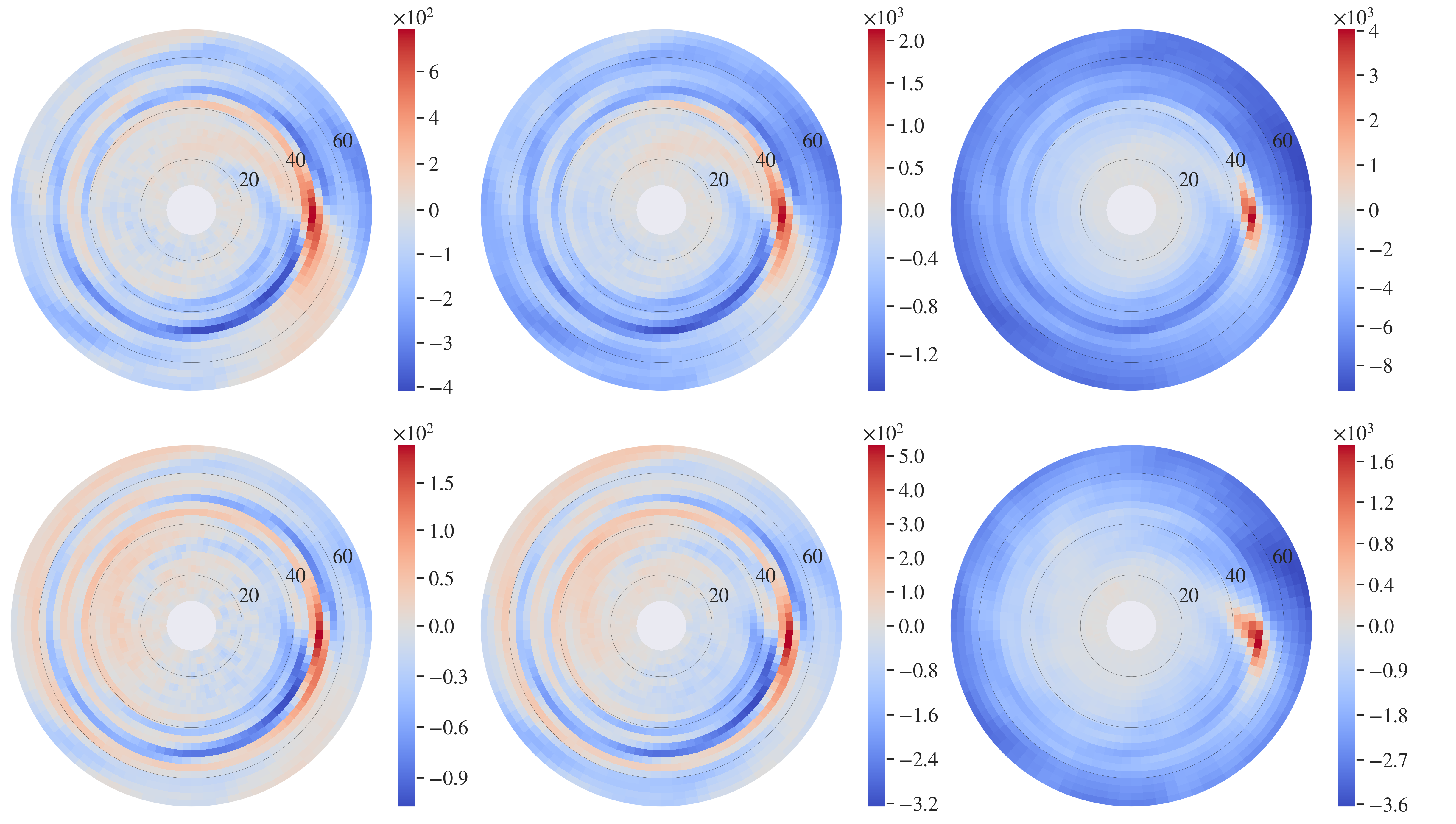}

    \end{minipage}

    \caption{$\Delta \chi^2$ heatmaps of a synthetic observation from our fiducial set of a model disc ($\tau_0=3$, $\beta=10$, $M_\text{p}=1\,M_\text{th}$, right-handed spiral) for $\tau_0=3$ spiral templates. $M_\text{p}/M_\text{th}=0.1,0.3,1$ from left to right, and $\beta=0,10$ from bottom to top. The heatmap for the matching spiral template is therefore in the top right. These heatmaps demonstrate that the method can recover the planet with non- 
 matching spirals.}
    \label{fig:mismatches}
\end{figure*}

\begin{figure*}
    \includegraphics[width=\textwidth]{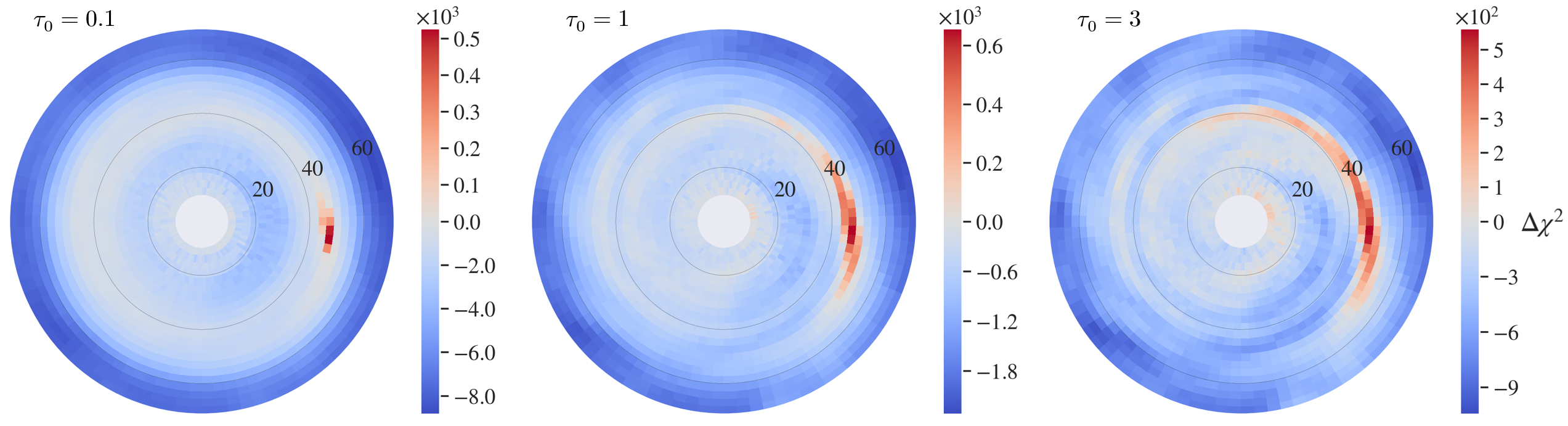}
    \caption{$\Delta \chi^2$ heatmaps of a synthetic observation from our fiducial set of a model disc ($\tau_0=0.3$, $\beta=10$, $M_\text{p}=1\,M_\text{th}$, right-handed spiral) for wrong optical depth, but otherwise matching, spiral templates. $\tau_0=0.1,1,3$ from left to right (the heatmap for the matching $\tau_0=0.3$ template appears almost identical to the $\tau_0=1$ one). These heatmaps demonstrate that the method can recover the planet without having to accurately predict the disc's optical thickness.}
    \label{fig:tau0}
\end{figure*}

\begin{figure*}
    \includegraphics[width=\textwidth]{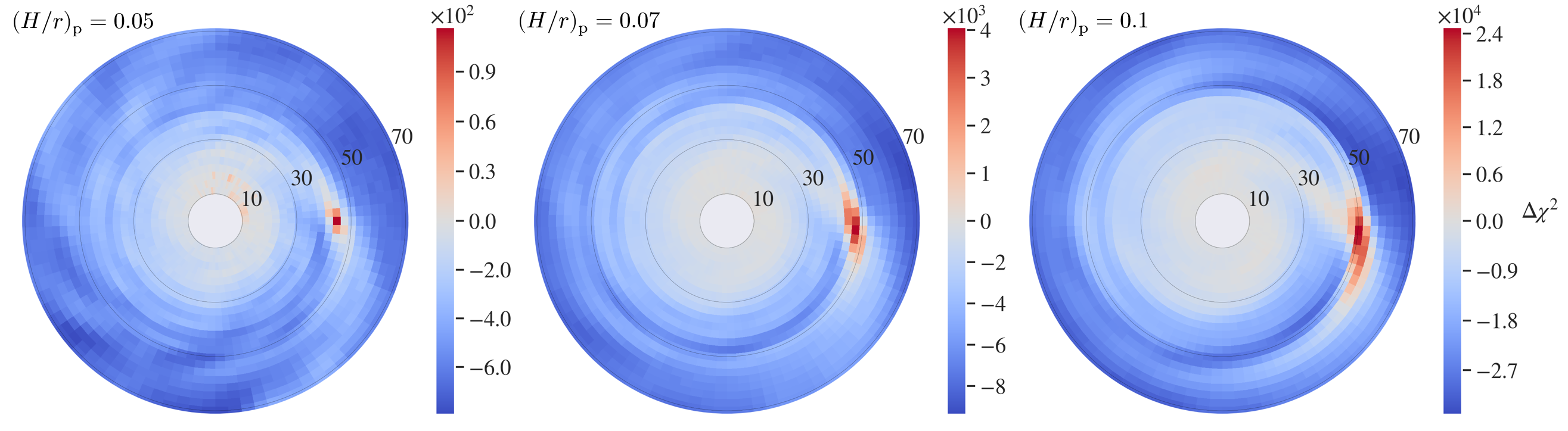}
    \caption{$\Delta \chi^2$ heatmaps of model discs with $ (H/r)_\text{p}=0.05, 0.07, 0.1$ (left, centre, right), and otherwise the same parameters: $\tau_0=3$, $\beta=10$, $M_\text{p}=1\,M_\text{th}$, and a right-handed spiral. The discs are synthetically observed with our fiducial set-up. The $\Delta \chi^2$ heatmaps are generated with the $ (H/r)_\text{p}=0.07$, and otherwise matching, spiral template in each case. This demonstrates that the method can recover the planet without having to accurately predict the disc aspect ratio.}
    \label{fig:H/r}
\end{figure*}

One would expect the greatest improvement in fit (i.e., the largest $\Delta \chi^2>0$) to occur for the matching spiral template, and indeed this is the case. A useful result is that the right-handed spiral (the correct one) shows a much larger maximum$\Delta \chi^2$ than the left-handed spiral (Fig.~\ref{fig:handedness test}). We can use this as a preliminary test for whether a spiral template is actually fitting a spiral or some other asymmetry that is approximately azimuthally symmetric, such as an arc; if it is fitting such an asymmetry, we would expect the fit improvement to be similar for a right- or left-handed spiral. 

When the wrong planet mass is used in the spiral template, we see that the lower planet mass models still show improved fits compared with the axisymmetric model when the planets are placed at a similar location to the disc's true planet location, although the improvement in $\chi^2$ is inferior to that achieved with the correct planet mass (Fig.~\ref{fig:mismatches}). In contrast, templates with too large a planet mass (\(3\,M_{\text{th}}\)) do not show any improved fits. 
Using the wrong cooling timescale decreases the improvement, but still recovers the planet, even for the smaller masses (Fig.~\ref{fig:mismatches}). 
Similarly, a poorly estimated optical thickness does not prevent the planet from being recovered in most cases (Fig.~\ref{fig:tau0}) (a strongly underestimated optical thickness can sometimes limit recovery).
Decreasing the spiral template's optical thickness increases its fractional brightness amplitude with a relatively modest impact on the spiral morphology, as indicated by a corresponding increase in $\Delta\chi^2$ amplitudes in its associated heatmap with little change in its structure. This means that for templates with amplitudes that are too low (e.g., due to a planet mass that is too low), decreasing their optical thickness can sometimes improve recovery, and vice-versa for templates with amplitudes that are too high. This also indicates a degree of degeneracy between the optical thickness and planet mass (this may be resolvable through alternative avenues to estimate the optical thickness, e.g., from the disc's brightness temperature).
A general theme is that when a template's amplitude is much too high, such as when the planet mass is too large, the $\Delta \chi^2$ heatmaps lose their ability to recover the planet. Whereas, when a template's amplitude is too low, there is often still improvement near the planet's location and the planet can be recovered.

The disc aspect ratio $H/r$ affects the spiral morphology, with spiral arms tending to be less tightly wound for higher $H/r$ \citep{rafikovNonlinearPropagation2002, baePlanetdrivenSpiral2018}. In our method we take a fixed $(H/r)_\text{p}=0.07$ for all of our spiral templates. Therefore, to investigate the effect of disc aspect ratio on our method's ability to recover spirals, we generate two discs with alternative ratios, $ (H/r)_\text{p}=0.05$ and $0.1$,
and compare the fits derived using our $ (H/r)_\text{p}=0.07$ spiral templates to the fit for the matching $ (H/r)_\text{p}=0.07$ disc. As can be seen in Figure~\ref{fig:H/r}, the $\Delta\chi^2$ heatmaps recover the planet in each case. 
In fact, the $ (H/r)_\text{p}=0.1$ disc displays a larger peak $\Delta \chi^2$ than the matching $ (H/r)_\text{p}=0.07$ disc. This is because, by increasing the aspect ratio, we increase the disc temperature (these parameters are coupled in the simulations), and therefore the disc brightness. This effectively increases the signal-to-noise of the observation. The same effect compounds the decrease in peak $\Delta \chi^2$ for the $ (H/r)_\text{p}=0.05$ disc. Therefore, the method is clearly sensitive to $H/r$, but does not rely on using an accurate value.
Each disc fit also retains a strong handedness preference (like in Fig.~\ref{fig:handedness test}), confirming that the method is not merely sensitive to emission near the planet position. 
This result justifies our decision to model the spirals as being scale-free.
That is, given the method's relative insensitivity to $ (H/r)_\text{p}$ and that $H(r)/r$ only varies weakly with $r$, we adopt a single value for $ (H/r)_\text{p}\ (=0.07)$ and then apply an overall scaling with $r_\text{p}$.

In complement to this section, in Appendix~\ref{Sec:Effect of adding a spiral}, we explore the effect on the $\chi^2$ amplitude of spiral templates with different parameters. Instead of using the discs considered in this section (which are face-on and would have a simple power law surface density and temperature profile in the absence of their planets), we consider discs at a range of inclinations and with more complex axisymmetric brightness profiles, informed by our Taurus disc sample analysed in \S\ref{sec:results Taurus}. The outcomes are consistent with the findings of this section.

\subsection{Visibility-space fitting versus image residuals} \label{sec:maps vs residuals}

Here we compare the spiral detection ability of \texttt{CLEAN} image residuals, generated by subtraction of the azimuthally averaged disc brightness, to our visibility-space fitting method. 
We examine our fiducial and higher resolution sets, both of which have a requested sensitivity of $35\,\mu\text{Jy/beam}$ (the lowest considered in \citetalias{speedieSpirals2022}, meaning the easiest observations to make and the hardest spirals to recover) and resolutions of $0.061''$ and $0.028''$ respectively.
For comparison, the Taurus sample discs analysed in \S\ref{sec:results Taurus} were observed at an angular resolution $\sim 0.12''$ and mean sensitivity $\sim 50\,\mu\text{Jy/beam}$ \citep{longCompactDisks2019}. Therefore, the synthetic observations with the lowest sensitivity and resolution considered in this section are still more sensitive and higher-resolution than the Taurus sample analysed in \S\ref{sec:results Taurus}, but exhibit lower sensitivities and resolutions than many ALMA observations (e.g., DSHARP survey, \citealp{andrewsDiskSubstructures2018}; and several case studies of individual systems, such as: TW Hya, \citealp{andrewsRINGEDSUBSTRUCTURE2016}; HD 169142, \citealp{perezDustUnveils2019}; HL Tau, \citealp{almapartnership2015}, and MP Mus, \citealp{ribasALMAView2023}). 

\begin{figure*}
    \begin{subfigure}{\textwidth}
        \centering
        \begin{minipage}[c]{0.07\textwidth}
            \centering
            \huge{(a)}
        \end{minipage} \hspace{-1.5cm}
        \begin{minipage}[c]{1.0\textwidth}
            \centering
            \includegraphics[width=0.84\textwidth]{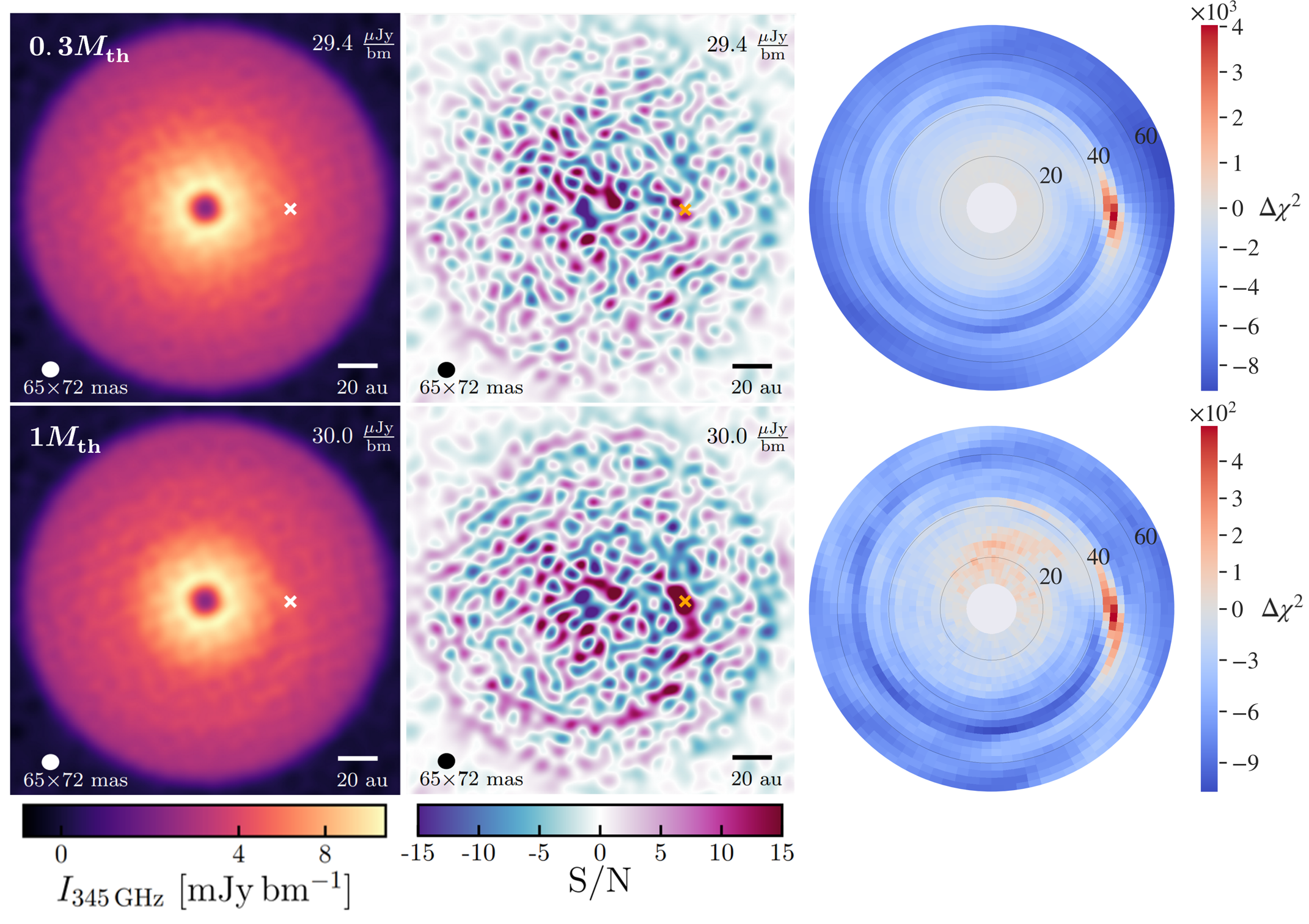} \hspace{-0.8cm}
            \label{fig:compare1}
        \end{minipage}
    \end{subfigure}

    \vspace{0.095cm} 

    \begin{subfigure}{\textwidth}
        \centering
        \begin{minipage}[c]{0.07\textwidth}
            \centering
            \huge{(b)}
        \end{minipage} \hspace{-1.5cm}
        \begin{minipage}[c]{1.0\textwidth}
            \centering
            \includegraphics[width=0.84\textwidth]{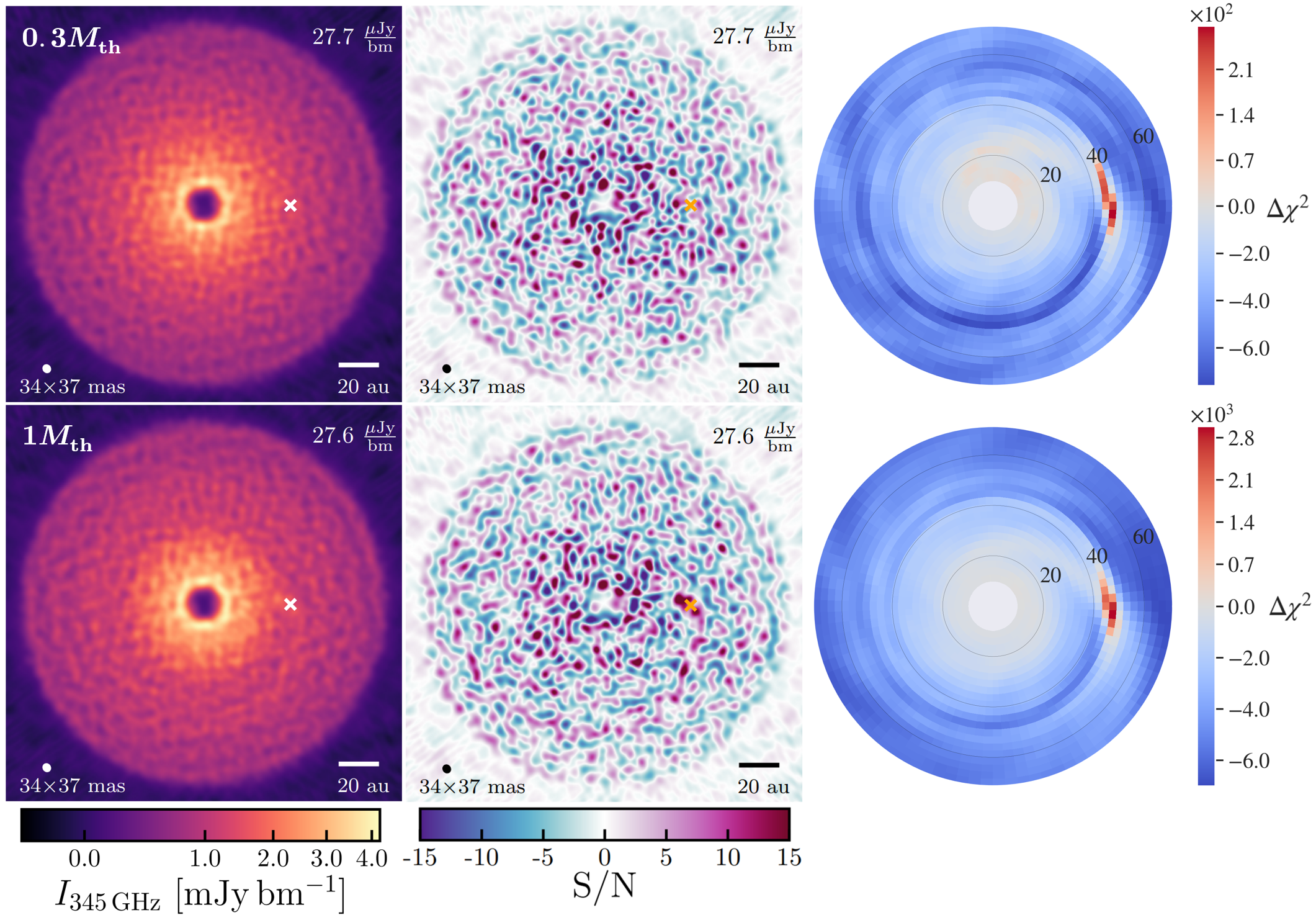} \hspace{-0.8cm}
            \label{fig:compare2}
        \end{minipage}
    \end{subfigure}

    \caption{Planet recovery for $\beta=10$, $\tau_0=3$ discs from our (a) (top two rows) fiducial set ($0.061''$ resolution) and (b) (bottom two rows) higher resolution set ($0.028''$ resolution). Each set has a requested sensitivity of $35\,\mu\text{Jy/beam}$ (40\,mins of on-source time), and contains a planet of mass $0.3\,M_\text{th}$ (top) or $1\,M_\text{th}$ (bottom). Left: \texttt{CLEAN} image. Centre: corresponding image residuals (generated by subtraction of the azimuthally averaged disc brightness). Right: corresponding \( \Delta \chi^2 \) heatmaps. Despite the spiral only being noticeable in one of the image residuals, the $\Delta\chi^2$ heatmaps clearly recover the planet in all four cases. In the synthetic observations and image residuals: the cross indicates the planet position (50\,au), the measured rms noise in each observation after \texttt{CLEAN}ing is written in the top right corner of each panel (and in each case is $\sim 10\text{--}20\%$ less than the requested sensitivity), and the synthesized beam is shown in the bottom left corner. The radial ticks in the heatmaps are in au.}
    \label{fig:combined_compare}
\end{figure*}

A series of comparisons between image residuals and $\Delta\chi^2$ heatmaps for the two resolutions are shown in Figure~\ref{fig:combined_compare}. In the image residuals, we see lower spiral signal-to-noise for the higher resolution observations. This result occurs because the spiral signal is larger for a larger beam (due to the beam's larger area). This means that, if noise (in $\text{Jy/beam}$) is independent of beam size, a lower angular resolution (larger beam) observation will have a higher spiral signal-to-noise ratio, provided it can resolve the distance between spirals. 

This is reflected in the $\Delta\chi^2$ values, for which the higher resolution observations give somewhat lower amplitudes, indicating that the spiral perturbations generally exhibit larger amplitudes at shorter baselines (larger spatial scales). 
However, the higher angular resolution observations do result in a tighter fit around the planet position and reduced `false fitting' ($\Delta\text{BIC}>0$ in disc regions far from the planet's true position). 

Both images and visibilities, therefore, exhibit a similar trade-off between spiral `strength' (signal-to-noise / spiral model support) and spiral `clarity' (how unambiguously spiral-like the asymmetry is), with higher resolution (at fixed sensitivity) yielding a weaker but clearer spiral signal.

Of the synthetic observations considered in Figure~\ref{fig:combined_compare}, only the $0.061''$, $1\,M_\text{th}$ one ((a), bottom row) shows a noticeable spiral in the image residuals, and even here it is an incomplete spiral. We see a much stronger detection in the $\Delta\chi^2$ heatmaps. For the same observations, they recover spirals that are completely washed out by noise in the residuals, and clearly locate the planet.

\subsection{Gaps versus spirals} \label{sec:gaps vs spirals}

Here we compare the detectability of gaps and spirals as signals of the embedded planets. For the gap method, we use the radial brightness profile extracted from the \texttt{CLEAN} image (generated with the \texttt{radial\_profile} function of \texttt{GoFish}; \citealp{teagueGoFishFishing2019}). We use \texttt{CLEAN} image profiles instead of \texttt{frank} fits for two reasons: first, because this aligns with conventional approaches, and second, out of necessity -- the model discs have sharp edges at 10 and 110\,au, which cause problems for \texttt{frank} fits \citep{jenningsFrankensteinProtoplanetary2020} (the \texttt{frank} fits are still closer to the true disc profiles than the \texttt{CLEAN} image profiles in many cases, but the oscillatory artefacts [due to Gibbs phenomenon] can produce gap false positives). In these radial profiles we use gap depth as a measure of detection strength, which we define as 
\begin{equation}
    \delta = I_\text{ring}/I_\text{gap}
\end{equation}
where $I_\text{gap}$ is the brightness minimum in the gap and $I_\text{ring}$ is the brightness maximum in the ring at the gap edge (as in \citealp{zhangDiskSubstructures2018}). 
 
For spirals, we use our visibility-space fitting method and consider our two tests: (i) the spiral model support, indicated by the maximum $\Delta\chi^2$ achieved, and (ii) how clearly spiral-like the asymmetry is, inferred by visual inspection of the $\Delta\chi^2$ heatmaps. Our qualitative criteria for such inferences, along with heatmaps for our fiducial set, are given in Appendix~\ref{sec:supplementary heatmaps}.

\begin{figure*}
    
    \begin{minipage}[t]{0.425\textwidth}
        \Large{\qquad\qquad\quad\quad Gaps}
    \end{minipage}
    \begin{minipage}[t]{0.425\textwidth}
        \Large{\qquad\qquad\;\; Spirals}
    \end{minipage}
    
    \centering
    \begin{minipage}[t]{0.425\textwidth}
        \centering
        \includegraphics[width=\linewidth]{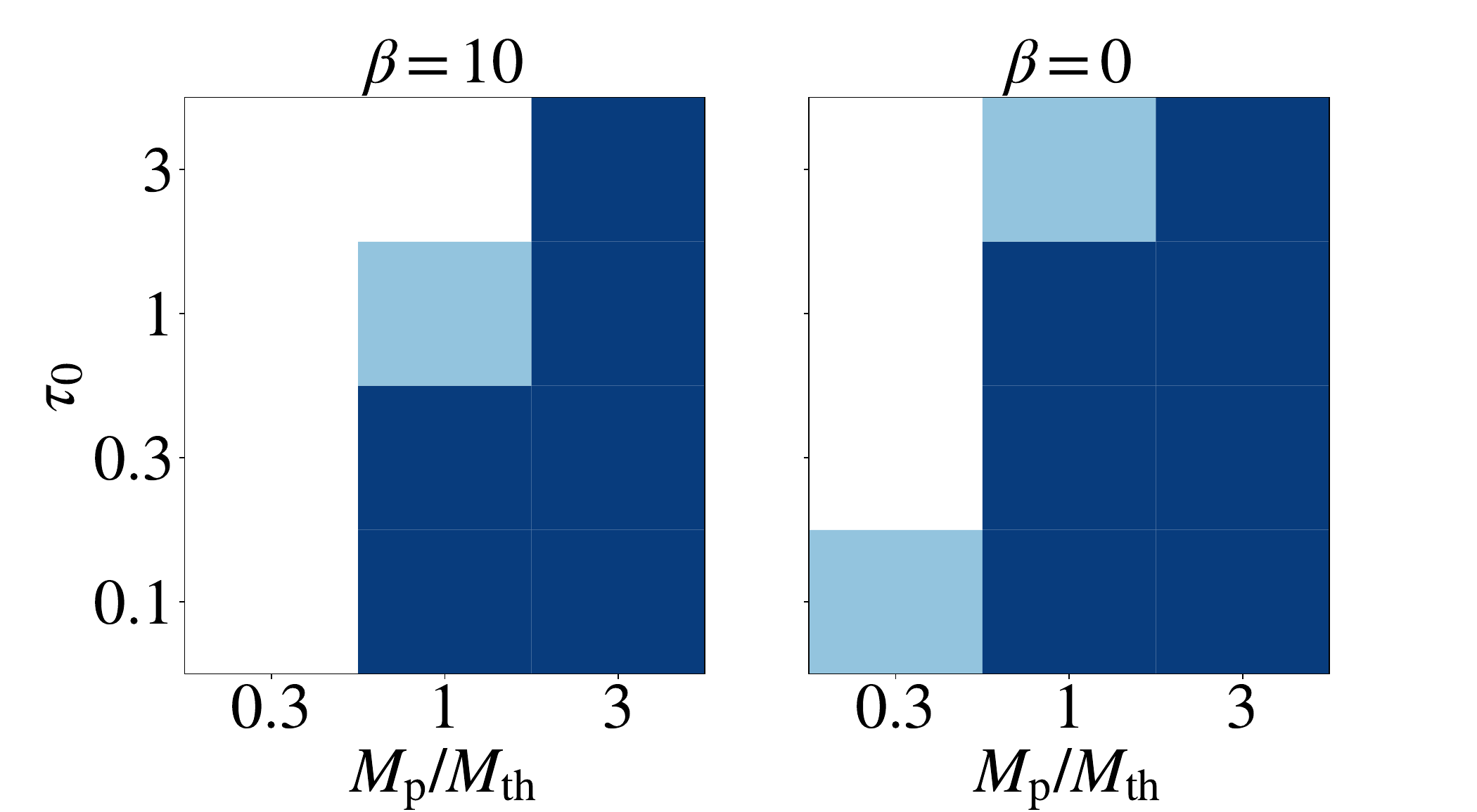}
    \end{minipage}%
    \hspace{-0.02\textwidth} 
    \begin{minipage}[t]{0.425\textwidth}
        \centering
        \includegraphics[width=\linewidth]{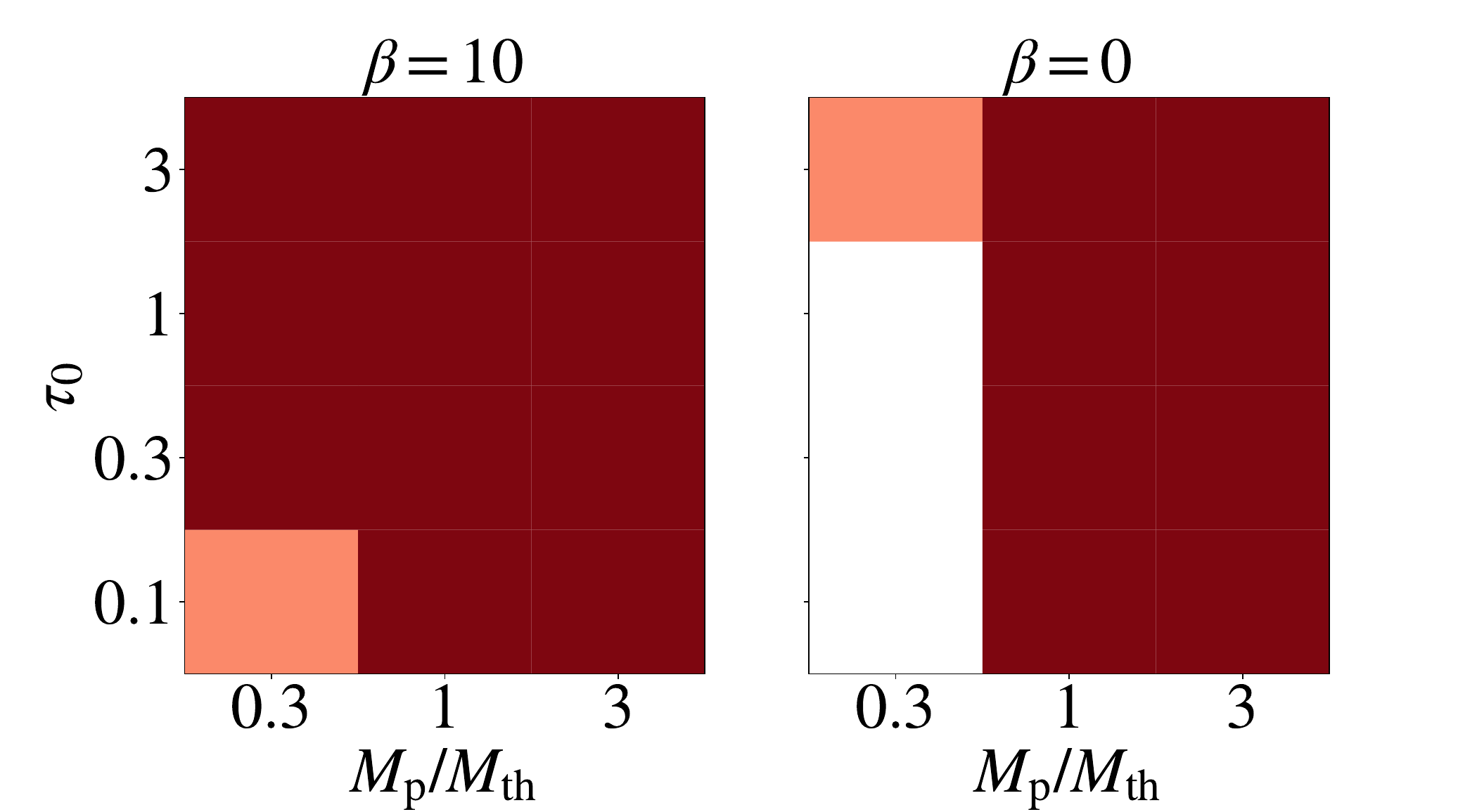}
    \end{minipage}%
    \hspace{-0.02\textwidth}
    \begin{minipage}[t]{0.14\textwidth}
        \vspace{-10\baselineskip} 
        {\raggedright{\Large{Fiducial set:\\[2pt]}\large{$\theta_{AR}=0.061''$\\[2pt]$\sigma=35\mu\text{Jy/beam}$\\[2pt]$\lambda_\text{obs}=0.87\,\text{mm}$}}}
    \end{minipage}

    \vspace{0.12cm}
        \centering
    \begin{minipage}[t]{0.425\textwidth}
        \centering
        \includegraphics[width=\linewidth]{images/C4C7_gap.pdf}
    \end{minipage}%
    \hspace{-0.02\textwidth} 
    \begin{minipage}[t]{0.425\textwidth}
        \centering
        \includegraphics[width=\linewidth]{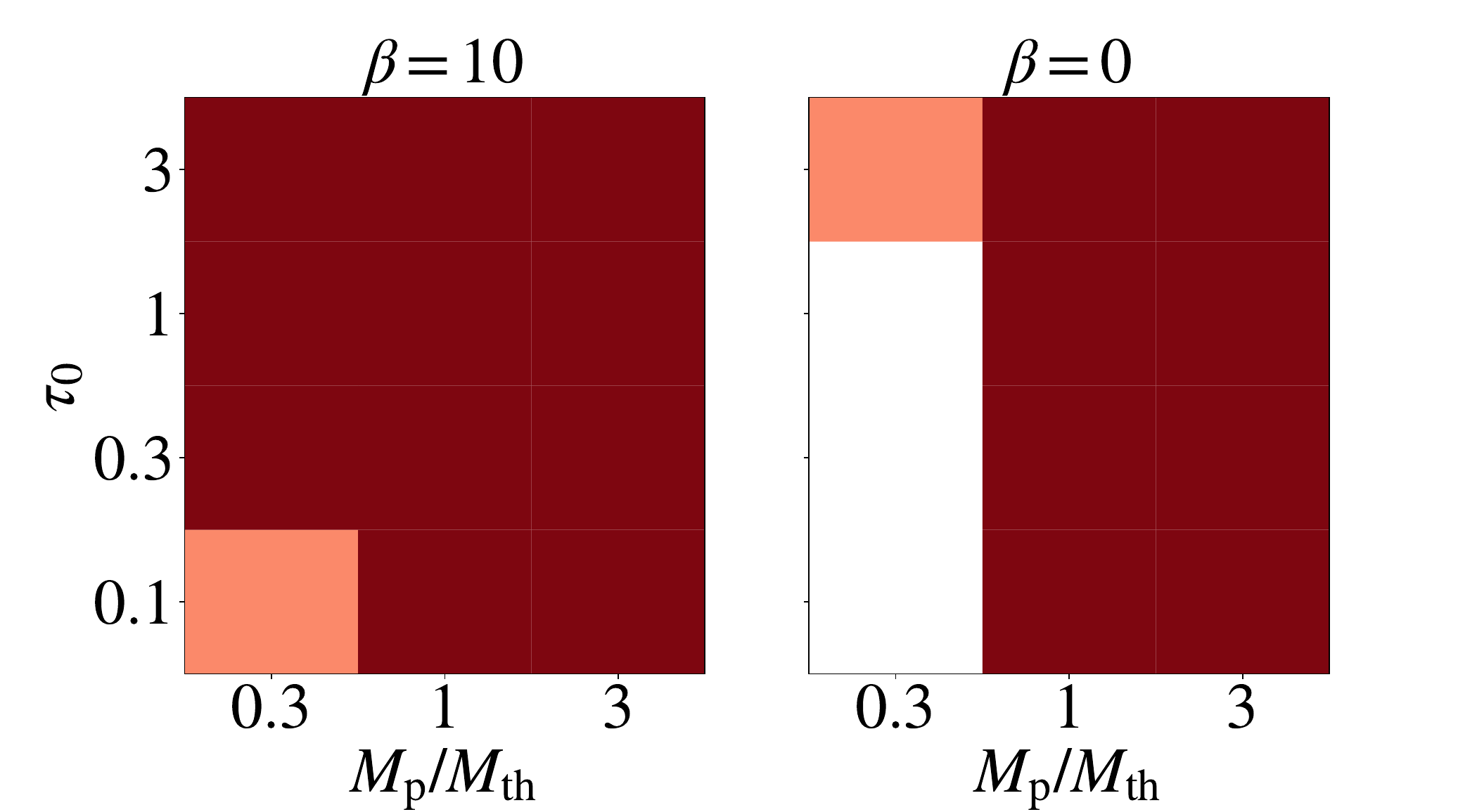}
    \end{minipage}%
    \hspace{-0.02\textwidth}
    \begin{minipage}[t]{0.14\textwidth}
        \vspace{-10\baselineskip} 
        {\raggedright{\Large{Higher \\ sensitivity set:\\[2pt]}\large{$\theta_{AR}=0.061''$\\[2pt]$\sigma=10\mu\text{Jy/beam}$\\[2pt]$\lambda_\text{obs}=0.87\,\text{mm}$}}}
    \end{minipage}

    \vspace{0.12cm}
        \centering
    \begin{minipage}[t]{0.425\textwidth}
        \centering
        \includegraphics[width=\linewidth]{images/C4C7_gap.pdf}
    \end{minipage}%
    \hspace{-0.02\textwidth} 
    \begin{minipage}[t]{0.425\textwidth}
        \centering
        \includegraphics[width=\linewidth]{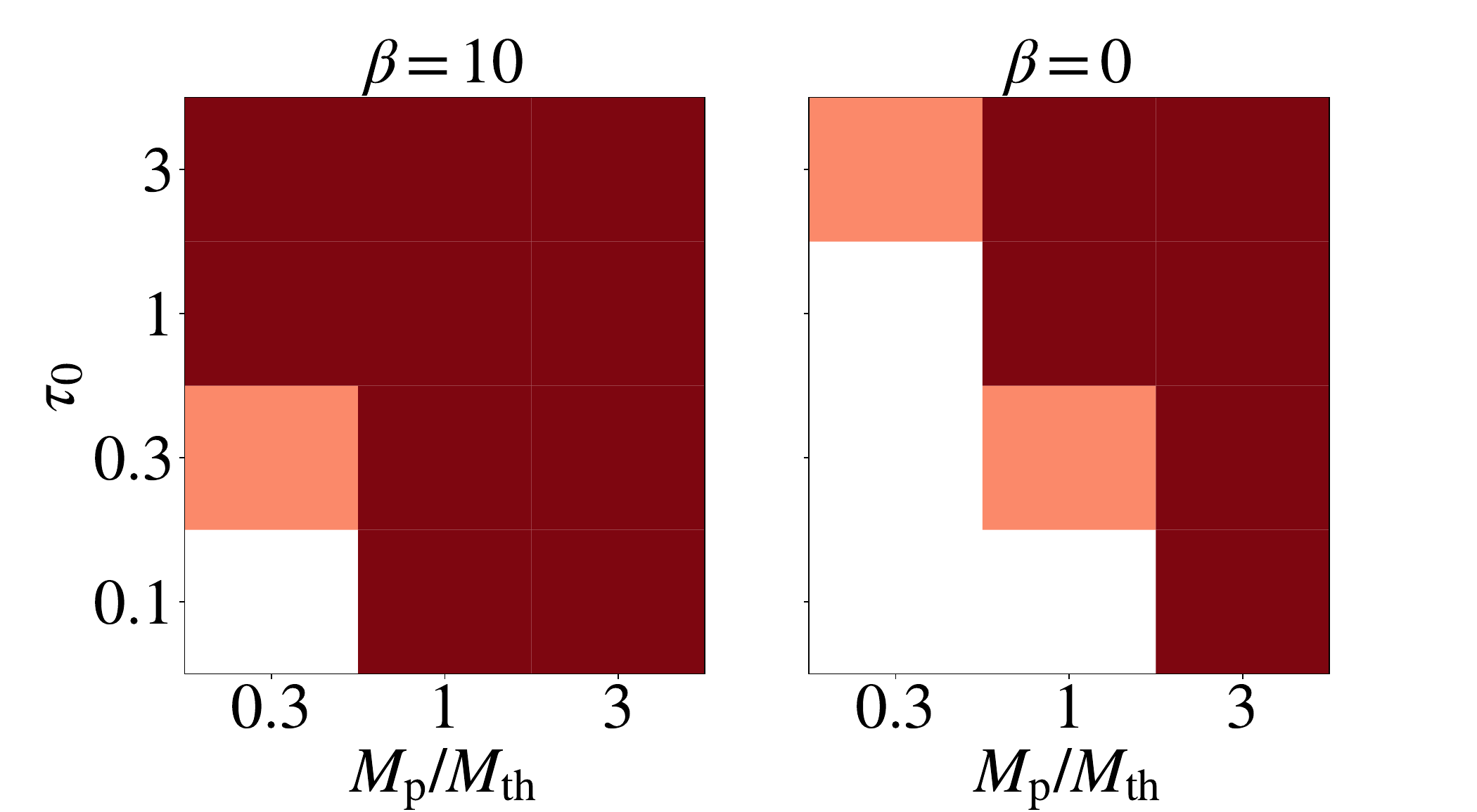}
    \end{minipage}%
    \hspace{-0.02\textwidth}
    \begin{minipage}[t]{0.14\textwidth}
        \vspace{-10\baselineskip} 
        {\raggedright{\Large{Lower \\ sensitivity set:\\[2pt]}\large{$\theta_{AR}=0.061''$\\[2pt]$\sigma=50\mu\text{Jy/beam}$\\[2pt]$\lambda_\text{obs}=0.87\,\text{mm}$}}}
    \end{minipage}

    \vspace{0.12cm}
    \centering
    \begin{minipage}[t]{0.425\textwidth}
        \centering
        \includegraphics[width=\linewidth]{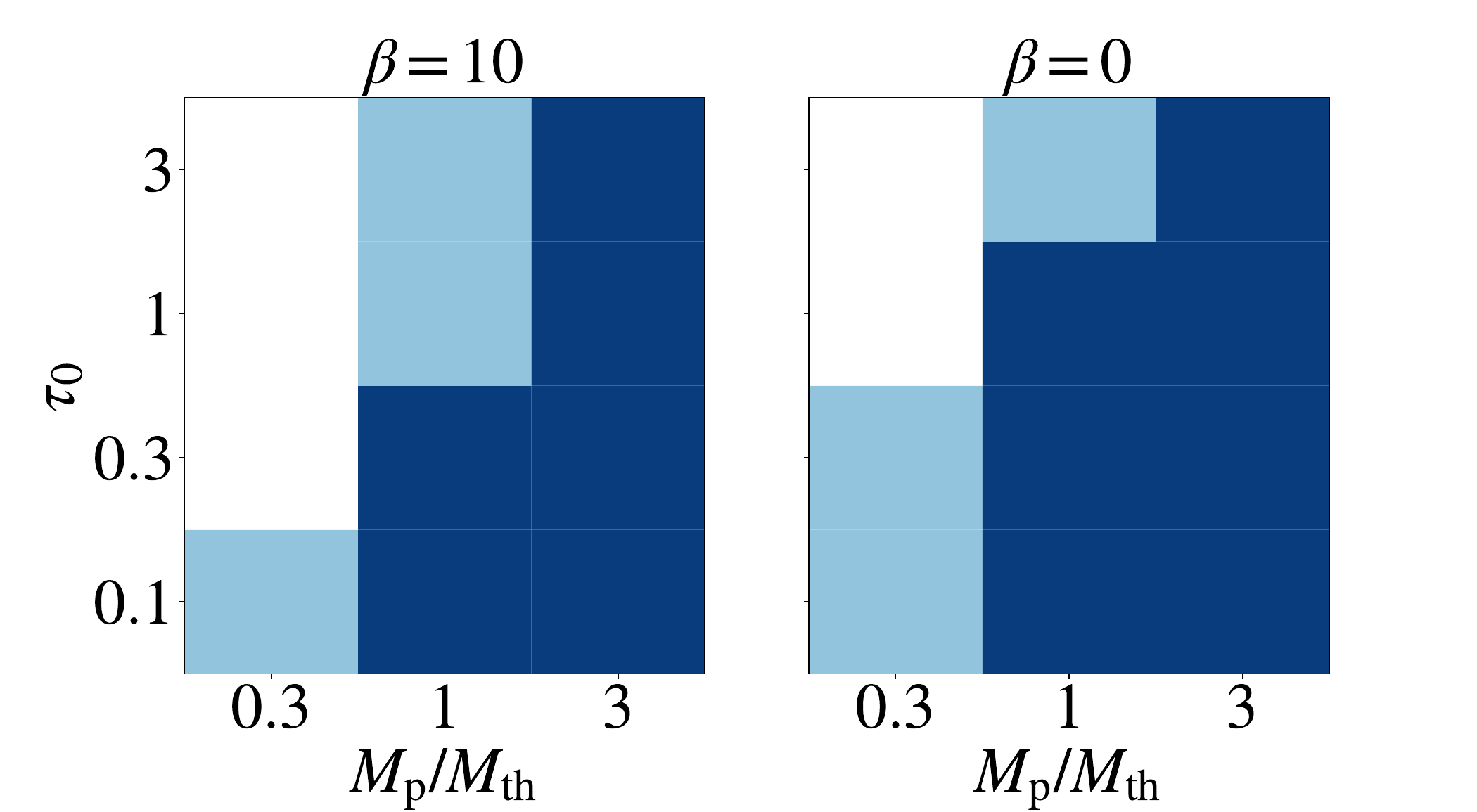}
    \end{minipage}%
    \hspace{-0.02\textwidth} 
    \begin{minipage}[t]{0.425\textwidth}
        \centering
        \includegraphics[width=\linewidth]{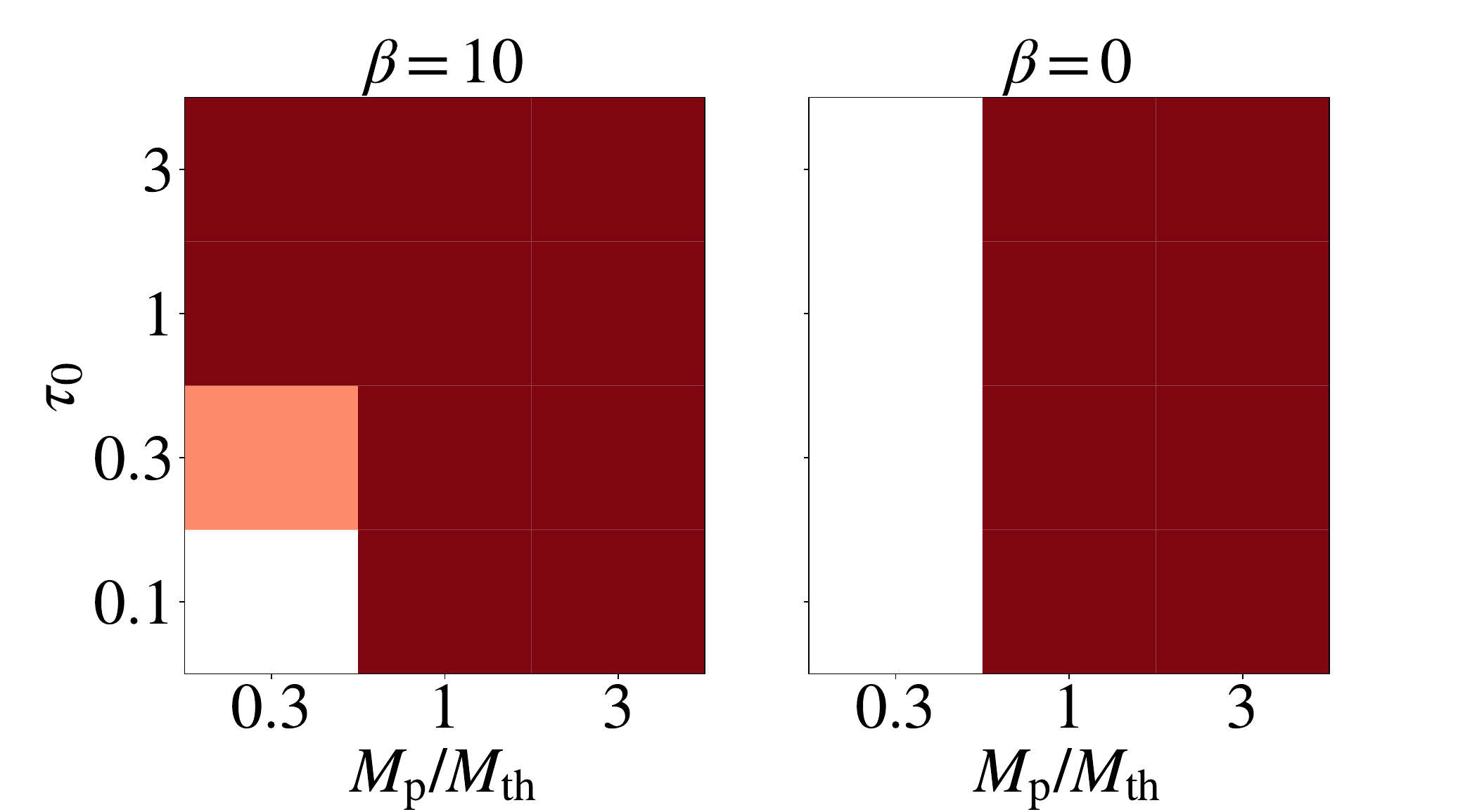}
    \end{minipage}%
    \hspace{-0.02\textwidth}
    \begin{minipage}[t]{0.14\textwidth}
        \vspace{-10\baselineskip} 
        {\raggedright{\Large{Higher \\ resolution set:\\[2pt]}\large{$\theta_{AR}=0.028''$\\[2pt]$\sigma=35\mu\text{Jy/beam}$\\[2pt]$\lambda_\text{obs}=0.87\,\text{mm}$}}}
    \end{minipage}

    \vspace{0.12cm}
    
    \centering
    \begin{minipage}[t]{0.425\textwidth}
        \centering
        \includegraphics[width=\linewidth]{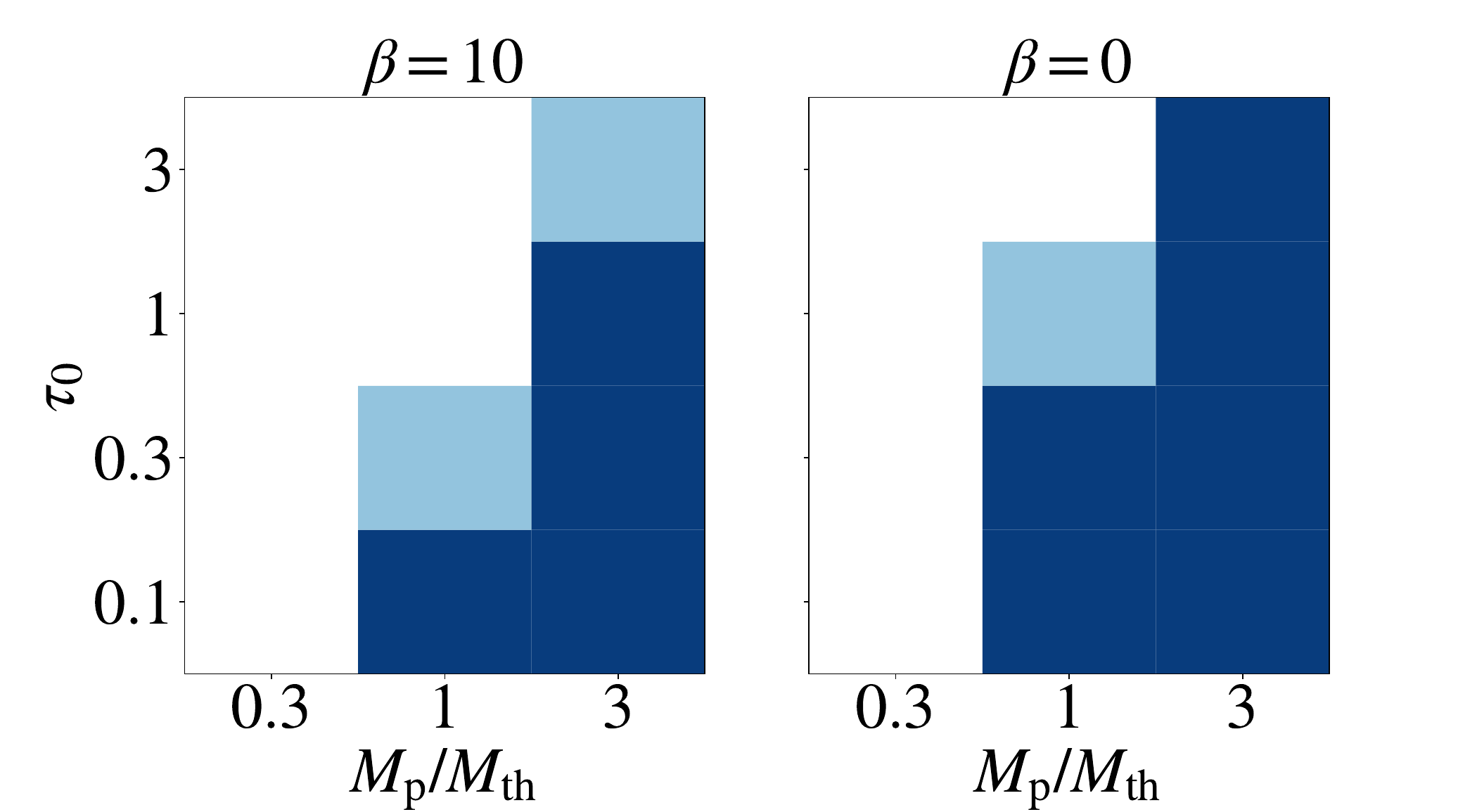}
    \end{minipage}%
    \hspace{-0.02\textwidth} 
    \begin{minipage}[t]{0.425\textwidth}
        \centering
        \includegraphics[width=\linewidth]{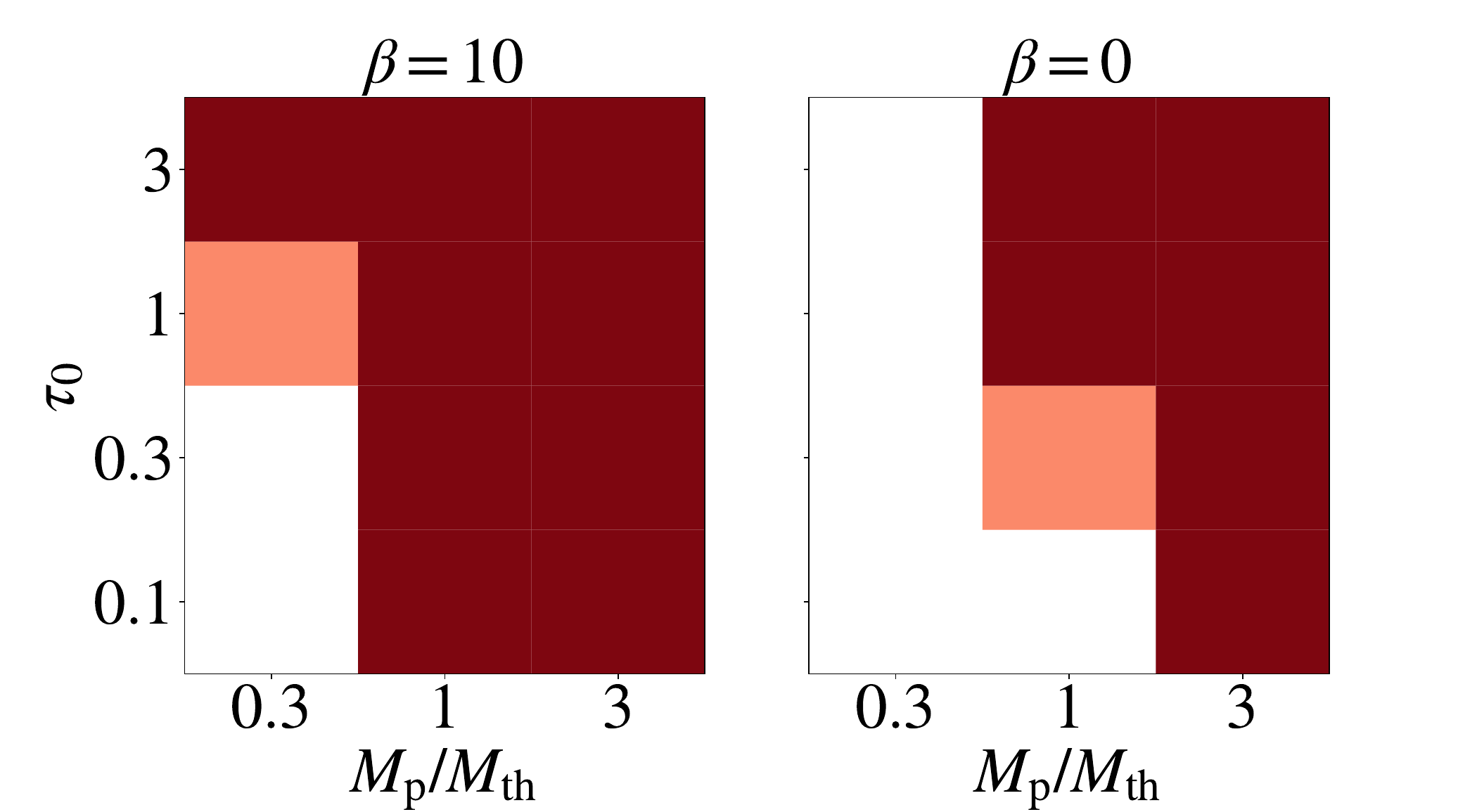}
    \end{minipage}%
    \hspace{-0.02\textwidth}
    \begin{minipage}[t]{0.14\textwidth}
        \vspace{-10\baselineskip} 
        {\raggedright{\Large{Taurus \\ comparison set:\\[2pt]}\large{$\theta_{AR}=0.13''$\\[2pt]$\sigma=50\mu\text{Jy/beam}$\\[2pt]$\lambda_\text{obs}=1.33\,\text{mm}$}}}
    \end{minipage}
    
    \caption{Planet recovery across the spiral simulation parameters (cooling time $\beta$, optical thickness $\tau_0$, and planet mass $M_\text{p}$) for gaps (left) and spirals (right), using \texttt{CLEAN} image profiles and visibility-space fitting respectively. White squares indicate no detection, light blue/red indicates a marginal gap/spiral detection, and a dark blue/red indicates a clear detection. The rows correspond to our various observation sets, labelled on the right by their descriptor from Table~\ref{tab:ALMA set-ups} along with their angular resolution, sensitivity, and observational wavelength. The radial profiles and $\Delta\chi^2$ heatmaps for our fiducial set are given in Appendix~\ref{sec:supplementary heatmaps}.}
    \label{fig:combined_figure}
\end{figure*}

Based on these considerations, we divide the strength of planet recovery into three levels: no recovery, marginal recovery, and clear recovery. Marginal recovery expresses that there is a detectable signal, but that potentially small decreases in observation quality, or slightly less favourable model parameters, could be enough to render the signal undetectable. Since, for a given set of observation/model parameters, the synthetic observations present best case scenarios for detection -- known planet location, disc models matching our spiral templates, and no interference from other disc substructure -- the observations with marginal planet recovery represent optimistic detection limits for the respective methods. For gaps, we take marginal recovery to be $1<\delta<1.5$, with values below and above this range indicating no recovery and clear recovery respectively. For spirals we take it to mean clearly passing one of our tests, but only marginally passing the other, i.e., $0<\Delta\text{BIC}<10$ or a suggestive, but not unambiguous, structure in the $\Delta\chi^2$ heatmap.

The recovery levels of the two methods are shown in Figure~\ref{fig:combined_figure} for various sets of observations. In terms of spiral simulation parameters, we see that: 
\begin{itemize}
    \item As expected, the recovery of both gaps and spirals improves with increasing planet mass. 
    \item For greater optical depths, spirals become easier to recover, whereas gaps become harder to recover.
    \item For the longer cooling time, gaps are harder to recover, whereas spirals are much easier to recover. 
    \item There are clear regions of the parameter space where gaps are not recovered, but spirals are.
\end{itemize}
Similar trends for planet mass, optical depth, and cooling time are seen for spiral recovery in image residuals (see e.g., Fig.~7 of \citetalias{speedieSpirals2022}).
The main reason for the spiral recovery trend for optical depth is likely that decreasing optical depth leads to a fainter disc and therefore a lower absolute amplitude of the spirals.
At some point, further increases in optical depth should lead to a decreased spiral signal, as the spiral perturbation becomes small relative to the background emission (especially for the isothermal case, where there is no temperature perturbation). For image residuals, this limit appears to be reached between $\tau_0=1\text{--}3$ in some cases (e.g., $\beta=0$, $M_\text{p}=3\,M_\text{th}$); however, for visibility-space fitting, this limit is not reached in any of the cases ($\tau_0 \leq 3$) we consider. 
The main reason for spiral recovery declining for shorter cooling times is likely that the spiral temperature perturbation decreases (disappearing for our isothermal case).

The opposite gap and spiral recovery trends for optical thickness and cooling time may also partly indicate that deeper/wider gaps interfere with the spiral signal. This is certainly true for images, for which deeper/wider gaps reduce the brightness/amount of disc area over which the spiral can be traced (see Fig.~10 of \citetalias{speedieSpirals2022}). This effect appears replicated in the visibilities, although it is somewhat diminished, meaning our visibility-space approach may be less vulnerable to interference from gaps.


While we have only sparsely explored the parameter space here, and some additional caveats should be kept in mind (\S\ref{sec:Discussion}), our inferences suggest that looking for spirals to evidence embedded planets would be most useful in optically thicker discs with longer cooling times, 
and that detectable spirals may be present in regions of the parameter space with no detectable gaps in the image.\footnote{Some of these gaps may be detectable with super-resolution fits to the radial profile such as with \texttt{GALARIO} \citep{tazzariGALARIOGPU2018} and \texttt{frank} \citep{jenningsFrankensteinProtoplanetary2020}.}

To explore the effect of observational parameters (angular resolution and sensitivity), we tested five different ALMA set-ups, which are described in Table~\ref{tab:ALMA set-ups}. The recovery levels across the spiral simulation parameters for each are shown in Figure~\ref{fig:combined_figure}. The complete set of radial profiles and $\Delta\chi^2$ heatmaps used in determining the recovery levels of our fiducial set are shown in Figures~\ref{fig:C4C7_gofishies}-\ref{fig:C4C7_isos_L_combined}.
\begin{itemize}
    \item \textbf{Sensitivity.} For gaps, we see that sensitivity has a negligible effect on the \texttt{CLEAN} image profiles, with no change in recovery levels from the lower sensitivity to fiducial to higher sensitivity sets. 
    This is unsurprising since, once one achieves a reasonable sensitivity, the estimate of mean intensity in a given radial bin will already well-approximate the true mean, and further increases in sensitivity will have little effect.

    For spirals, we find that increasing sensitivity increases spiral model support (evidenced by larger $\Delta\text{BIC}$ values), but has negligible effect on the spatial structure of $\Delta\chi^2$ seen in the heatmaps.
    This is because the noise at each visibility point is largely independent of observing time, and the $\Delta\chi^2$ at each visibility point characterises the strength of the spiral relative to the axisymmetric background. Therefore, for a given ALMA configuration and spiral model instance, $\Delta\chi^2$ will scale linearly with the number of visibility points (or equivalently, the observing time), increasing model support but leaving the spatial structure of $\Delta\chi^2$ unchanged.
    For the lower sensitivity set, spiral model support is the limiting factor for spiral detection. Hence, going from the lower sensitivity to fiducial set results in improved recovery levels. For the fiducial set, however, the limiting factor for spiral detection becomes identifying the asymmetry (that is providing that spiral model support) as clearly spiral-like. Since we rely on the spatial structure of $\Delta\chi^2$ for this, going from the fiducial to higher sensitivity set brings no improvement in recovery levels.

    This is in stark contrast to the results for images in \citetalias{speedieSpirals2022}, which show a strong increase in spiral recovery for sensitivities beyond (our fiducial set's) $35\mu\text{Jy/beam}$ (see, e.g., their Fig.~7). However, only for our higher sensitivity set (which offer the best spiral recovery in images of the observations considered here or in \citetalias{speedieSpirals2022}) does the spiral recovery in images approach that of visibility-space fitting.

    \item \textbf{Resolution.} For gaps, our higher resolution set does slightly better than our fiducial set, with a few additional marginal recoveries, due to better resolving the gap.

    For spirals, our higher resolution set gives  $\Delta\chi^2$ heatmaps with spatial structures that indicate spirals slightly more clearly than our fiducial set. However, for this improvement they trade a decrease in $\Delta\chi^2$ magnitude, giving weaker spiral model support. This demonstrates the trade-off between spiral strength and clarity with resolution at fixed sensitivity mentioned in \S\ref{sec:maps vs residuals}. In this case, as for image residuals \citepalias{speedieSpirals2022}, the balance slightly favours a lower resolution (fiducial set).
\end{itemize}

\noindent Our Taurus comparison set's observational parameters are chosen to align closely with the Taurus disc sample analysed in \S\ref{sec:results Taurus}. They are observed at a slightly longer wavelength of 1.33\,mm, and are our lowest resolution and lowest sensitivity set (joint with our lower sensitivity set). 
As for the fiducial and higher sensitivity sets, the limiting factor for spiral detection in this set is the strength of the spiral signal. 
Comparing the detections in this set with the lower sensitivity set (which has the same sensitivity but higher resolution) suggests that, in this case, the trade-off between spiral strength and clarity favours a higher resolution, in contrast to the fiducial set versus higher resolution set comparison.

For these Taurus comparison observations, we also note a peculiar result: the spirals in the $M_\text{p}=3\,M_\text{th}$, $\beta=0$ discs are not clearly recoverable with their \textit{matching} spiral templates, with only the $\tau_0=0.1$ disc showing marginal spiral recovery, and the rest showing no recovery. All the planets can be clearly recovered, however, using \textit{non-matching} templates with lower amplitudes, achieved by increasing $\tau_0$ or decreasing $M_\text{p}$. For example, the planet in the $\tau_0=1$ disc can be recovered using the otherwise matching $M_\text{p}=1\,M_\text{th}$ template (as shown in the bottom right of Fig.~\ref{fig:band6_dusts}).
The reason for these template amplitudes being too high for their matching model discs may be explained by errors in the \texttt{frank} profiles. The \texttt{frank} fits produce profiles that have slightly brighter rings and darker gaps than the true profiles of the discs, an error likely due to their sharp inner and outer edges \citep{jenningsFrankensteinProtoplanetary2020}. The bright ring results in the template inducing a spiral amplitude that is too large in the ring's vicinity. As a result, templates using a lower amplitude reproduce the data more accurately. While this issue could arise when fitting real observations, real discs are unlikely to have sharp edges and thus the effect should be weaker. 

Our key conclusions from this section are as follows: planets are likely easier to detect in discs with shorter cooling times and lower optical depths via their gap signature. Many such gaps have already been detected in the field and are often interpreted as evidencing a planet. In contrast, in discs with longer cooling times and higher optical depths, planets may be easier to detect via their spiral signature. However, with typical image-space approaches, these spiral detections could require very high sensitivities (e.g., the $10\,\mu\text{Jy/beam}$ of our higher sensitivity set, requiring 8\,hrs of on-source time), which could partly explain the paucity of detections. A visibility-space fitting approach may be able to achieve sufficient detection capabilities with much more moderate sensitivities (e.g., the $35\,\mu\text{Jy/beam}$ of our fiducial set, requiring only 40\,mins of on-source time). 

In complement to this section, in Appendix~\ref{sec:supplementary heatmaps} we provide the full set of radial profiles and $\Delta\chi^2$ heatmaps for our fiducial set of observations, and provide interpretations of the heatmaps that we made in arriving at our spiral recovery levels (middle row of Figure~\ref{fig:combined_figure}).

\section{Results: Taurus disc sample} \label{sec:results Taurus}

\begin{figure}
\includegraphics[width=\columnwidth]{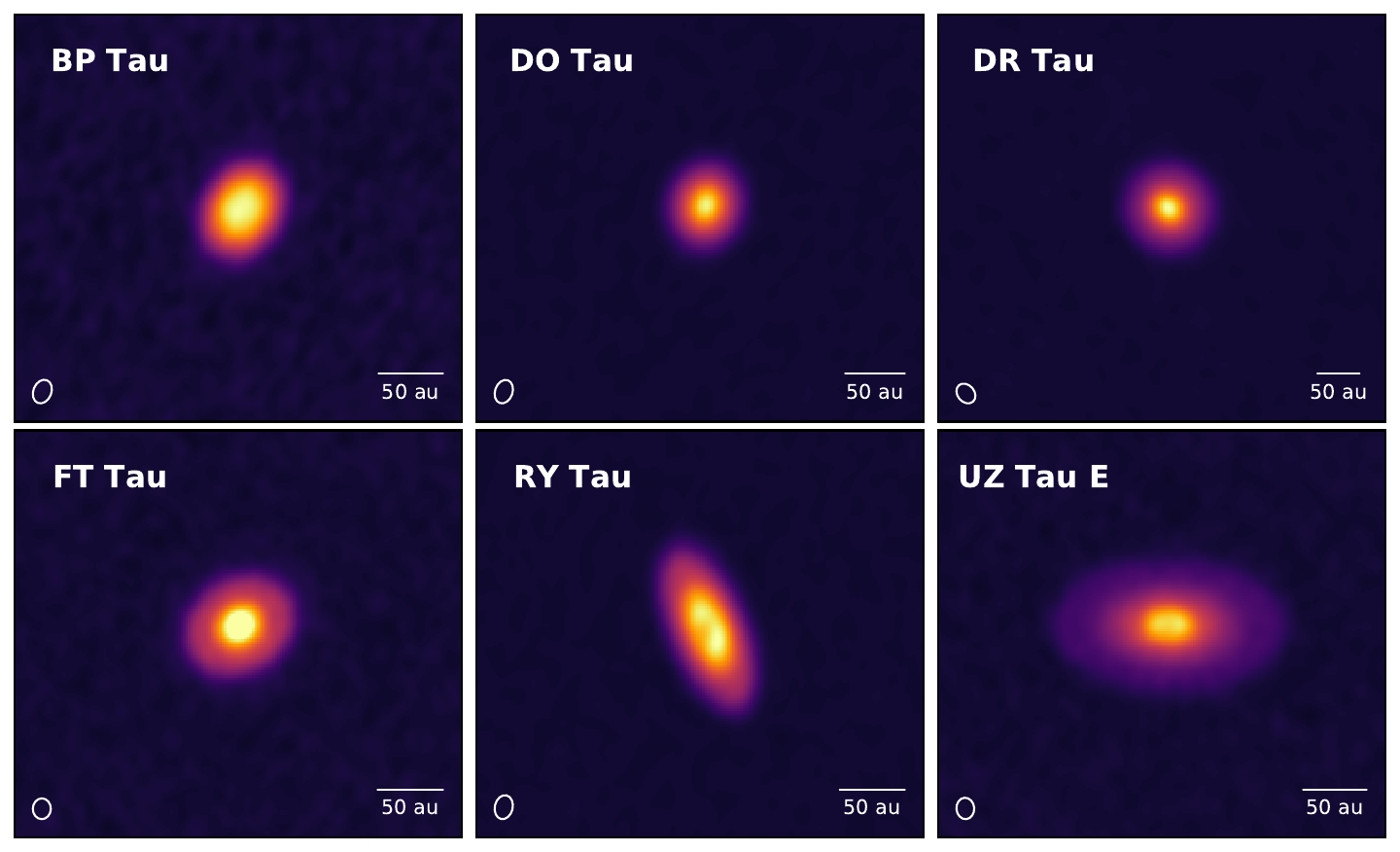}
    \caption{The 1.33\,mm \texttt{CLEAN} images for our disc sample. Each panel is $2.4''\,\times\,2.4''$, with the synthesized beam shown in the bottom left corner of each panel.}
    \label{fig:clean_panels}
\end{figure}

In this section we apply our method to six `smooth' (relative lack of annular substructure) discs from the ALMA Taurus survey presented in \citet{longCompactDisks2019} (Table~\ref{tab:disc properties}, Fig.~\ref{fig:clean_panels}). The Taurus survey observations have an angular resolution $\sim 0.12''$ and a mean sensitivity $\sim 50\,\mu\text{Jy/beam}$ at 1.33\,mm (Band 6), resulting from on-source times between 4 and 10\,mins in the C-6 configuration. To generate the Taurus discs' radial brightness profiles in our method, we use the same \texttt{frank} hyperparameters as for the synthetic observations. Again, the resulting profiles are quite insensitive to variations within the recommended ranges.

\begin{table} 
    \centering
    \caption{Parameters of our Taurus sample discs. $d$ is the distance to the source (using Gaia DR2 measurements from \citet{bailer-jonesEstimatingDistance2018}). Total specific flux $F_\text{1.33\,mm}$ and $R_{\text{eff,95\%}}$ (the disc radius containing $95\%$ of the flux) are derived from the \texttt{frank} profiles for each disc. The inclination (inc.) and position angle ($PA$) are from \citet{longCompactDisks2019}.}
    \label{tab:disc properties}
    \renewcommand{\arraystretch}{1.5}
    \begin{tabular}{ 
        p{1.05cm}  
        >{\centering\arraybackslash}p{0.78cm} 
        >{\centering\arraybackslash}p{1.0cm}   
        >{\centering\arraybackslash}p{1.0cm} 
        >{\centering\arraybackslash}p{1.11cm} 
        >{\centering\arraybackslash}p{1.11cm}}  
        \hline
        \parbox[c][0.8cm]{1cm}{Disc} & \;$d$\; [pc] & $F_\text{1.33\,mm}$ [Jy] & \(R_{\text{eff,95\%}}\) [$''$] & \;\;inc.\;\; $[^\circ]$ & \;\;\,$\text{PA}$\;\;\; [$^\circ$] \\
        \hline
        \multicolumn{6}{c}{Compact Discs} \\
        \hline
        BP Tau & 129 & 0.045 & \(0.321\) & \(38.2_{-0.5}^{+0.5}\) & \(151.1_{-1.0}^{+1.0}\) \\
        DO Tau & 139 & 0.12 & \(0.263\) & \(27.6_{-0.3}^{+0.3}\) & \(170.0_{-0.9}^{+0.9}\) \\
        DR Tau & 195 & 0.13 & \(0.276\) & \(5.4_{-2.6}^{+2.1}\) & \(3.4_{-8.0}^{+8.2}\) \\
        FT Tau & 127 & 0.091 & \(0.357\) & \(35.5_{-0.4}^{+0.4}\) & \(121.8_{-0.7}^{+0.7}\) \\
        \hline
        \multicolumn{6}{c}{Extended Discs} \\
        \hline
        RY Tau & 128 & 0.21 & \(0.509\) & \(65.0_{-0.1}^{+0.1}\) & \(23.1_{-0.1}^{+0.1}\) \\
        UZ Tau E & 131 & 0.13 & \(0.667\) & \(56.1_{-0.4}^{+0.4}\) & \(90.4_{-0.4}^{+0.4}\) \\
        \hline
    \end{tabular}
\end{table}

\begin{figure*}
    \begin{tabular}[b]{@{}c@{}} 
        \rotatebox{90}{\parbox{3.6cm}{\centering\large $\beta=10$}} \\ [0.9cm]
        \rotatebox{90}{\parbox{3.6cm}{\centering\large $\beta=0$}} \\ 
        \hspace{0.6cm}
        
    \end{tabular}%
    \begin{minipage}[b]{0.95\linewidth}
        \begin{minipage}{0.27\linewidth}
            \centering
            \hspace{-.1cm} \large$M_\text{p}=0.3\,M_\text{th}$
        \end{minipage}\hfill
        \begin{minipage}{0.3\linewidth}
            \centering
            \hspace{1.1cm} \large$M_\text{p}=1\,M_\text{th}$
        \end{minipage}\hfill
        \begin{minipage}{0.43\linewidth}
            \centering
            \large$M_\text{p}=3\,M_\text{th}$
            \hspace{0.2cm}
        \end{minipage}
        
        \includegraphics[width=\linewidth]{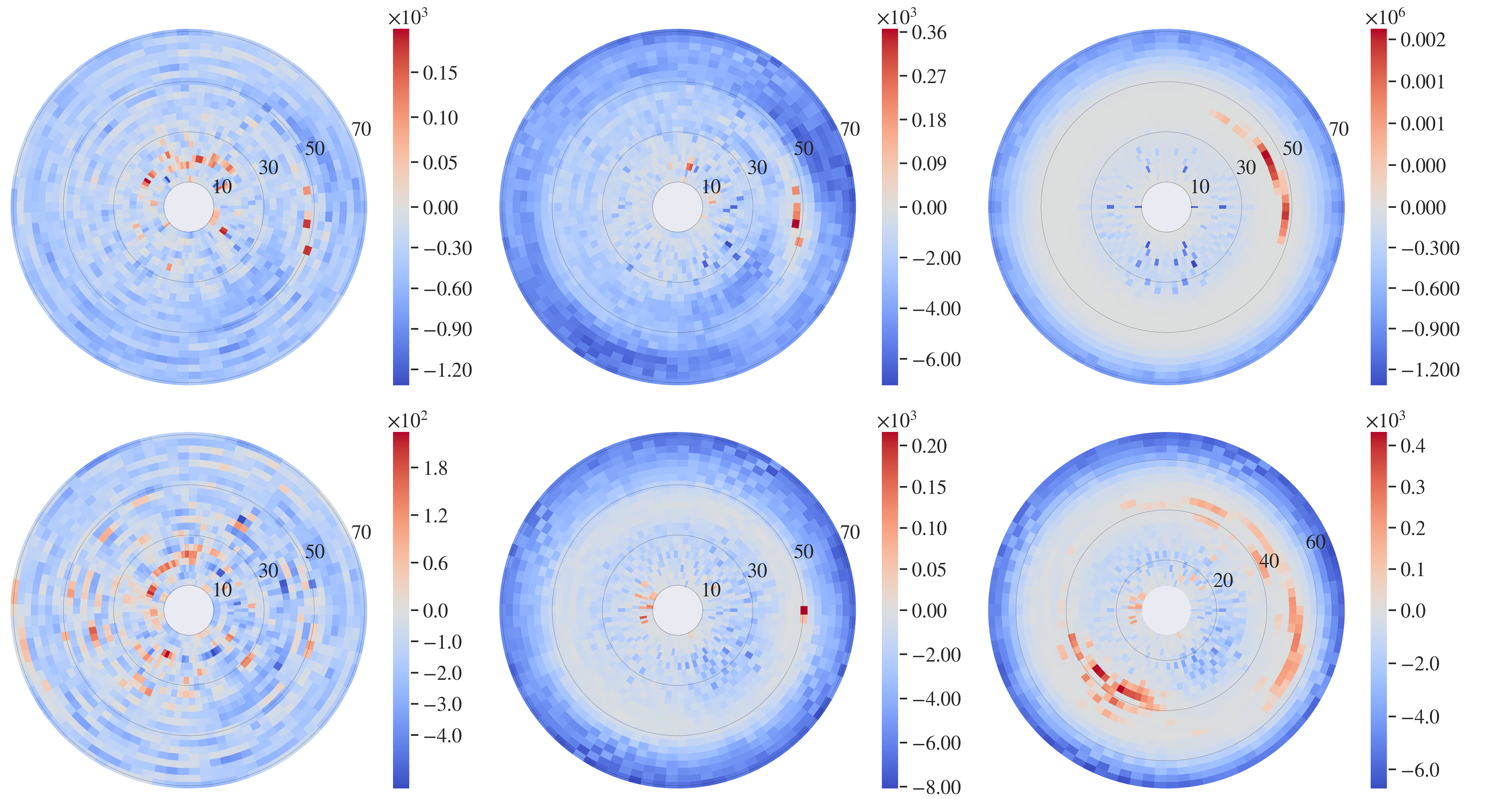}

    \end{minipage}

    \caption{$\Delta\chi^2$ heatmaps of $\tau_0=1$, right-handed model discs with their matching spiral templates (except for the $\beta=0$, $M_\text{p}=3\,M_\text{th}$ disc, whose planet is not recovered with its matching template, but is with the otherwise matching $M_\text{p}=1\,M_\text{th}$ template (among others). $M_\text{p}/M_\text{th}=0.3,1,3$ from left to right, and $\beta=0,10$ from bottom to top. The discs are observed at $1.33\,\text{mm}$ with an angular resolution of $0.13''$ and sensitivity of $50\,\mu\text{Jy/beam}$ (ALMA C-6 configuration with 6\,mins of on-source time).} 
    \label{fig:band6_dusts}
\end{figure*}

Our Taurus sample's mean observational parameters (wavelength, resolution and sensitivity) align closely with our Taurus comparison set of synthetic observations (Fig.~\ref{tab:ALMA set-ups}, bottom row of Figure~\ref{fig:combined_figure}). Their total fluxes, derived from the \texttt{frank} profiles, range from $\approx 0.05$ to $0.2\,\text{Jy}$ (BP Tau to RY Tau). Those of our Taurus comparison set are generally somewhat brighter, and range from $\approx 0.2$ to $1.3\,\text{Jy}$ ($\tau_0=0.1$ to $3$, with some small variations with planet mass and cooling time). The synthetic observations are therefore effectively somewhat more sensitive than our Taurus observations. The effective radii of our Taurus sample discs (mean $R_{\text{eff,95\%}} \approx 55\,\text{au}$) are also somewhat smaller than that of our model discs ($\approx 95\,\text{au}$). 
Overall, the Taurus sample and comparison synthetic observations are broadly comparable, and we can use the successful spiral recoveries achieved in those synthetic observations to inform what a spiral-like signal should look like in the $\Delta\chi^2$ heatmaps of the Taurus sample. The  heatmaps of an illustrative set of these synthetic observations is shown in Figure~\ref{fig:band6_dusts}. Recall that $\Delta\chi^2 = \chi^2_{AO} - \chi^2_{SO}$, i.e., the difference in goodness-of-fit between the axisymmetric model and spiral model instance. Positive (negative) $\Delta \chi^2$ values represent an improvement (decline) in fit with the spiral model instance. 
Since, of the simulation parameters, optical thickness has the smallest effect on spiral morphology (mostly just acting to increase or decrease the amplitude of the spiral perturbation), we present the heatmaps in this section at a fixed optical thickness ($\tau_0=1$) for simplicity.

\subsection{Spiral fits} \label{sec: spiral fits to Taurus discs}

Of the six discs in our Taurus sample (Table~\ref{tab:disc properties}, Fig.~\ref{fig:clean_panels}), BP Tau, DO Tau, FT Tau, and UZ Tau E show little asymmetry, each giving $\Delta \text{BIC}<-10$, and therefore strongly favouring the axisymmetric model. In contrast, DR Tau and RY Tau exhibit consistent localised regions of $\Delta \chi^2 \sim 100$ in the inner disc (e.g., Fig.~\ref{fig:DR/RY}), strongly favouring the spiral model and passing our first test (\S\ref{sec:spiral fits}). They do not, however, pass our second test. Both fail to clearly distinguish handedness, with left- and right-handed templates achieving similar peak $\Delta\chi^2$. Hence, we do not find evidence for planet-driven spirals in these discs. Nevertheless, the $\Delta\chi^2$ heatmaps provide insights into the nature of the observations and the method, and the possibility of planets hiding in the discs. 

\begin{figure*}
    \centering
    {\LARGE{FT Tau}} \\[10pt]  
    \includegraphics[width=0.99\textwidth]{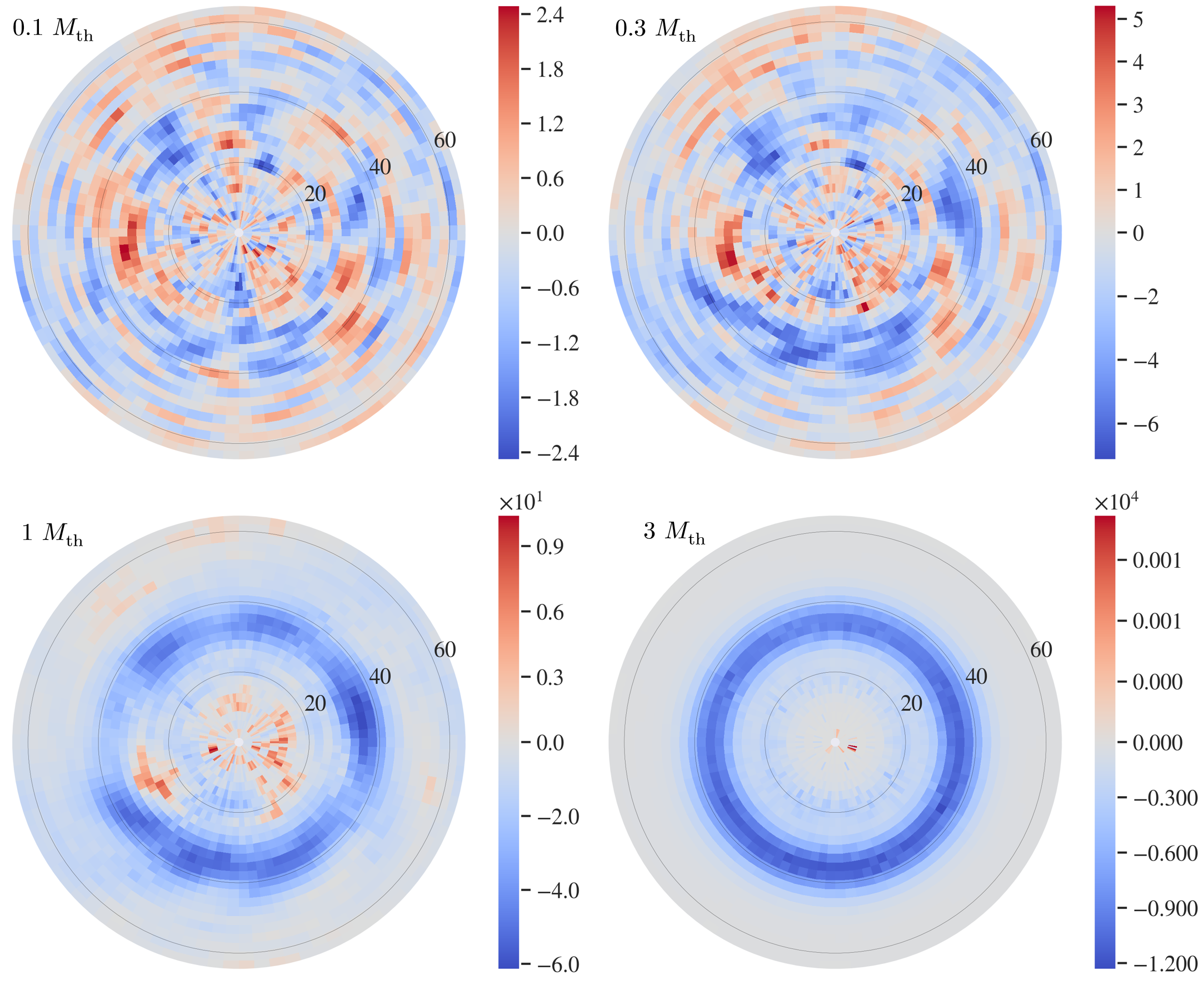}
    \caption{$\Delta \chi^2$ heatmaps of FT Tau for $\beta=10, \tau_0=1$, and right-handed spiral templates. The heatmap spatially represents the deprojected disc, with the disc's minor axis projected onto the horizontal axis ($\phi_\text{p} = 0$). The radial ticks are in au.}
    \label{fig:4circles}
\end{figure*}

\begin{figure*}
    \includegraphics[width=\textwidth]{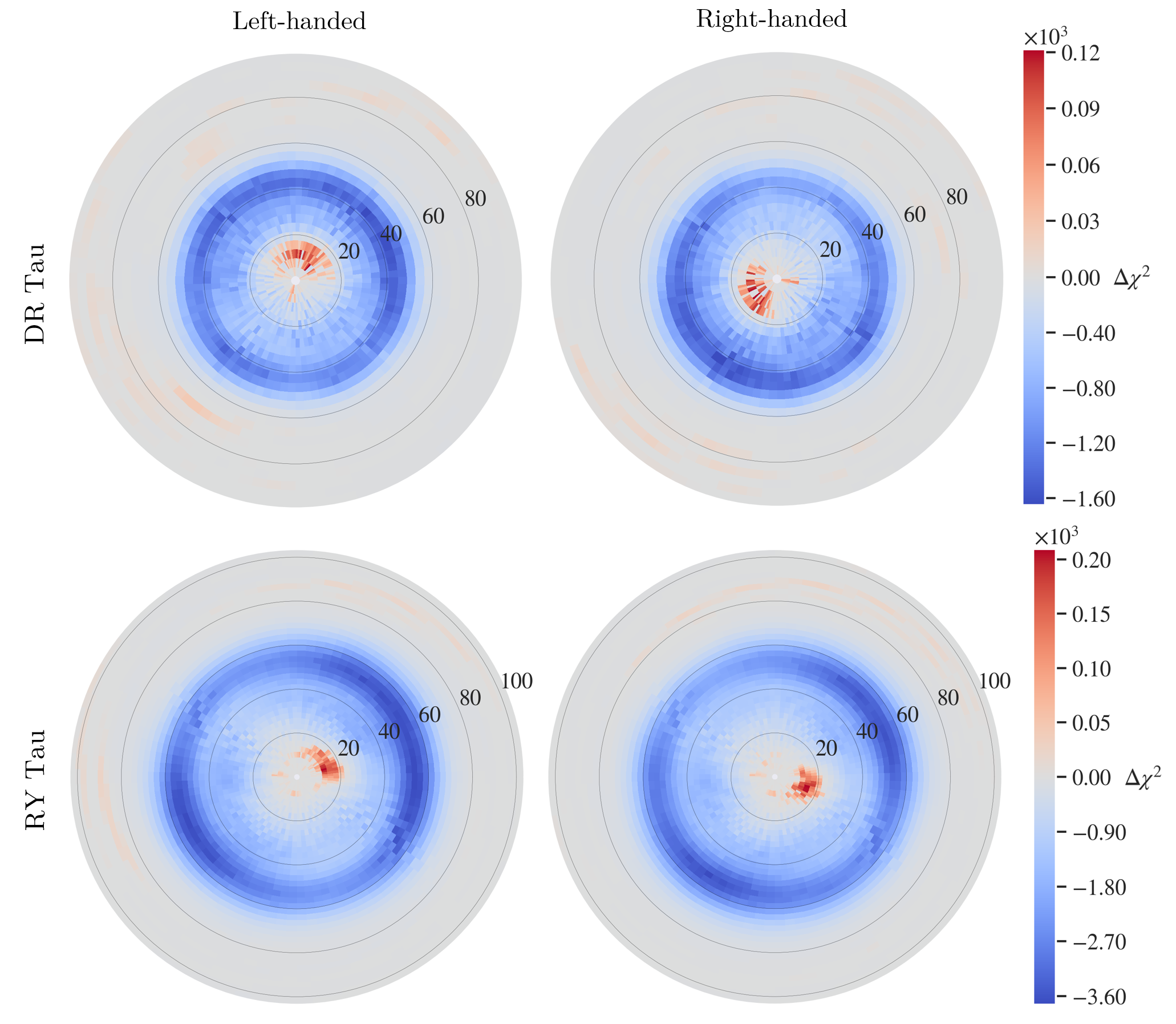}
    \caption{$\Delta \chi^2$ heatmaps of DR Tau (top row) and RY Tau (bottom row) for $\beta=10$, $\tau_0=3$, $M_\text{p}=3\,M_\text{th}$ left- and right-handed spiral templates. DR Tau and RY Tau show maximum $\Delta\chi^2 \approx 100$ and $200$ respectively, with similar results for both hands. The colourbars apply to each row of heatmaps.}
    \label{fig:DR/RY}
\end{figure*}

Using FT Tau as an example of the low-asymmetry discs, the $\Delta\chi^2$ heatmaps in Figure~\ref{fig:4circles} for the high mass ($1$ and $3\,M_\text{th}$) planets exhibit large negative $\Delta \chi^2$ ($\sim -50$ for the $1\,M_\text{th}$ case and $\sim -10^4$ for the $3\,M_\text{th}$ case) over significant regions of the disc. These planets therefore clearly have a significant effect on the disc structure and would be detectable if they were present, as suggested by the Taurus comparison set.
Considering this, one can rule out the presence of these higher-mass planets over the majority of the radial extent of FT Tau. For example, it is evident from the \(\Delta \chi^2\) in the \(3 \, M_{\text{th}}\) heatmap (bottom right of Fig.~\ref{fig:4circles}) that a \(3 \, M_{\text{th}}\) planet residing between 10 and 60\,au is unlikely. It is more difficult to draw conclusions about planets in the inner disc, where there are some \(\Delta \chi^2>0\) planet positions and results are inherently unreliable due to the spiral contributing on scales smaller than the angular resolution of the observations. Similarly, conclusions for large \(r_\text{p}\) are limited by low disc brightness. We apply this reasoning more fully in \S\ref{Sec:which planets hiding}. The two blue rings (at $\approx 18$ and $34\,\text{au}$) visible in the \(3 \, M_{\text{th}}\) heatmap of Figure~\ref{fig:4circles} coincide with radii of elevated brightness (and are consistent with the radial \(\chi_{SA}^2\) plots; see Fig.~\ref{fig:squiggles}). 

For the low mass ($0.1$ and $0.3\,M_\text{th}$) planets, the large negative $\Delta \chi^2$ disappear. Instead, the heatmaps appear `noisy' and have small $\Delta \chi^2$ values ($|\Delta \chi^2|<10$), roughly symmetrically distributed about zero. 
This is suggestive of a detectability limit in the observations; if a low mass planet were present in the disc, its spiral would likely be indistinguishable from observational noise. The equivalent heatmaps of RY Tau and UZ Tau E (the extended discs) generally show slightly more structure than the compact discs, possibly due to their larger size and brightness allowing for effectively a greater resolution and sensitivity. Both sets of discs, however, suggest a minimum planet mass for detectability of between $0.3$ and $1\,M_\text{th}$ across the disc radii considered, which is broadly similar to the range suggested by the Taurus comparison set (as indicated in the bottom right of Fig.~\ref{fig:combined_figure}).

The \(\Delta \chi^2\) analysis for the two more asymmetric discs (DR Tau and RY Tau) shares many of the same features as for the low-asymmetry discs: noisy heatmaps suggesting the low detectability of low-mass planets, and blue rings and consistent \(\Delta \chi^2<0\) suggesting a lack of high mass planets across most of the disc. However, both of these discs exhibit much larger $\Delta\chi^2>0$ at inner radii (\(0.04'' \leq r_p \leq 0.08''\) for DR Tau and \(0.08'' \leq r_p \leq 0.18''\) for RY Tau, which has an inner cavity) for the high mass planets, suggesting a much stronger asymmetry in these discs.
These regions of improvement are localised and consistent across different spiral templates.\footnote{They are also in reasonable agreement with where one might expect an asymmetry to be from the corresponding imaged \texttt{frank} residuals in Figures~6 and 11 of \citealp{jenningsSuperresolutionTrends2022}.}

However, as can be seen in Figure~\ref{fig:DR/RY}, spiral templates of either handedness result in similarly large peak $\Delta\chi^2$ values. This indicates that the templates are probably not fitting genuine spirals, but some other asymmetry. The small position offsets between the peak left- and right-handed $\Delta\chi^2$ for RY Tau suggest that these high mass planet templates could be fitting some arc-like asymmetry in the disc. 
In fact, higher-resolution scattered light observations for RY Tau \citep{francisDustdepletedInner2020} show such an arc, extending over \(\sim 180^\circ\) in azimuth, at the same location as indicated by the \( \Delta \chi^2 \) heatmaps. Since RY Tau has a significant inner cavity and is highly inclined (\(65.0^\circ\)), this could even be from the viewing angle of the hotter disc wall at the edge of the cavity (Ribas et al., submitted), rather than any asymmetry in the disc density. 

The fitting in the equivalent heatmaps of DR Tau show a slightly more `spotty' structure that broadly resembles the false fitting seen for errors in the disc phase centre (Appendix~\ref{sec:errors}). \citet{jenningsSuperresolutionTrends2022} find the same asymmetry in their imaged \texttt{frank} residuals and show that it is not due to an incorrect determination of the disc phase centre (see their Appendix~A), instead concluding that there is genuine unresolved inner disc structure. DR Tau is also only slightly inclined (\(5.4^\circ\)), hence the viewing angle of a disc wall is not a likely explanation. Other possible explanations include an azimuthal dust trap (such as in Oph-IRS 48; \citealp{brudererGasStructure2014}), or a vortex (such as in HD 34282; \citealp{marrAppearanceVortices2022}). Such structures could of course point to the presence of a planet in the disc, but the evidence would be separate to the spirals considered in this work.

\subsection{Which planets can we rule out?} \label{Sec:which planets hiding}

Extending the reasoning in the previous section, for a given planet mass and location within a disc, we infer that there is unlikely to be a planet of that mass or greater at that location if all the plausible spiral templates which account for such a planet yield $\Delta\chi^2$ values which give $\Delta\text{BIC}<-10$.\footnote{The BIC is strictly only valid for the maximum $\Delta\chi^2$ achieved by the spiral model, but it can still serve as a guide to which $\Delta\chi^2>0$ values are plainly insignificant.} We also require that these planet masses are \textit{capable} of inducing a significant $\Delta \chi^2$, which is indicated by a large negative $\Delta \chi^2$ at other comparable locations in the disc. The first condition is a requirement for there being no evidence for the planet, and the second is a requirement for the planet to be detectable.

Hence, using the $\Delta\chi^2$ heatmaps, we can suggest planets to reject over ranges of disc radii. Table~\ref{tab:exclusion ranges} provides suggestions for the \( r_\text{p} \) ranges over which one can rule out planets greater than 1 and \(3 \, M_{\text{th}}\) ($\approx 0.3$ and $0.9\,M_\text{Jup}$ at 50\,au) based on these considerations. 

\begin{figure}
    \includegraphics[width=0.98\columnwidth]{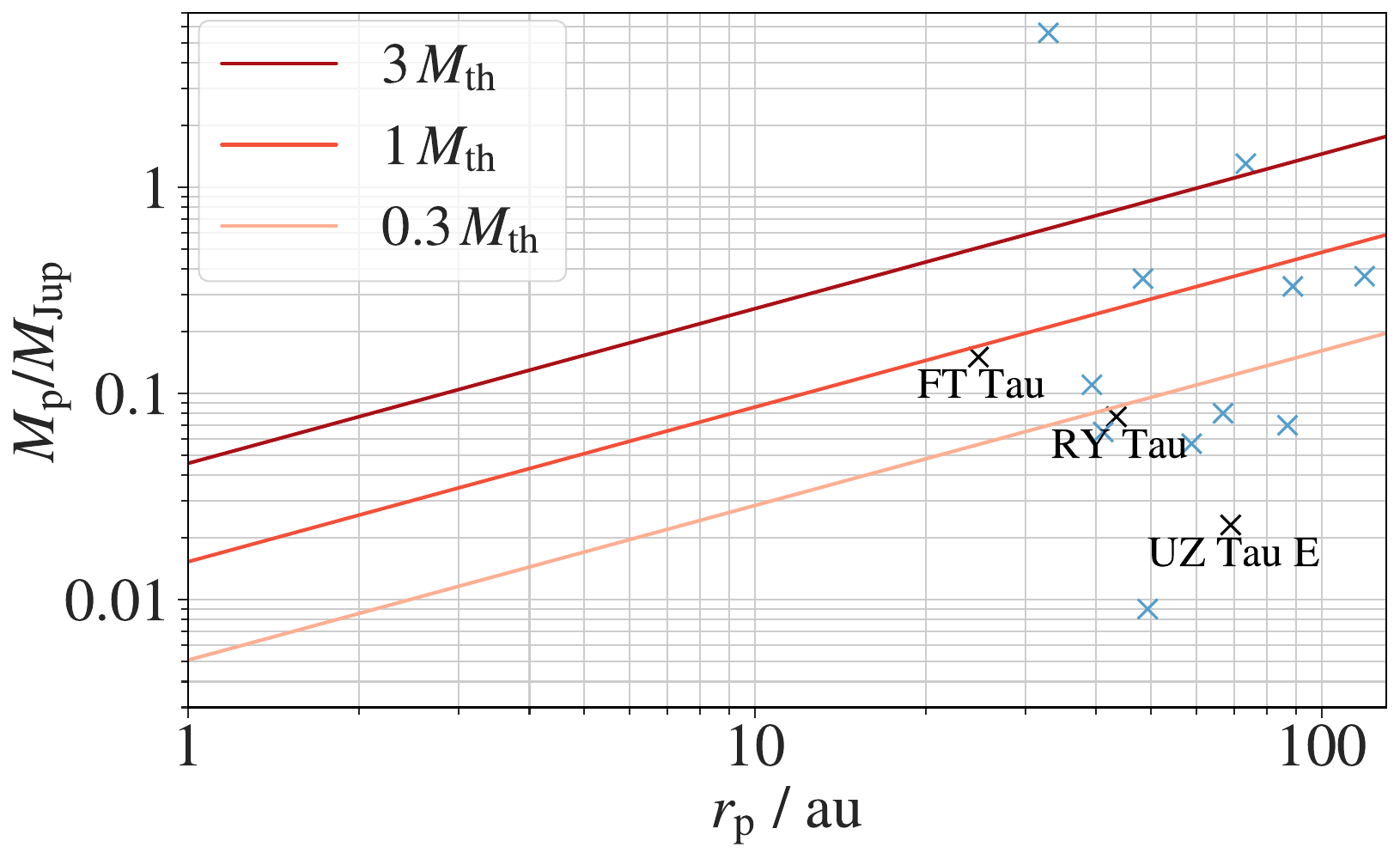}
    \caption{Planet mass against orbital radius for $0.3$, $1$ and $3$ thermal masses ($M_\text{th} \propto r_\text{p}^{3/4}$). The minimum mass for driving a detectable spiral is between 0.3 and $1\,M_\text{th}$ for disc observations like those presented in \citet{longCompactDisks2019}. The crosses indicate the planet masses and orbital radii inferred by \citet{lodatoNewbornPlanet2019} for the disc sample in \citet{longGapsRings2018} (excluding the $15.7\,M_\text{Jup}$ planet inferred for CI Tau). The labelled black crosses indicate discs that we also consider in this work.}
    \label{fig:mass_r}
\end{figure}

From Table~\ref{tab:exclusion ranges}, we can place an upper mass limit of approximately \( 3 \, M_{\text{th}} \) for planets at intermediate radii of the discs, with conclusions at the innermost and outermost radii being limited by insufficient resolution and low disc brightness, respectively. In each case, the outermost radii are notably beyond the effective `edges' of the discs in \(1.33 \, \text{mm}\). We can drop this limit to approximately \( 1 \, M_{\text{th}} \) over most of DO Tau, and smaller annuli (centred on bright rings) in FT Tau and UZ Tau E. Recalling that our models have \( M_\text{th} \propto r_\text{p}^{3/4} \), we note that these thermal mass limits correspond to different absolute masses at different orbital radii. The mapping between thermal masses and Jupiter masses is shown in Figure~\ref{fig:mass_r}, and the resulting \( M_\text{p} \) ranges are included in Table~\ref{tab:exclusion ranges}.

\begin{table*}
    \centering
    \caption{Approximate orbital radii ranges for which we can rule out planets of masses greater than 3 and $1~M_\text{th}$ for our Taurus sample discs. The corresponding planet mass range for the 3 and $1~M_\text{th}$ lower limits are given in the adjacent columns to the right. $R_{\text{eff,95\%}}$ is the disc radius containing $95\%$ of the total flux. The final two columns give the planet orbital radius and mass inferred from the gap present in these discs (Table~1 of \citealp{lodatoNewbornPlanet2019}).}
    \label{tab:exclusion ranges}
    \renewcommand{\arraystretch}{1.3}
    \begin{tabular}{l c c c c c !{\VRule[0.4pt]}c c}
        & \multicolumn{5}{c!{\VRule[0.4pt]}}{} & \multicolumn{2}{c}{From \citet{lodatoNewbornPlanet2019}} \\         
        \hline
         Disc & 
        \parbox[c][1.4cm]{1.5cm}{\centering $R_\text{eff,95\%}$ [au]} & 
        \parbox{2cm}{\centering \( r_\text{p} \) exclusion range for \( M_\text{p} \geq 3 M_{\text{th}} \) [au]} & 
        \parbox{2cm}{\centering \( M_\text{p} = 3 M_{\text{th}} \) mass range [\( M_{\text{Jup}} \)]} & 
        \parbox{2cm}{\centering \( r_\text{p} \) exclusion range for \( M_\text{p} \geq 1 M_{\text{th}} \) [au]} & 
        \parbox{2cm}{\centering \( M_\text{p} = 1 M_{\text{th}} \) mass range [\( M_{\text{Jup}} \)]} & 
        \parbox{1.7cm}{\centering Inferred \( r_\text{p} \) [au]} & 
        \parbox{1.7cm}{\centering Inferred \( M_\text{p} \) [\( M_{\text{Jup}} \)]} \\
        \hline
        BP Tau & 41 & (10,\ 54) & (0.26,\ 0.91) & - & - & - & - \\
        DO Tau & 37 & (8,\ 61) & (0.22,\ 1.00) & (11,\ 61) & (0.09,\ 0.33) & - & - \\
        DR Tau & 54 & (23,\ 62) & (0.48,\ 1.01) & - & - & - & - \\
        FT Tau & 45 & (10,\ 64) & (0.26,\ 1.04) & (33,\ 43) & (0.21,\ 0.26) & 25 & 0.15 \\
        RY Tau & 65 & (23,\ 82) & (0.48,\ 1.25) & - & - & 43 & 0.077 \\
        UZ Tau E & 87 & (13,\ 97) & (0.31,\ 1.42) & (55,\ 63) & (0.31,\ 0.34) & 69 & 0.023 \\
        \hline
    \end{tabular}
\end{table*}

One can compare our locus of upper mass limits to the masses and orbital radii \citet{lodatoNewbornPlanet2019} inferred from the Taurus disc sample in \citet{longGapsRings2018} (Figure~\ref{fig:mass_r}). Their focus is on radially structured discs, whereas our sample consists of relatively smooth discs. Only FT Tau, RY Tau and UZ Tau E from our sample are included in their analysis, with their inferred embedded planet masses sitting well below the masses that we rule out. RY Tau and UZ Tau E's inferred planets also sit notably below our suggested minimum mass locus for driving a detectable spiral ($0.3\text{--}1\,M_\text{th}$ across the disc radii considered), with FT Tau's sitting in the zone between detectable and undetectable. Therefore, if such planets are present, they are unlikely to be detectable via their spiral signature in these observations. It is worth noting that \citet{lodatoNewbornPlanet2019} did not impose a minimum gap depth to make these inferences, and the gaps appear quite shallow in the discs' parametric profiles (UZ Tau E's in particular) that they used (from \citealp{longGapsRings2018}).\footnote{They remain shallow in the discs' \texttt{frank} profiles and do not appear in their \texttt{CLEAN} image profiles. See Figures~6 and 11 of \citealp{jenningsSuperresolutionTrends2022} for comparisons of these three profiles for each of the discs.} 

A number of the gaps considered in \citet{lodatoNewbornPlanet2019} (from various sources) do yield inferred planet masses sitting above the lower limit for detectability. This limit is also likely lower for discs observed at higher sensitivities and resolutions than the sample from \citet{longGapsRings2018}, such as the DSHARP discs whose gap widths were measured in \citet{zhangDiskSubstructures2018}. The wider gaps in these disc observations will give us a better chance to support or oppose planets as their progenitors.

\section{Discussion} \label{sec:Discussion} 

The method described in this work is the first implementation of a visibility-space spiral fitting approach, and it faces a number of issues and uncertainties. Rather than offering a ready-to-go visibility-space method for finding planet-driven spirals, this paper is intended as a proof-of-concept for this approach and a foundation for a more sophisticated method based upon it. Here we give a couple key issues that the approach faces, and offer suggestions for how they could be addressed in future work. 

\paragraph*{Misleading signals.} As we saw for DR Tau and RY Tau in \S\ref{sec: spiral fits to Taurus discs}, non-spiral asymmetries such as arcs can be fit by the spiral model (to a statistically significant degree). In order to reliably recover a planet-driven spiral, it is necessary to distinguish between these misleading signals and true signals. The handedness test (Fig.~\ref{fig:handedness test}) is an example of an initial attempt at this. This can be readily improved upon, for instance, by introducing models of non-spiral asymmetries (e.g., bright spots or arcs) and comparing their signal to that of the spiral model. Distinguishing between the morphology of spirals driven by planets, and those driven by other mechanisms (e.g., stellar fly-bys or gravitational instability) may also be worthwhile. 

\paragraph*{Alternative spiral morphologies.} The model discs present more favourable observations for our method than would be the case for real discs because of the perfect correspondence between their spiral and the method's spiral templates. The multitude of assumptions made in generating these spiral simulations will undoubtedly result in a deviation from the planet-driven spiral structure in real discs. Our tests in \S\ref{sec:effect of spiral model parameters} suggest that this may not be a big problem for planet recovery as the method can often pick out the presence of a spiral when using templates that do not exactly match the model disc's (e.g., Figs.~\ref{fig:mismatches} and \ref{fig:H/r}). 
It is possible, however, that certain parameters assumed constant here, or neglected physics, could have a stronger effect on planet recovery. Examples include: the assumption of the planet being on a circular, co-planar, and non-migrating orbit \citep{quillenDrivingSpiral2005, duffellECCENTRICJUPITERS2015}, viscosity \citep{fungHOWEMPTY2014}, dust scattering \citep{sierraEffectsScattering2020}, vertical disc structure \citep{krapp3DDust2022}, disc inclination \citepalias{speedieSpirals2022}, MHD winds \citep{wafflard-fernandezPlanetdiskwindInteraction2023}, and interfering disc substructures (e.g., spirals/gaps from other planets). 
Related to this, the degeneracy in fit across cooling time, optical depth, and planet mass (particularly optical depth and planet mass), as well as simulation assumptions and neglected physics, make it difficult to infer planet mass from a spiral signal (a similar degeneracy occurs for images; \citetalias{speedieSpirals2022}), although with constraints on the cooling timescale and optical thickness from other data this would improve. Such constraints would lead to a similar improvement in our ability to rule out planets (as done in \S\ref{Sec:which planets hiding}).

The extent of these limitations depends on the coverage of possible spiral morphologies provided by the spiral simulations we use in the fitting approach. The method which we developed here employs the pre-generated hydrodynamical spiral simulations of \citetalias{speedieSpirals2022}. Whilst these `first-principles' simulations have many advantages, they are computationally expensive and slow. A shift towards faster simulation methodologies would enable a larger and finer exploration of the parameter space, allowing for a greater coverage of possible spiral morphologies (including possibly other non-spiral asymmetries) and tighter constraints on planet masses.

One possibility for achieving this is machine learning models. For instance, \texttt{PPDONet} \citep{maoPPDONetDeep2023} can produce 2D gas surface density perturbation maps for arbitrary combinations of planet mass, disc viscosity, and disc aspect ratio in "less than a second on a laptop". Alternatively, one could take the (semi-)analytic route; \texttt{wakeflow} \citep{bollatiTheoryKinks2021,hilderWakeflowPython2023} can swiftly generate 2D gas surface density perturbation heatmaps for a range of planet and disc properties, and has the added advantage of interfacing with \texttt{MCFOST} \citep{pinteMonteCarlo2006,pinteBenchmarkProblems2009} to perform the necessary radiative transfer.

\section{Summary and conclusions} \label{sec:Conclusions} 

In this paper we have explored the planet detection capabilities of fitting planet-driven dust spirals to protoplanetary disc observations in visibility space. We devised a method for this approach in \S\ref{Sec:Methods} which combines an axisymmetric model of the given disc with a series of planet-driven spiral simulations to create a series of spiral-containing model discs. The model disc images are transformed to visibility space and fit to the observed visibilities using $\chi^2$ statistics.
Through our tests on synthetic disc observations from an extended version of the set generated in \citet{speedieSpirals2022} (S22), we explored how different model parameters affect the fitting results, and compared its ability to identify planets with looking for spirals in \texttt{CLEAN} image residuals and gaps in \texttt{CLEAN} image profiles.
We also applied our method to six smooth discs from the ALMA Taurus survey presented in \citet{longCompactDisks2019} (\S\ref{sec:results Taurus}). Our main conclusions can be summarised as follows:
\begin{enumerate}
    \item The method developed in this work significantly outperforms image residuals in recovering spiral-driving planets from the synthetic observations, showing clear signals for spirals well beyond the detection limits of image residuals (Fig.~\ref{fig:combined_compare}). The planet recovery ability of the method may also be less vulnerable to interference from annular gaps than image-space methods, and could potentially be used to support or oppose a planet origin for such observed gaps. Taken together, the discovery of a spiral co-located with a gap would constitute strong evidence for a planet.
    
    \item The trends in the detectability of gaps and spirals with cooling timescale and optical depth are opposite, with spirals being easier to detect for longer cooling times and higher optical depths, while gaps are more easily opened for shorter cooling times and lower optical depths. In the synthetic observations, spirals were recovered in significant regions of the disc/observational parameter space where gaps were not (Fig.~\ref{fig:combined_figure}).

    \item Our sample of Taurus disc observations and the synthetic Taurus comparison set suggest planets with masses $> 0.3\text{--}1\,M_{\text{th}}$ can drive detectable spirals in such observations. This corresponds to an absolute mass range of $\approx 0.03$ to $0.5\,M_{\text{Jup}}$ over orbital radii of 10 to $100\,\text{au}$.
    
    \item In our sample of Taurus discs we find evidence of inner disc asymmetries in DR Tau and RY Tau, but no planet-driven spirals (\S\ref{sec: spiral fits to Taurus discs}). Higher resolution/sensitivity disc observations (of which many already exist), and more structured discs, may offer a better chance of detection.
    
    \item For our sample of Taurus discs, we can reasonably rule out the existence of any planets with masses \( \gtrsim 3 \, M_{\text{th}} \) over orbital radii of 20 to $60\,\text{au}$, corresponding to a mass range of $\approx$ \(0.5\) to \(1 \, M_{\text{Jup}}\). The exclusion of \( \gtrsim 1 \, M_{\text{th}} \) planets is possible over more limited radii ranges for some of the discs (Table~\ref{tab:exclusion ranges}).
    
    \item The method developed in this work is the first implementation of a visibility-space spiral fitting approach, and it has a large scope for improvement. Despite this, it demonstrated a promising ability to recover planet-driven spirals in the synthetic observations. This suggests that such an approach warrants further investigation and may be able to find planet-driven spirals in disc observations that have not yet been found with existing methods. 
\end{enumerate}


\section*{Acknowledgements}


ETS thanks the Leverhulme Centre for Life in the Universe for their support via an internship.
AR has been supported by the UK Science and Technology Facilities Council (STFC) via the consolidated grant ST/W000997/1 and by the European Union’s Horizon 2020 research and innovation programme under the Marie Sklodowska-Curie grant agreement No. 823823 (RISE DUSTBUSTERS project).
RAB thanks the Royal Society for their support through a University Research Fellowship. This research was enabled in part by support provided by the Digital Research Alliance of Canada (alliancecan.ca).

\section*{Data Availability}
The data underlying this article are available at \href{https://doi.org/10.6084/m9.figshare.25464109}{doi.org/10.6084/m9.figshare.25464109}.





\bibliographystyle{mnras}
\bibliography{plnt_formn}




\appendix

\section{Effects of errors in the fitted disc geometry} \label{sec:errors}

In our analysis of the synthetic observations, we assume that the disc phase centre, inclination and position angle are known. However, incorrect determination of these parameters can produce asymmetry artefacts that could potentially disrupt spiral recovery. Therefore, here we briefly examine how such errors manifest in the $\Delta \chi^2$ heatmaps (see Appendix~A of \citet{andrewsLimitsMillimeter2021} for a similar discussion for image residuals) and how they affect spiral recovery.

\begin{figure}
    \includegraphics[width=\columnwidth]{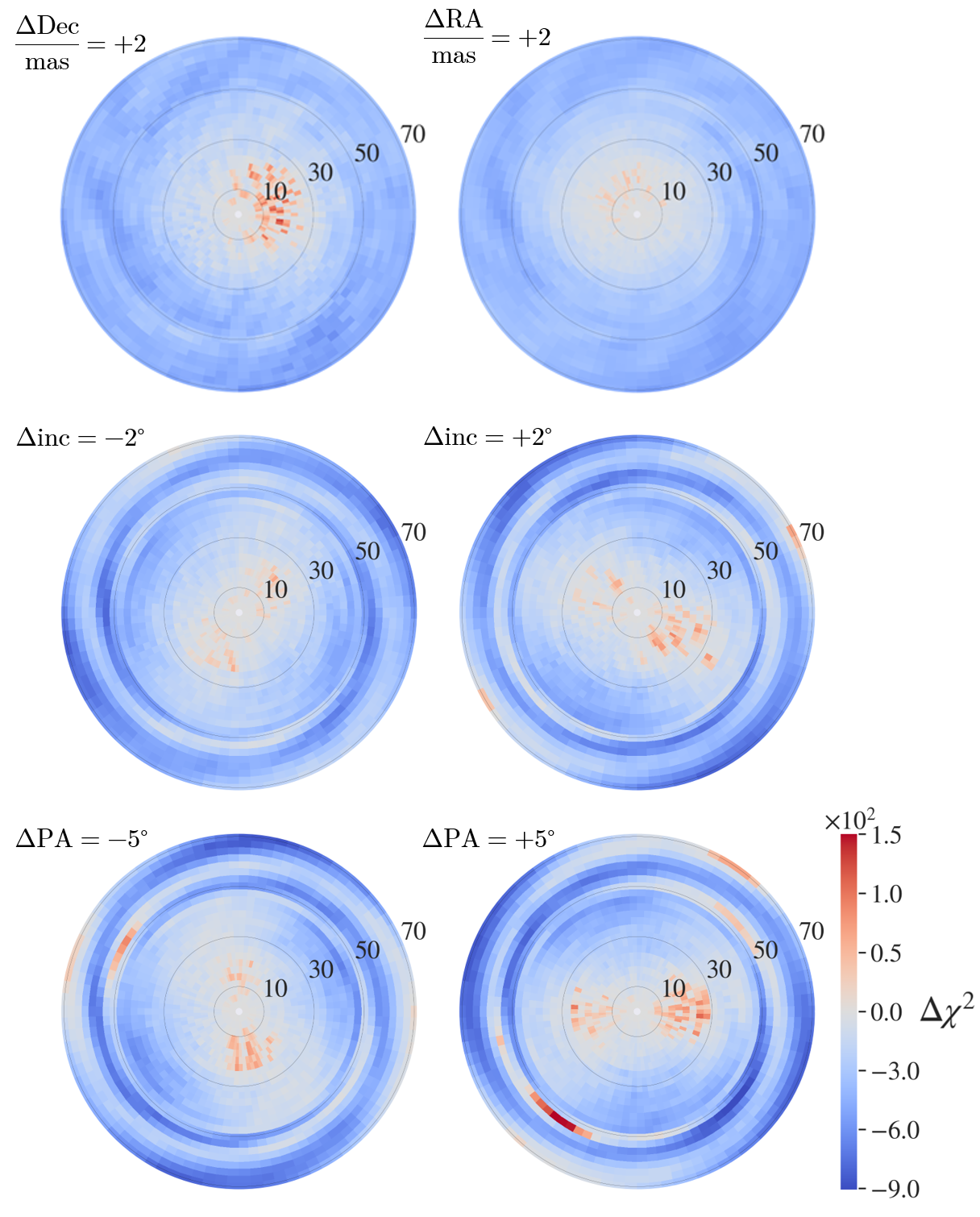}
    \caption{$\Delta\chi^2$ heatmaps for a ($\beta=10$, $\tau_0=1$, $M_\text{p}=0.3\,M_\text{th}$) spiral template applied to a ($\beta=10$, $\tau_0=1$) planet-less disc. The disc is fitted with deviations from its correct geometric parameters (in the text) noted in the upper left corner of each panel. The colourbar in the bottom right applies to all the heatmaps.}
    \label{fig:errors_zero}
\end{figure}

\begin{figure}
    \includegraphics[width=\columnwidth]{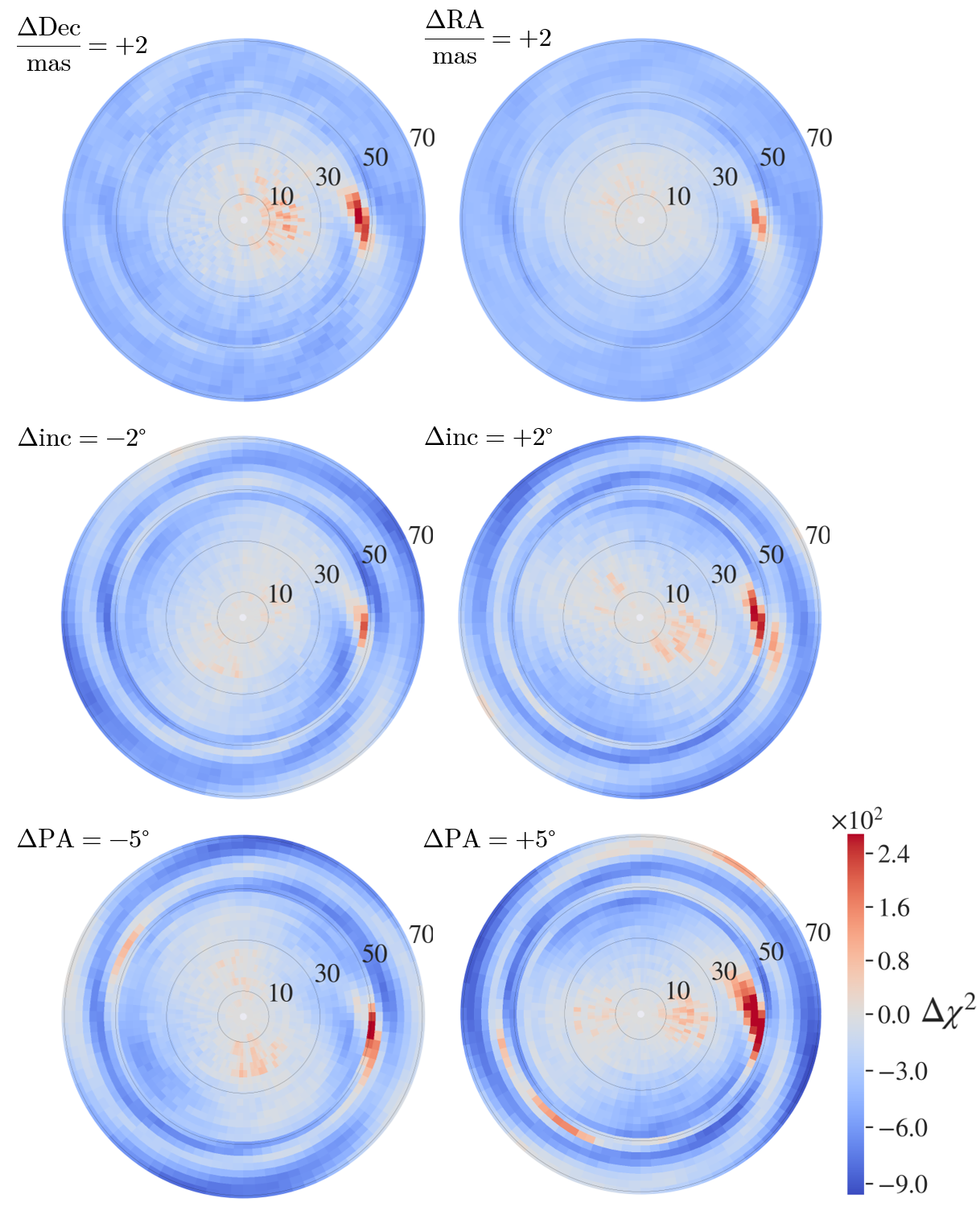}
    \caption{$\Delta\chi^2$ heatmaps as in Figure~\ref{fig:errors_zero}, except the disc contains a $0.3\,M_\text{th}$ planet.}
    \label{fig:errors_03Mth}
\end{figure}

We consider two example synthetic observations of a disc with geometric parameters $(\Delta \text{Dec}, \Delta \text{RA})=(0\,\text{mas},0\,\text{mas})$, $\text{inc.}=30^\circ$, $\text{PA}=90^\circ$. Both discs have $\beta=10$, $\tau_0=1$ and are observed with our fiducial set-up. One contains no planet, and the other contains a $0.3\,M_\text{th}$ planet (at $r_\text{p} = 50\,\text{au}$, $\phi_\text{p} = 0^\circ$ [along the projected disc's minor axis], and producing a right-handed spiral). We attempt to fit both with a $\beta=10$, $\tau_0=1$, $M_\text{p}=0.3\,M_\text{th}$ spiral template (i.e., the template matching the parameters of the planet-containing disc). The $\Delta\chi^2$ heatmaps for the planet-less disc observation allow us to see the false fitting ($\Delta\text{BIC}>0$ planet positions) that errors in disc geometry can introduce (Fig.~\ref{fig:errors_zero}), and those of the planet-containing disc allow us to see how this impacts recovery for our smallest planet case (Fig.~\ref{fig:errors_03Mth}). We expect the recovery of any larger mass planets to be less affected.

First we note that the effect on spiral recovery of the disc's inclination is small: compared to the face-on version, (Fig.~\ref{fig:C4C7_adis_R_combined}) the peak $\Delta\chi^2$ is reduced by $\sim 10\,\%$, and the $\Delta \chi^2$ heatmaps appear structurally similar. The effect is larger for higher inclinations, and is reduced if the planet is closer to the major axis of the deprojected disc. For errors in disc geometry, we find that:

\begin{itemize}
    \item False fitting due to phase centre errors is most prominent in the inner disc. We see more significant false fitting for errors in $\Delta\text{Dec}$ than $\Delta\text{RA}$. This is likely due to the projected disc's minor axis being aligned with the Declination axis in this example, meaning that errors in $\Delta\text{Dec}$ represent a larger fraction of the projected disc radius compared to errors in $\Delta\text{RA}$, which are aligned with the major axis. 

    \item False fitting due to errors in inclination and position angle also appear prominently in the inner disc, though can additionally appear as streaks in the rest of the disc. These streaks are perhaps the most problematic for spiral recovery as they somewhat resemble a spiral signal. However, the other characteristics of these errors (multiple streaks, two clusters of inner disc fits) should help distinguish them from a true spiral signal. We generally see more significant false fitting for errors in inclination than position angle.

    \item Phase centre errors give one cluster of false fits on one side of the assumed disc centre, whereas orientation errors give two clusters on opposite sides of the disc centre, which can help distinguish between them.

    \item For the given errors in disc geometry, the $0.3\,M_{\text{th}}$ planet is still clearly recoverable by its spiral signature. They serve as a cautious (and usually readily achievable) upper limits for attempting spiral detection with our approach.
\end{itemize}

\section{The visibility-space effect of adding a spiral to an axisymmetric disc} 
\label{Sec:Effect of adding a spiral}

To explore how strong of an effect planet-driven spirals can have on visibilities, we try adding them to discs that are perfectly axisymmetric, but projected and with realistic radial structure. We obtain such discs from the axisymmetric models of our Taurus discs. Specifically, we take the visibilities of the Axisymmetric disc model ($A$) and the Spiral model instance ($S$) and quantify their difference using $\chi_{SA}^2  = \sum_k w_k \lvert V_{P,k} - V_{Q,k} \rvert^2$. $\chi_{SA}^2$ is always a positive quantity, with its magnitude representing the strength of effect on the visibilities the spiral is having.

\begin{figure}
	\includegraphics[width=\columnwidth]{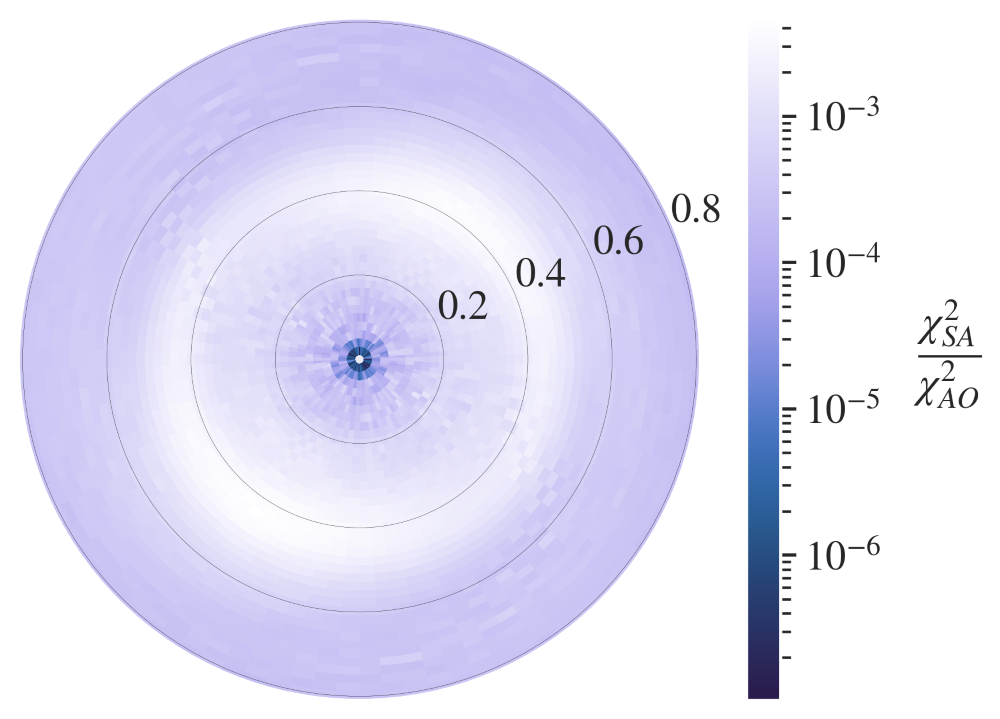}
    \caption{\( \chi_{SA}^2 \) heatmap, in units of \( \chi_{AO}^2 \), of RY Tau (\( \text{inclination} = 65.0^\circ \)) for a $\beta=10$, $\tau_0=3$, \(M_\text{p} = 1\,M_{\text{th}} \), right-handed spiral template. The heatmap spatially represents the deprojected disc, with the disc's minor axis projected onto the horizontal axis ($\phi_\text{p} = 0$). Radial ticks are in arcseconds.}
    \label{fig:purple}
\end{figure}

We calculate $\chi_{SA}^2$ across four of the spiral template parameters, neglecting to vary $\phi_\text{p}$ or spiral handedness as we expect these to have little impact on the results due to the axisymmetry of the disc. 
To justify this neglect, we must consider that the disc orientation will affect the strength of expression of a spiral from a planet placed at different azimuthal positions in the disc. 
To demonstrate the magnitude of this effect, we include an example $\chi_{SA}^2$ heatmap for RY Tau (Fig.~\ref{fig:purple}), which is the most inclined disc in our Taurus sample. Even in this highly inclined case, the effect is small, with the heatmap appearing very close to axisymmetric. Therefore, in order to facilitate comparison between spiral templates, we let all spiral templates take $\phi_\text{p}=0$. We also let all spiral templates be right-handed as the effect of handedness should be negligible for the same reason.
We can then compare the effects of the remaining spiral model parameters: orbital radius, planet mass, optical depth, and cooling time. 

\begin{figure}
    \centering
         {\large $\tau_0=1$, $\beta=10$}\\[6pt]  
    \begin{subfigure}{\columnwidth}

        \includegraphics[width=\linewidth]{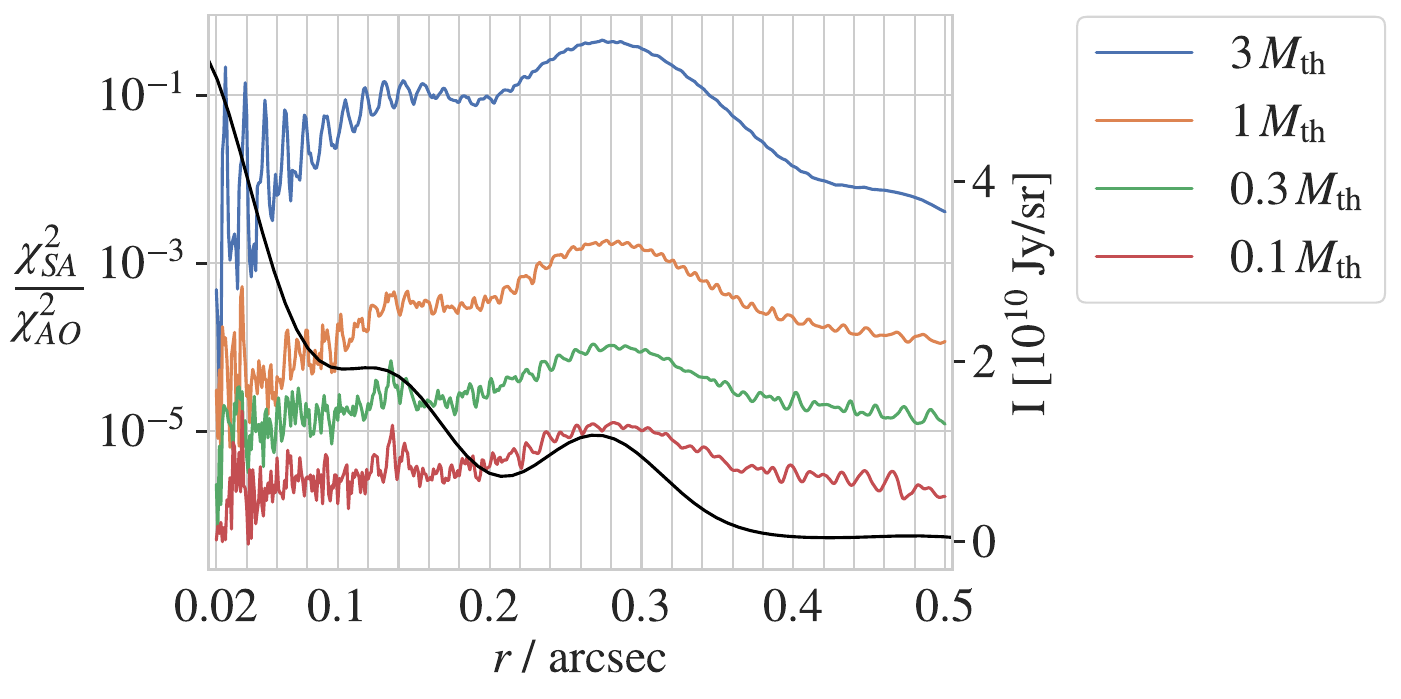}

    \end{subfigure}

    \vspace{0.4cm}
    
    {\large $M_\text{p}=1\,M_\text{th}$}\\[6pt]  
    \begin{subfigure}{\columnwidth}
            
        \includegraphics[width=\linewidth]{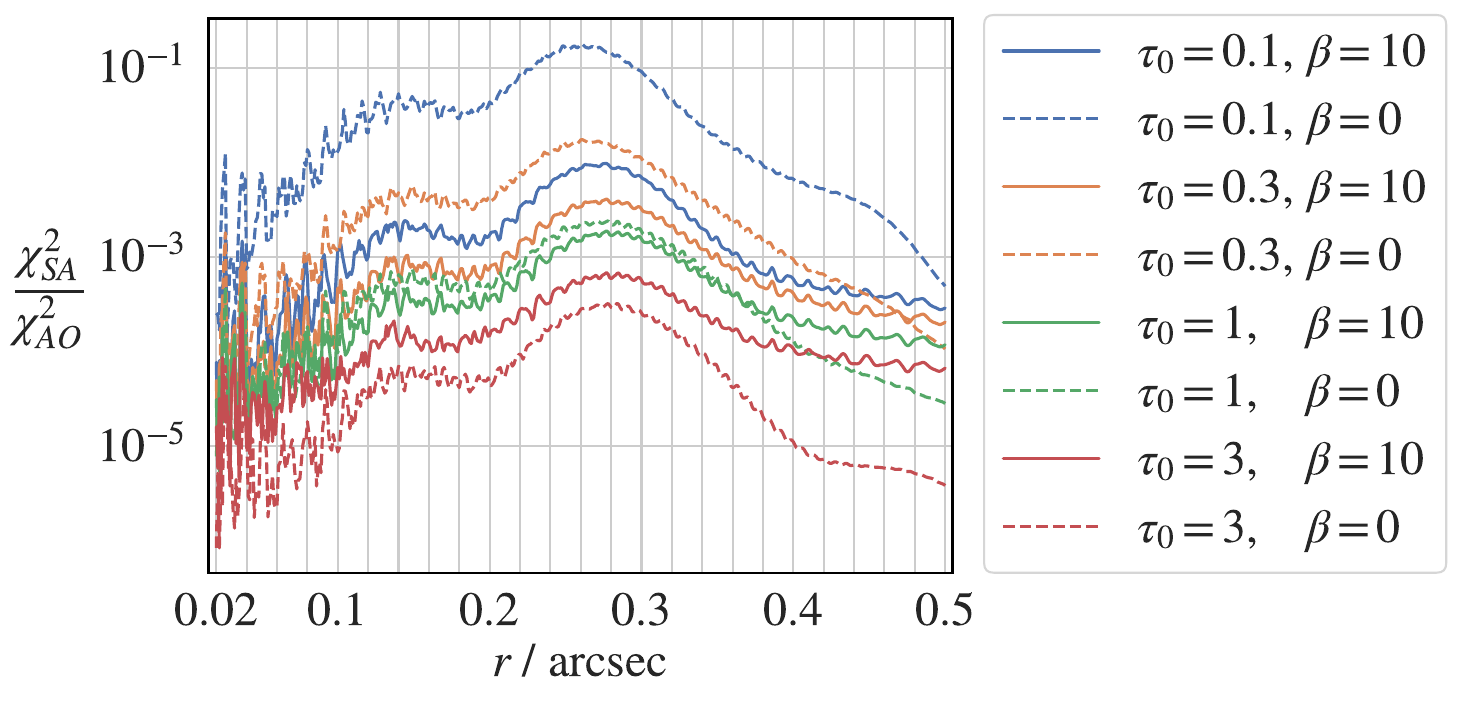}
        
    \end{subfigure}

    \caption{$\chi_{SA}^2$ plots, in units of $\chi_{AO}^2 = 2.48\times10^5$, of FT Tau for right-handed, $\phi_\text{p}=0$ spiral templates. Top: variation with planet mass for adiabatic, \( \tau_0=1 \) templates. The \texttt{frank} brightness profile also is shown for comparison. Bottom: variation with cooling time and optical depth for \(M_\text{p}=1\,M_{\text{th}} \) templates.}
    \label{fig:squiggles}

\end{figure}

We explore this in Figure~\ref{fig:squiggles} for FT Tau, a typical compact disc from our Taurus sample.  We see significant radial variations in $\chi_{SA}^2$, with intermediate radii having the greatest effect across the discs in our Taurus sample. This is unsurprising as large $r_\text{p}$ place the strongest parts of the spiral in faint reaches of the disc, and small $r_\text{p}$ limit the strongest parts of the spiral to a small area of the disc. We also see peaks and troughs that coincide with radii of elevated and reduced brightness respectively. 
This is to be expected because the spiral templates represent \textit{fractional} residual brightness maps; thus, placing the planet in a brighter region will lead to a stronger effect. At low $r_\text{p}$, $\chi_{SA}^2$ becomes increasingly noisy. This is likely due to the fact that the spiral is contributing on scales smaller than the angular resolution of the observations. This high $\chi^2$ variability at inner disc radii also appears in many of our $\Delta\chi^2$ heatmaps.

Of the simulation parameters, planet mass has the largest effect, with higher mass planets unsurprisingly having greater effects on the disc visibilities. Optically thinner spiral templates also show larger magnitude effects, which is opposite to what is seen in the corresponding image residuals (Fig.~6 of \citetalias{speedieSpirals2022}). 
The reason for this is that, in images, decreasing optical depth leads to a decrease in overall disc brightness, whereas, since our method incorporates the spiral as a fractional residual brightness perturbation, there is no change in overall disc brightness (the overall disc brightness is instead set by the axisymmetric \texttt{frank} model). Hence, only fractional change matters, and the fractional change is larger for optically thinner models. The radial structure of $\chi^2_{SA}$ changes little between different optical depths, consistent with its modest impact on spiral morphology. 

The isothermal templates tend to exhibit a slightly greater dependence on orbital radius than the (approximately) adiabatic ones. Additionally, the isothermal templates appear more sensitive to optical depth variations than the adiabatic ones, with the strongest spiral template being the $\tau_0=0.1$ isothermal one across most radii, and the weakest usually being the $\tau_0=3$ isothermal one. This is somewhat counterintuitive since the temperature contrast of the adiabatic templates generally results in a greater spiral brightness contrast than in the corresponding isothermal templates. 
However, the isothermal spiral simulations drive a deeper/wider gap in their discs and this manifests as a higher amplitude brightness perturbations around the planet's position in the spiral templates ($I_S \propto 1/\left<I_\text{sim}\right>_{\!\phi}$ [Eq.~\ref{eq:I_S}], and for a deep gap $\left<I_\text{sim}\right>_{\!\phi}$ is small).
This appears to have a greater effect than the temperature perturbation of the adiabatic templates for the optically thin cases, although this reverses as optical depth increases.

\section[Supplementary radial profiles and chi2 heatmaps]{Supplementary radial profiles and $\Delta\chi^2$ heatmaps} \label{sec:supplementary heatmaps}

\begin{figure*}
    \begin{tabular}[b]{@{}c@{}} 
        \rotatebox{90}{\parbox{4.4cm}{\centering\large $\tau_0=3$}} \\ [0.3cm]
        \rotatebox{90}{\parbox{4.4cm}{\centering\large $\tau_0=1$}} \\ [0.3cm]
        \rotatebox{90}{\parbox{4.0cm}{\centering\large $\tau_0=0.3$}} \\ 
        \rotatebox{90}{\parbox{5.2cm}{\centering\large $\tau_0=0.1$}} \\ 
        
    \end{tabular}%
    \begin{minipage}[b]{0.95\linewidth}
        \begin{minipage}{0.28\linewidth}
            \centering
            \hspace{1.4cm} \large$M_\text{p}=0.3\,M_\text{th}$
        \end{minipage}\hfill
        \begin{minipage}{0.28\linewidth}
            \centering
            \hspace{2.3cm} \large$M_\text{p}=1\,M_\text{th}$
        \end{minipage}\hfill
        \begin{minipage}{0.41\linewidth}
            \centering
            \large$M_\text{p}=3\,M_\text{th}$
            \hspace{-1.4cm}
        \end{minipage}

        \includegraphics[width=\linewidth]{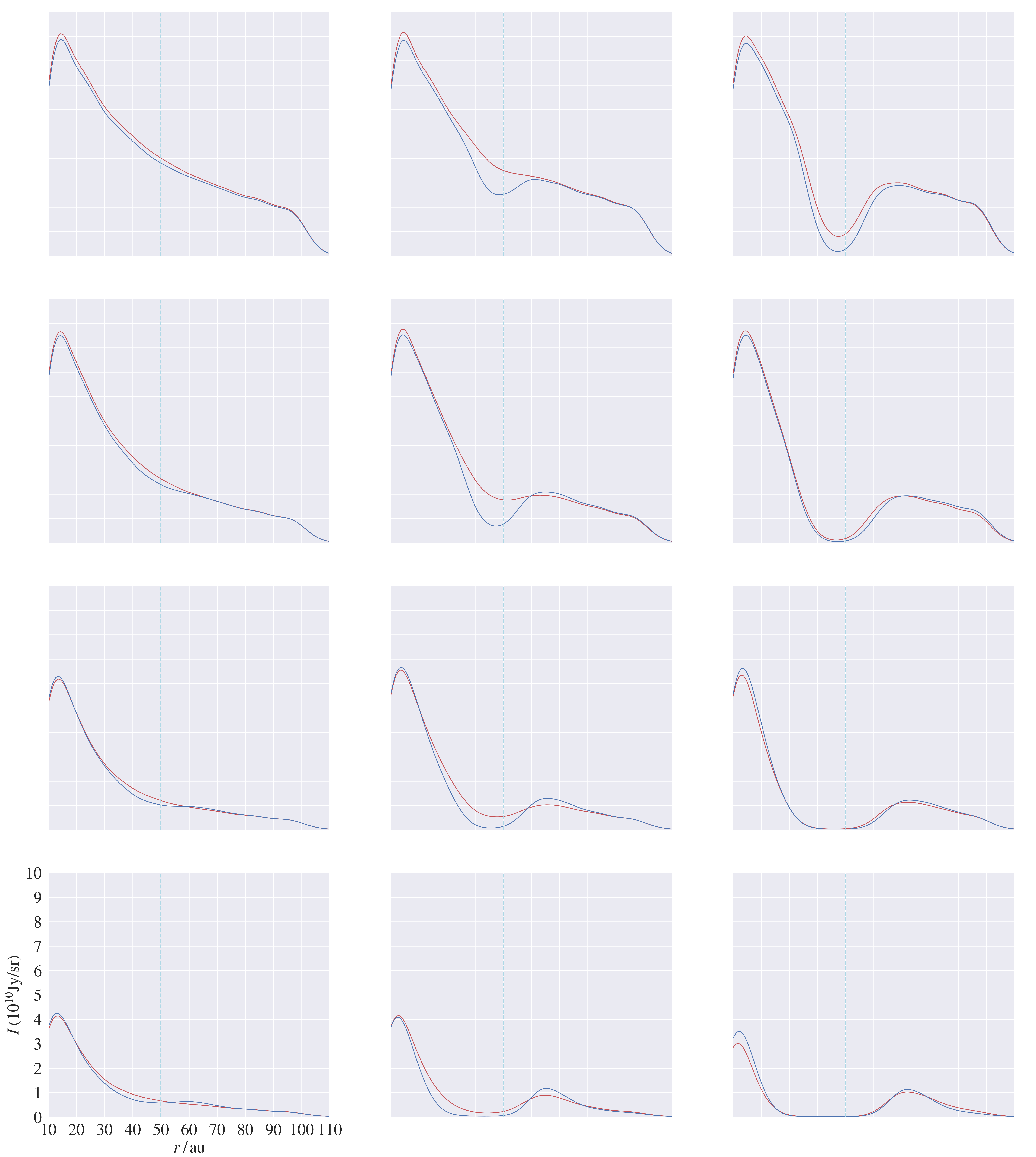}
    \end{minipage}

    \caption{Radial profiles for adiabatic ($\beta=10$) (red lines) and isothermal ($\beta=0$) (blue lines) model discs. The axes shown in the bottom left profile apply to all the profiles. The planet radial position (50\,au) is indicated by a blue dashed line. The gaps in the isothermal profiles are generally deeper than in the adiabatic profiles.
    $M_\text{p}/M_\text{th}=0.3, 1, 3$ from left to right, and $\tau_0=0.1, 0.3, 1, 3$ from bottom to top. The discs are observed with our fiducial set-up ($0.061''$ angular resolution and $35\,\mu\text{Jy/beam}$ sensitivity, from ALMA configuration pair C-4 + C-7 with 40\,mins of on-source time).}
    \label{fig:C4C7_gofishies}
\end{figure*}

\begin{figure*}
    \begin{tabular}[b]{@{}c@{}} 
        \rotatebox{90}{\parbox[t][0.5cm]{4.2cm}{\centering\large $\tau_0=3$}} \\ [0.35cm]
        \rotatebox{90}{\parbox[t][0.5cm]{4.2cm}{\centering\large $\tau_0=1$}} \\ [0.35cm]
        \rotatebox{90}{\parbox[t][0.5cm]{4.2cm}{\centering\large $\tau_0=0.3$}} \\ [0.2cm]
        \rotatebox{90}{\parbox[t][0.5cm]{4.4cm}{\centering\large $\tau_0=0.1$}} \\ 
        
    \end{tabular}%
    \begin{minipage}[b]{0.95\linewidth}
        \begin{minipage}{0.27\linewidth}
            \centering
            \hspace{-.1cm} \large$M_\text{p}=0.3\,M_\text{th}$
        \end{minipage}\hfill
        \begin{minipage}{0.3\linewidth}
            \centering
            \hspace{1.1cm} \large$M_\text{p}=1\,M_\text{th}$
        \end{minipage}\hfill
        \begin{minipage}{0.43\linewidth}
            \centering
            \large$M_\text{p}=3\,M_\text{th}$
            \hspace{-0.2cm}
        \end{minipage}

        \includegraphics[width=\linewidth]{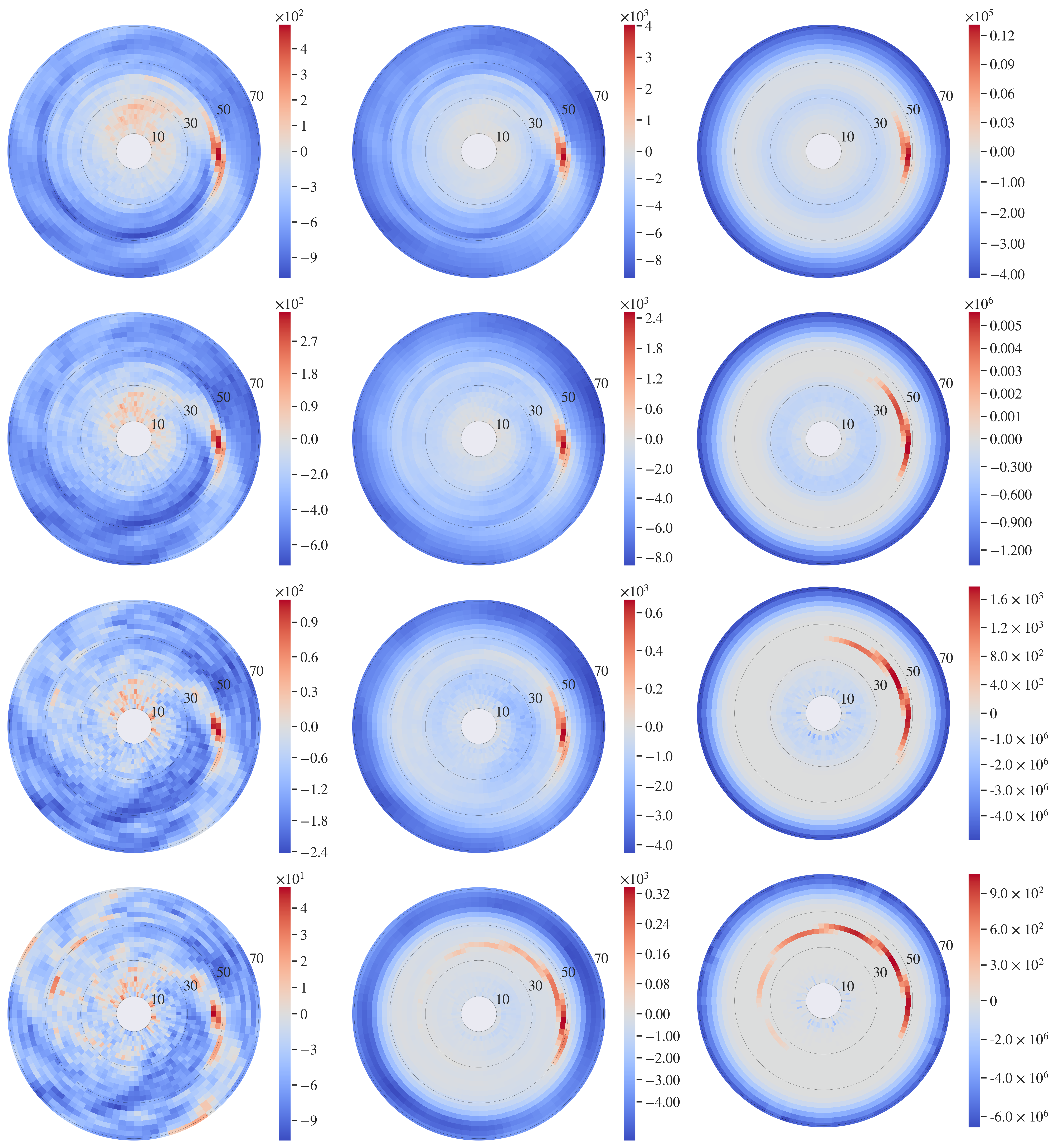}

    \end{minipage}
    \caption{$\Delta\chi^2$ heatmaps of adiabatic ($\beta=10$), right-handed model discs with their matching spiral templates. The discs host their planet at 50\,au along the right horizontal axis, and are observed with our fiducial set-up ($0.061''$ angular resolution and $35\,\mu\text{Jy/beam}$ sensitivity). For these observations, $\Delta\chi^2(\Delta\text{BIC}=0)=63$, as determined by the number of visibility points. All the adiabatic discs, except the $\tau_0=0.1, M_\text{p}=M_\text{th}$ case, achieve a maximum $\Delta\chi^2$ significantly larger than this, indicating a very strong preference for the spiral model.} 
    \label{fig:C4C7_adis_R_combined}
\end{figure*}

\begin{figure*}
    \begin{tabular}[b]{@{}c@{}} 
        \rotatebox{90}{\parbox[t][0.5cm]{4.2cm}{\centering\large $\tau_0=3$}} \\ [0.35cm]
        \rotatebox{90}{\parbox[t][0.5cm]{4.2cm}{\centering\large $\tau_0=1$}} \\ [0.35cm]
        \rotatebox{90}{\parbox[t][0.5cm]{4.2cm}{\centering\large $\tau_0=0.3$}} \\ [0.2cm]
        \rotatebox{90}{\parbox[t][0.5cm]{4.4cm}{\centering\large $\tau_0=0.1$}} \\

    \end{tabular}%
    \begin{minipage}[b]{0.95\linewidth}
        \begin{minipage}{0.27\linewidth}
            \centering
            \hspace{-.1cm} \large$M_\text{p}=0.3\,M_\text{th}$
        \end{minipage}\hfill
        \begin{minipage}{0.3\linewidth}
            \centering
            \hspace{1.1cm} \large$M_\text{p}=1\,M_\text{th}$
        \end{minipage}\hfill
        \begin{minipage}{0.43\linewidth}
            \centering
            \large$M_\text{p}=3\,M_\text{th}$
            \hspace{-0.2cm}
        \end{minipage}

        \includegraphics[width=\linewidth]{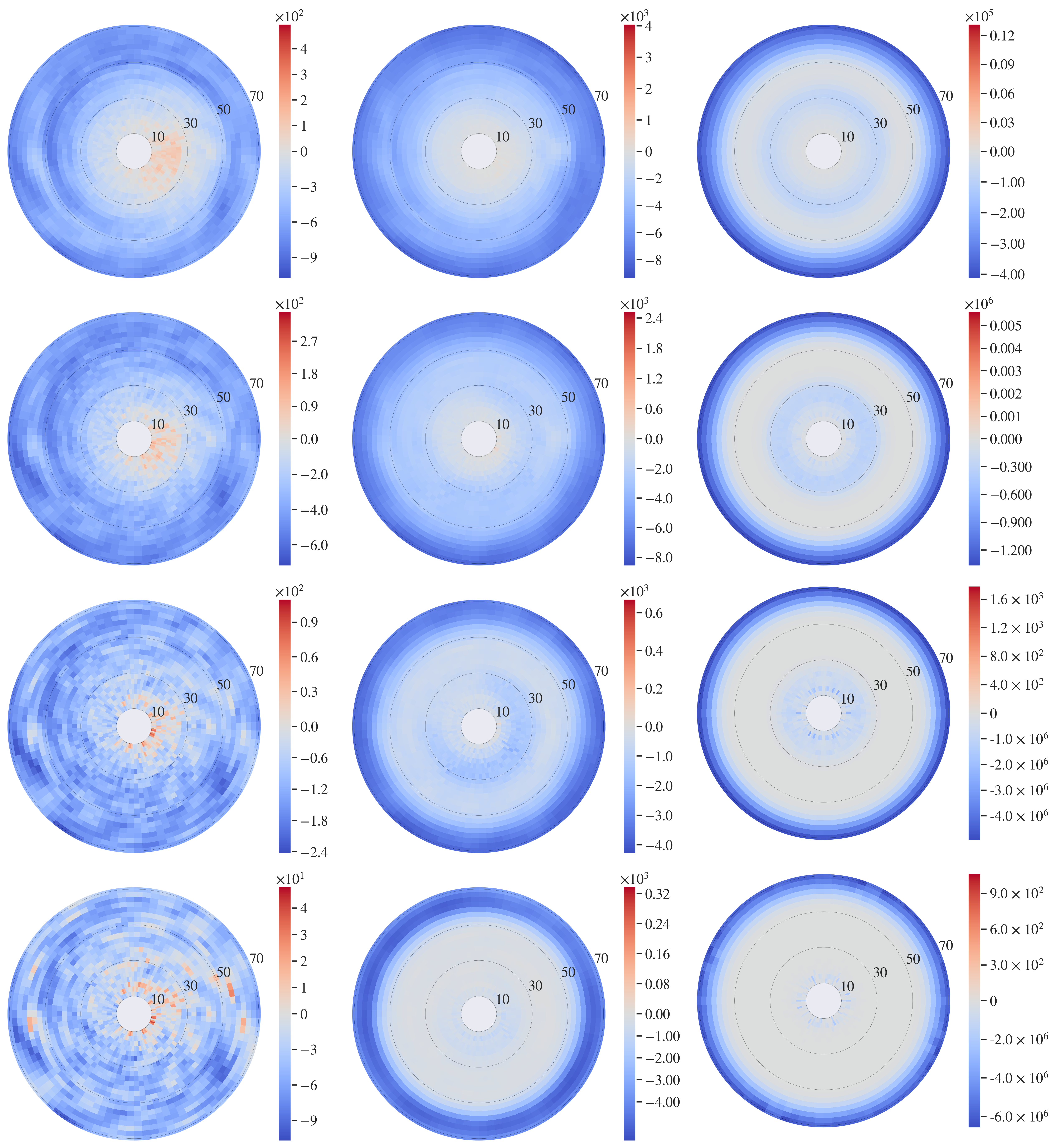}

    \end{minipage}
    \caption{$\Delta\chi^2$ heatmaps of adiabatic ($\beta=10$), right-handed model discs with their left-handed, but otherwise matching, spiral templates. Because these spirals are of the wrong handedness, low $\Delta\chi^2>0$ values (compared to the corresponding right-handed template in Fig.~\ref{fig:C4C7_adis_R_combined}) are interpreted as giving a clearer spiral recovery. The colourbars for each heatmap are the same as for the corresponding right-handed template in Figure~\ref{fig:C4C7_adis_R_combined} to aid comparison.} 
    \label{fig:C4C7_adis_L_combined}
\end{figure*}

\begin{figure*}
    \begin{tabular}[b]{@{}c@{}} 
        \rotatebox{90}{\parbox[t][0.5cm]{4.1cm}{\centering\large $\tau_0=3$}} \\ [0.4cm]
        \rotatebox{90}{\parbox[t][0.5cm]{4.1cm}{\centering\large $\tau_0=1$}} \\ [0.43cm]
        \rotatebox{90}{\parbox[t][0.5cm]{4.0cm}{\centering\large $\tau_0=0.3$}} \\ [0.22cm]
        \rotatebox{90}{\parbox[t][0.5cm]{4.6cm}{\centering\large $\tau_0=0.1$}} \\ 
        
    \end{tabular}%
    \begin{minipage}[b]{0.95\linewidth}
        \begin{minipage}{0.27\linewidth}
            \centering
            \hspace{-.1cm} \large$M_\text{p}=0.3\,M_\text{th}$
        \end{minipage}\hfill
        \begin{minipage}{0.3\linewidth}
            \centering
            \hspace{1.1cm} \large$M_\text{p}=1\,M_\text{th}$
        \end{minipage}\hfill
        \begin{minipage}{0.43\linewidth}
            \centering
            \large$M_\text{p}=3\,M_\text{th}$
            \hspace{-0.2cm}
        \end{minipage}

        \includegraphics[width=\linewidth]{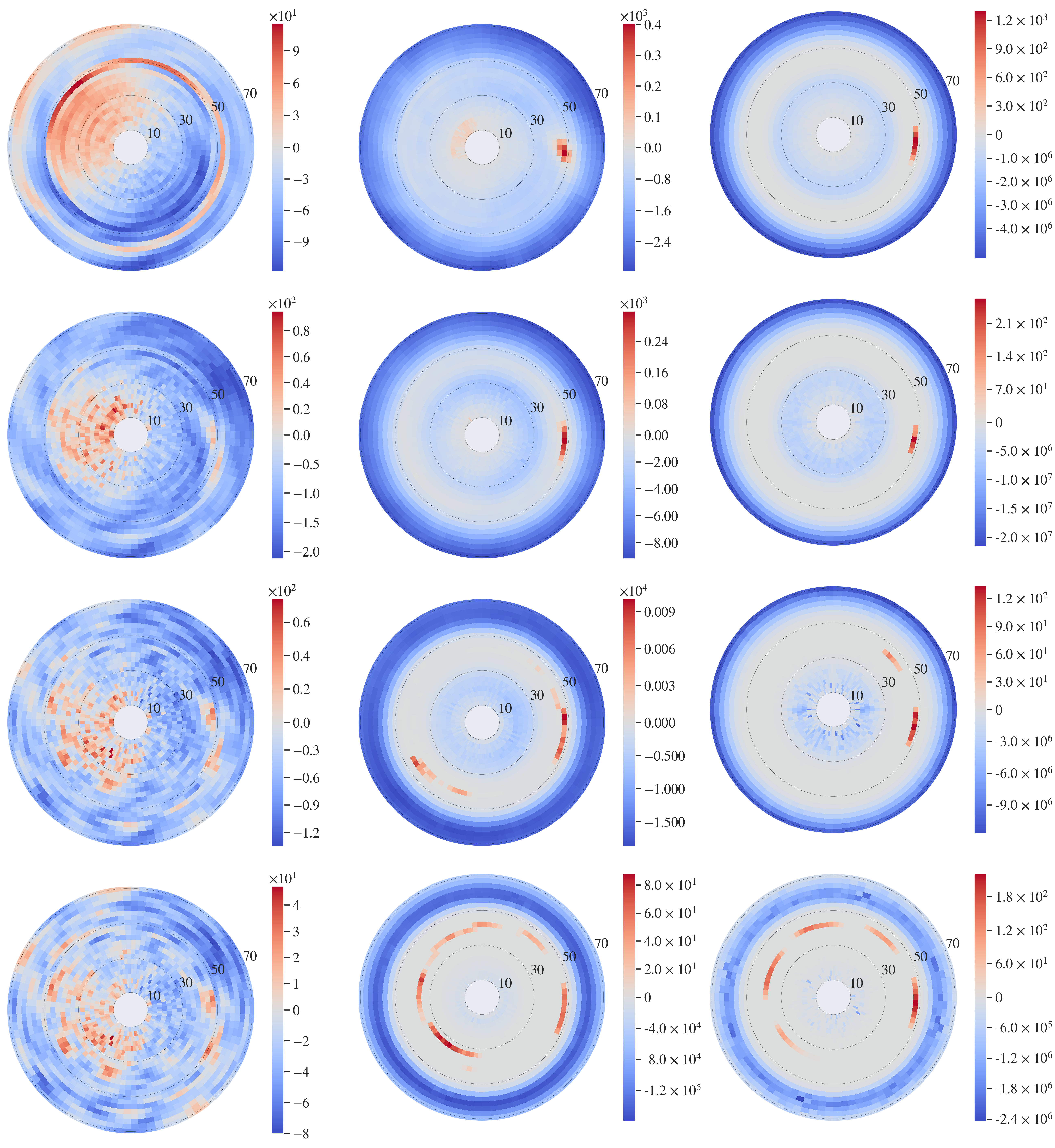}

    \end{minipage}
    \caption{$\Delta\chi^2$ heatmaps of isothermal ($\beta=0$), right-handed model discs with their matching spiral templates. Like in Figure~\ref{fig:C4C7_adis_R_combined}, the discs host their planet at 50\,au along the right horizontal axis, and are observed with our fiducial set-up ($0.061''$ angular resolution and $35\,\mu\text{Jy/beam}$ sensitivity).}
    \label{fig:C4C7_isos_R_combined}
\end{figure*}

\begin{figure*}
    \begin{tabular}[b]{@{}c@{}} 
        \rotatebox{90}{\parbox[t][0.5cm]{4.2cm}{\centering\large $\tau_0=3$}} \\ [0.35cm]
        \rotatebox{90}{\parbox[t][0.5cm]{4.2cm}{\centering\large $\tau_0=1$}} \\ [0.35cm]
        \rotatebox{90}{\parbox[t][0.5cm]{4.2cm}{\centering\large $\tau_0=0.3$}} \\ [0.2cm]
        \rotatebox{90}{\parbox[t][0.5cm]{4.4cm}{\centering\large $\tau_0=0.1$}} \\ 
        
    \end{tabular}%
    \begin{minipage}[b]{0.95\linewidth}
        \begin{minipage}{0.27\linewidth}
            \centering
            \hspace{-.1cm} \large$M_\text{p}=0.3\,M_\text{th}$
        \end{minipage}\hfill
        \begin{minipage}{0.3\linewidth}
            \centering
            \hspace{1.1cm} \large$M_\text{p}=1\,M_\text{th}$
        \end{minipage}\hfill
        \begin{minipage}{0.43\linewidth}
            \centering
            \large$M_\text{p}=3\,M_\text{th}$
            \hspace{-0.2cm}
        \end{minipage}
        
        \includegraphics[width=\linewidth]{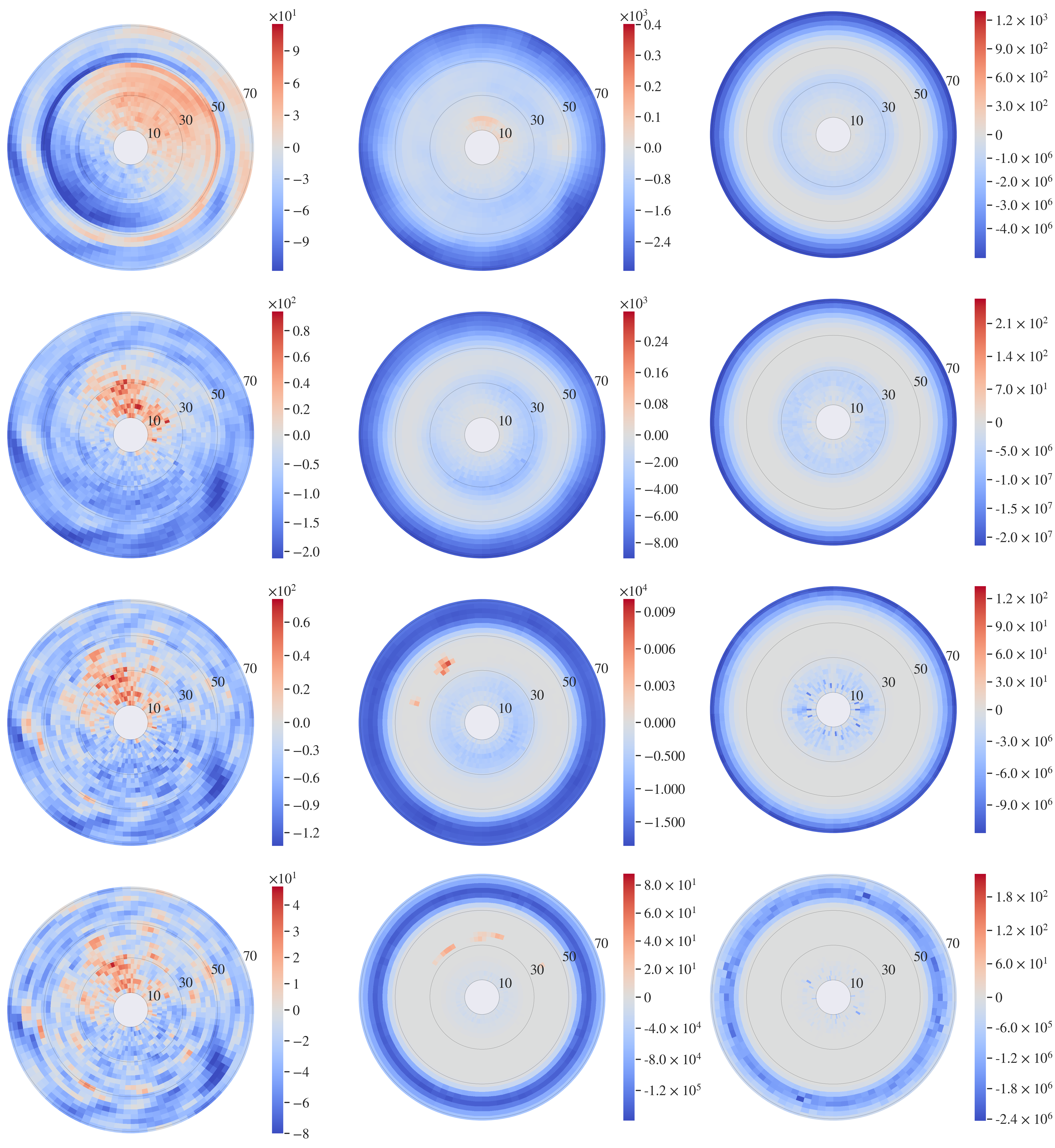}

    \end{minipage}
    \caption{$\Delta\chi^2$ heatmaps of isothermal ($\beta=0$), right-handed model discs with their left-handed, but otherwise matching, spiral templates. Because these spirals are of the wrong handedness, low $\Delta\chi^2>0$ values (compared to the corresponding right-handed template in Fig.~\ref{fig:C4C7_isos_R_combined}) are interpreted as giving a clearer spiral recovery. The colourbars for each heatmap are the same as for the corresponding right-handed template in Figure~\ref{fig:C4C7_isos_R_combined} to aid comparison.}
    \label{fig:C4C7_isos_L_combined}
\end{figure*}

For our fiducial set of synthetic observations ($0.061''$ resolution and $35\,\mu\text{Jy/beam}$ sensitivity), we include here the full set of \texttt{CLEAN} image profiles (Fig.~\ref{fig:C4C7_gofishies}) used in determining the gap recovery grids, and $\Delta\chi^2$ heatmaps (Fig.~\ref{fig:C4C7_adis_R_combined}-\ref{fig:C4C7_isos_L_combined}) used in determining the spiral recovery grids (Fig.~\ref{fig:combined_figure}). 

All the model discs contain right-handed spirals; therefore, for the right-handed spiral templates (Figs.~\ref{fig:C4C7_adis_R_combined}, \ref{fig:C4C7_isos_R_combined}), a positive $\Delta\chi^2$ near $r_\text{p}=50\,\text{au}$, $\phi_\text{p}=0$ is interpreted as a true signal for the planet. By contrast, the left-handed templates (Figs.~\ref{fig:C4C7_adis_L_combined}, \ref{fig:C4C7_isos_L_combined}) serve as handedness tests for spiral recovery, and low $\Delta\chi^2$ values (compared to the corresponding right-handed template) are interpreted as giving a clearer spiral recovery, particularly for regions near the peak $\Delta\chi^2$ in the right-handed templates.

A `perfect' spiral signal is a `spot' around the planet position showing a significantly higher $\Delta\chi^2$ than anywhere else. However, since it is possible to partially fit the true spiral with an incorrect spiral, we tolerate some degree of false fitting ($\Delta\text{BIC}>0$) with the left-handed templates and incorrectly positioned right-handed templates. We interpret this false fitting as not significantly impeding spiral recovery if its spatial structure is clearly not noise-like and not consistent with other asymmetries. 
For instance, some heatmaps have false fitting that traces out part of a spiral arm, such as for the optically thin, high planet mass templates towards the bottom right of Figures~\ref{fig:C4C7_adis_R_combined} and \ref{fig:C4C7_isos_R_combined}. Whereas, for the $M_\text{p}=0.3\,M_\text{th}$ isothermal templates in the left column of Figure~\ref{fig:C4C7_isos_R_combined}, although there is clearly some spatial structure, it is not clearly differentiable from noise or non-spiral asymmetries (the $\tau_0=3$ case is interpreted as a marginal recovery, owing to the right-handed template's notably higher peak $\Delta\chi^2$ than the left-handed one).

\bsp	
\label{lastpage}
\end{document}